\documentclass[fleqn,usenatbib]{mnras}

\usepackage[T1]{fontenc}

\DeclareRobustCommand{\VAN}[3]{#2}
\let\VANthebibliography\thebibliography
\def\thebibliography{\DeclareRobustCommand{\VAN}[3]{##3}\VANthebibliography}

\usepackage{multirow}
\usepackage{amssymb, amsmath}

\usepackage{graphicx}	% Including figure files
\usepackage{multicol}   % Multi-column entries in tables
\usepackage{bm}		    % Bold maths symbols, including upright Greek
\usepackage{pdflscape}	% Landscape pages
\usepackage{orcidlink}
\usepackage{newtxtext,newtxmath}

\usepackage{arydshln}
\usepackage{footmisc}
\usepackage{threeparttable}

\title[SNRs as seen by the QUIJOTE-MFI wide survey]{QUIJOTE scientific results -- XIII. 
Intensity and polarization study of supernova remnants in the QUIJOTE-MFI wide survey: 
CTB\,80, Cygnus Loop, HB\,21, CTA\,1, Tycho and HB\,9}

\author[L\'opez-Caraballo C. H. et al.]
{
{C. H. L\'opez-Caraballo$^{\orcidlink{0000-0002-6439-5385}}$,$^{1,2}$} \thanks{E-mail: clopez@iac.es, lopezcaraballoch@gmail.com}
{B. Ruiz-Granados$^{\orcidlink{0000-0003-3229-2725}}$,$^{3}$} %{B. Ruiz-Granados$^{\orcidlink{0000-0003-3229-2725}}$,$^{1,2,14}$}
{R.~T. G\'{e}nova-Santos$^{\orcidlink{0000-0001-5479-0034}}$,$^{1,2}$}
\newauthor
{M. Fern\'{a}ndez-Torreiro$^{\orcidlink{0000-0002-6805-9100}}$,$^{1,2}$}
{J.~A. Rubi\~{n}o-Mart\'{\i}n$^{\orcidlink{0000-0001-5289-3021}}$,$^{1,2}$}
{M.~W. Peel$^{\orcidlink{0000-0003-3412-2586}}$,$^{4,1,2}$}
{F. Poidevin$^{\orcidlink{0000-0002-5391-5568}}$,$^{1,2}$}
\newauthor
{E. Artal$^{\orcidlink{0000-0002-2569-1894}}$,$^{5}$} %{E. Artal$^{\orcidlink{0000-0002-2569-1894}}$,$^{5}$}
{M. Ashdown$^{\orcidlink{0000-0003-2044-7523}}$,$^{6,7}$} %{M. Ashdown$^{\orcidlink{0000-0003-2044-7523}}$,$^{6,7}$}
{R.~B. Barreiro$^{\orcidlink{0000-0002-6139-4272}}$,$^{8}$} % {R.~B. Barreiro$^{\orcidlink{0000-0002-6139-4272}}$,$^{8}$}
{F.~J. Casas$^{\orcidlink{0000-0002-2217-5843}}$,$^{8}$} %{F.~J. Casas$^{\orcidlink{0000-0002-2217-5843}}$,$^{8}$}
{E. de la Hoz$^{\orcidlink{0000-0002-5066-816X}}$,$^{8,9,10}$} %{E. de la Hoz$^{\orcidlink{0000-0002-5066-816X}}$,$^{8,9,10}$}
\newauthor
{R. Gonz\'alez-Gonz\'alez$^{1,2}$} %
{F. Guidi$^{\orcidlink{0000-0001-7593-3962}}$,$^{11}$} %{F. Guidi$^{\orcidlink{0000-0001-7593-3962}}$,$^{11}$}
{D. Herranz$^{\orcidlink{0000-0003-4540-1417}}$,$^{8}$} %{D. Herranz$^{\orcidlink{0000-0003-4540-1417}}$,$^{8}$}
{R. Hoyland$^{\orcidlink{0000-0001-5346-0519}}$,$^{1,2}$} %{R. Hoyland$^{\orcidlink{0000-0001-5346-0519}}$,$^{1,2}$}
{A. Lasenby$^{\orcidlink{0000-0002-8208-6332}}$,$^{6,7}$} %{A. Lasenby$^{\orcidlink{0000-0002-8208-6332}}$,$^{6,7}$}
\newauthor
{E. Mart\'{i}nez-Gonz\'{a}lez$^{\orcidlink{0000-0002-0179-8590}}$,$^{8}$} %{E. Mart\'{i}nez-Gonzalez$^{\orcidlink{0000-0002-0179-8590}}$,$^{8}$}
L. Piccirillo,$^{12}$
{R. Rebolo$^{\orcidlink{0000-0003-3767-7085}}$,$^{1,2,13}$} %{R. Rebolo$^{\orcidlink{0000-0003-3767-7085}}$,$^{1,2,11}$}
D. Tramonte,$^{14,1,2}$
F. Vansyngel,$^{1,2}$
\newauthor
{P. Vielva$^{\orcidlink{0000-0003-0051-272X}}$,$^{8}$} %{P. Vielva$^{\orcidlink{0000-0003-0051-272X}}$,$^{8}$}
{R.~A. Watson$^{\orcidlink{0000-0002-5873-0124}}$,$^{12}$} %{R.~A. Watson$^{\orcidlink{0000-0002-5873-0124}}$,$^{12}$}
\\
$^{1}$Instituto de Astrof\'{i}sica de Canarias, E-38200 La Laguna, Tenerife, Spain\\
$^{2}$Departamento de Astrof\'{i}sica, Universidad de La Laguna, E-38206 La Laguna, Tenerife, Spain\\
$^{3}$Departamento de F\'{i}sica. Facultad de Ciencias. Universidad de C\'{o}rdoba. Campus de Rabanales, Edif. C2. Planta Baja.  E-14071 C\'{o}rdoba, Spain\\
$^{4}$
Imperial College London, Blackett Lab, Prince Consort Road, London SW7 2AZ, UK\\
$^{5}$Universidad de Cantabria, Departamento de Ingeniería de Comunicaciones, Edificio Ingenieria de Telecomunicación, \\ \quad Plaza de la Ciencia nº 1, 39005 Santander, Spain\\
$^{6}$Astrophysics Group, Cavendish Laboratory, University of Cambridge, 
J J Thomson Avenue, Cambridge CB3 0HE, UK\\
$^{7}$Kavli Institute for Cosmology, University of Cambridge, Madingley Road, Cambridge CB3 0HA, UK\\
$^{8}$Instituto de F\'{\i}sica de Cantabria (IFCA), CSIC-Univ. de Cantabria, Avda. los
Castros, s/n, E-39005 Santander, Spain\\
$^{9}$Departamento de F\'{\i}sica Moderna, Universidad de Cantabria,
Avda. de los Castros s/n, 39005 Santander, Spain\\
$^{10}$CNRS-UCB International Research Laboratory, Centre Pierre Binétruy, IRL2007, CPB-IN2P3, Berkeley, CA 94720, USA\\
$^{11}$Institut d'Astrophysique de Paris, UMR 7095, CNRS \& Sorbonne Universit\'e, 98 bis boulevard Arago, 75014 Paris, France\\
$^{12}$Jodrell Bank Centre for Astrophysics, Alan Turing Building, Department of Physics \& Astronomy, School of Natural Sciences,\\ \quad The University of Manchester, Oxford Road, Manchester, M13 9PL, U.K. \\
$^{13}$Consejo Superior de Investigaciones Cient\'{i}ficas, E-28006
Madrid, Spain\\
$^{14}$Department of Physics, Xi'an Jiaotong-Liverpool University, 111 Ren'ai Road, \\ \quad Suzhou Dushu Lake Science and Education Innovation District, Suzhou Industrial Park, Suzhou 215123, P.R. China.
}%
\date{Accepted XXX. Received YYY; in original form ZZZ}

\pubyear{2023}

\begin{document}
\label{firstpage}
\pagerange{\pageref{firstpage}--\pageref{lastpage}}
\maketitle

\begin{abstract}
We use the new QUIJOTE-MFI wide survey (11, 13, 17 and 19\,GHz) to produce spectral energy distributions (SEDs), on an angular scale of 1\,deg, of the  supernova remnants (SNRs) CTB\,80, Cygnus Loop, HB\,21, CTA\,1, Tycho and HB\,9. We provide new measurements of the polarized synchrotron radiation in the microwave range.
For each SNR, the intensity and polarization SEDs are obtained and modelled by combining QUIJOTE-MFI maps with ancillary data. In intensity, we confirm the curved power law spectra of CTB\,80 and HB\,21 with a break frequency $\nu_{\rm b}$ at 2.0$^{+1.2}_{-0.5}$\,GHz and 5.0$^{+1.2}_{-1.0}$\,GHz respectively; and spectral indices respectively below and above the spectral break of $-0.34\pm0.04$ and $-0.86\pm0.5$ for CTB\,80, and $-0.24\pm0.07$ and $-0.60\pm0.05$ for HB\,21.
In addition, we provide upper limits on the Anomalous Microwave Emission (AME), suggesting that the AME contribution is negligible towards these remnants.
From a simultaneous intensity and polarization fit, we recover synchrotron spectral indices as flat as $-0.24$, and the whole sample has a mean and scatter of $-0.44\pm0.12$. The polarization fractions have a mean and scatter of $6.1\pm1.9$\,\%. 
When combining our results with the measurements from other QUIJOTE studies of SNRs, we find that radio spectral indices are flatter for mature SNRs, and particularly flatter for CTB\,80 ($-0.24^{+0.07}_{-0.06}$) and HB\,21 ($-0.34^{+0.04}_{-0.03}$).
In addition,  the evolution of the spectral indices against the SNRs age is modelled with a power-law function, providing an exponent $-0.07\pm0.03$ and amplitude $-0.49\pm0.02$ (normalised at 10\,kyr), which are conservative with respect to previous studies of our Galaxy and  the Large Magellanic Cloud. 
\end{abstract}


%%%%%%%%%%%%%%%%%%%%%%%%%%%%%%%%%%%%%%%%%%%%%%%%%%

%%%%%%%%%%%%%%%%% BODY OF PAPER %%%%%%%%%%%%%%%%%%

\section{Introduction}
\label{sec:introduction}

In recent decades, observations devoted of the Cosmic Microwave Background \citep[CMB,][]{Penzias1965,Gamow1946} have allowed us to confirm the $\Lambda$--CDM paradigm successfully \citep{Corey1976,Lubin1983,Smoot1992,Fixsen1996} and have yielded cosmological parameters to the one percent level of precision \citep[][]{2020A&A...641A...1P,2013ApJS..208...20B}.
Naturally, characterization of physical mechanisms that act in the Galactic Interstellar Medium (ISM) has also deepened, resulting in increased knowledge of synchrotron radiation. Such mechanisms include cosmic-ray electrons accelerated by the Galactic magnetic 
field \citep{1965ARA&A...3..297G,1966ARA&A...4..245G}, free--free emission due to electron--ion collisions \citep{Oster1961,1961ApJ...134.1010O}, and thermal radiation originating in Galactic interstellar dust \citep{Hildebrand1983,Desert1990} and triggering the discovery of Anomalous Microwave Emission \citep[AME,][]{1996ApJ...464L...5K,1997ApJ...486L..23L}.
Despite this progress, Galactic emission remains a major hindrance for the study of CMB B-modes, which are induced by primordial gravitational waves and are hidden in CMB polarization patterns \citep{Starobinsky1979,Rubakov1982,Kamionkowski1997,Zaldarriaga1997}. A comprehensive characterization of Galactic emission on different scales is therefore mandatory in order to meet the challenge of current and upcoming studies of the CMB, as well as for improving our understanding of the ISM.
Supernova Remnants (SNRs) are Galactic regions of interest, not only because their microwave emission is dominated by one of the main Galactic foregrounds (synchrotron radiation), but also because AME detections towards some SNRs have been reported \citep[e.g.][]{2017MNRAS.464.4107G,mfi_widesurvey_w51}.
In this context, the new QUIJOTE-MFI Wide Survey \citep{mfi_widesurvey}, covering the range 10--20\,GHz, will provide new insight into intensity and polarization properties of supernova remnants and on the role played by AME in SNRs.

Observations of SNRs allow us to understand the physical process acting in these powerful astronomical objects and to model their evolution \citep[][and references therein]{2015A&ARv..23....3D,2020pesr.book.....V}.
The radio-to-microwave emission from SNRs is explained in terms of synchrotron emission from cosmic ray electrons spiralling around magnetic field lines, where the shock waves driven by the remnant could accelerate particles by diffusive shock acceleration \citep[DSA,][]{1978MNRAS.182..147B,1978ApJ...221L..29B}.
For energies between 10$^{10}$ and 10$^{20}$ eV, the cosmic-ray spectrum is described by a power law distribution $dN(E) = A E^{-\Gamma_{\rm p}}dE$, where $A$ and $E$ respectively represent a scale factor and the energy. $\Gamma_{\rm p}$ identifies the particle spectral index, which reaches typical values of $\Gamma_{\rm p} \approx$\,3 for Galactic cosmic rays \citep[e.g. see][]{1979rpa..book.....R,2020pesr.book.....V}.
The particle spectral index can be related to the radio spectral index via $\alpha_{\rm radio} = (1 - \Gamma_{\rm p})/2$ \citep[see e.g.,][]{1979rpa..book.....R}, where we adopt the convention $S_{\nu} \propto \nu^{\alpha_{\rm radio}}$ for the flux density spectrum. 
The DSA mechanism predicts $\alpha_{\rm radio}\approx-0.5$ for cases of strong compression shocks, as expected in SNRs. Radio observations of Galactic SNRs reveal an average radio 
spectral index of $-$0.5 for shell-type remnants \citep[Galactic SNRs catalog\footnote{\label{fn:green_catalog} Updated versions of the Galactic SNR catalog focusing on radio frequencies have been maintained by D. Green. The latest version is publicly available at VizieR \cite{2019yCat.7284....0G}: \url{http://vizier.cds.unistra.fr/viz-bin/VizieR?-source=VII/284}},][]{2019yCat.7284....0G}, although there is a significant spread requiring theoretical explanation \citep[][]{2015A&ARv..23....3D,2020pesr.book.....V}. 
Despite the spread in spectral indices \citep[]{2017ApJS..230....2B,2023arXiv230206593R}, observations suggest that young remnants are bright and have steep spectra (i.e.\ $\alpha$ from $-0.8$ to $-1$), while mature remnants are fainter with flatter spectral indices (i.e.\ \mbox{$\alpha \lesssim-0.5$)} \citep[e.g. see][]{1976MNRAS.174..267C,1962ApJ...135..661H,2019ApJ...874...50Z}. For the Large Magellanic Cloud (LMC), \cite{2017ApJS..230....2B} have established the correlation of the spectral index and age by using 29 LMC SNRs, including multi-epoch observations of the younger remnant SN1987A. The latter was crucial in confirming that evolved remnants have flatter spectral indices than younger ones.
To date, the debate about the evolution of spectral index of Galactic SNRs is still open and is mostly blurred by significant scatter of the spectral index distribution \citep[]{2017ApJS..230....2B,2023arXiv230206593R}.

Observations show lower polarization levels than expected. In an ideal case, the synchrotron emission is driven by a homogeneous magnetic field perfectly orientated with respect to the observer, yielding an intrinsic degree of polarization $\Pi_{o}$ that can be written in term of the radio spectral index $\alpha_{\rm radio}$ as follows \citep[e.g.,][]{1965ARA&A...3..297G}:
\begin{equation}
\Pi_{o}(\alpha_{\rm radio}) = \dfrac{1-\alpha_{\rm radio}}{5/3-\alpha_{\rm radio}},
\label{eq:intrinsic_pi_alpha}
\end{equation}
which is independent of the frequency. Therefore, for values of $\alpha=-0.5$ (predicted by the DSA) the maximum theoretical polarization degree is around 69\% towards SNRs.
However, owing to irregularities in the magnetic field, typical observed values range between 10 and 15\% \citep[e.g. see][]{2006A&A...457.1081K,Gao2011A&A...529A.159}, while only in some exceptional cases with remarkable homogeneity of the magnetic field does the observed polarization levels reach around 60\% \citep[e.g. SN\,1006,][]{2013AJ....145..104R}.
Low-angular resolution observations, e.g.\ QUIJOTE-MFI at scales of 1\,deg, are usually affected by beam depolarization, causing lower polarization fractions in SNRs; for instance, \cite{2017MNRAS.464.4107G} found values of around 15\% for a region containing W44.
In this context, there are a variety of effects dimming the polarization \citep[][]{1998MNRAS.299..189S}, spanning instrumental depolarization \citep[e.g.\ beam and bandwidth 
depolarizations,][]{1966ARA&A...4..245G} and physical conditions acting locally \citep[e.g. inhomogenious and/or non-uniform magnetic field,][]{1966MNRAS.133...67B,1973SvA....16..774S}.

Whereas the radio emission of SNRs is well explained by synchrotron radiation, the physical mechanism responsible for AME is still under discussion. Electric dipole radiation \citep[ED,][]{1998ApJ...508..157D} from very small ($N\leq 10^3$ atoms) rapidly rotating ($\sim1.5\times 10^{10} \mathrm{s}^{-1}$) dust grains in the ISM, known as \textit{spinning dust emission} (SD), appears to reproduce the observations well \citep[see the review by][and references therein]{2018NewAR..80....1D}. However, other physical mechanisms such as magnetic dipole (MD) emission from thermal fluctuations in magnetic dust grains \citep[][]{1999ApJ...512..740D,2013ApJ...765..159D} and thermal emission from amorphous dust grains 
\citep[][]{2020ApJ...900L..40N}, have been invoked to explain the intensity of AME behaviour.
The observational polarization properties of AME, however, are still being debated, the only constraints suggesting weak or negligible polarization \citep[]{2012AdAst2012E..40R,2018NewAR..80....1D}. Predictions from the SD yield a polarization fraction  $\Pi_{\rm AME}\leq$\,1\% at 20\,GHz \citep[for grains aligned via resonance paramagnetic relaxation,][]{2000ApJ...536L..15L}, and significantly unpolarized levels where there is quantum  alignment suppression \citep[$\ll$1\%,][]{2016ApJ...831...59D}.
To date the most restrictive upper limits fall in the range $\Pi_{\rm AME} \leq 1$--6\% \citep[see][]{2007ApJ...665..355K,2011MNRAS.418..888M,2011ApJ...729...25L,2011MNRAS.418L..35D,2012AdAst2012E..40R,2022arXiv220103530H}, the strongest being, QUIJOTE-MFI and WMAP data \citep[][]{2017MNRAS.464.4107G}, $\Pi_{\rm AME} \leq 0.2\%$ at 41\,GHz towards the molecular cloud W\,43. 
Therefore, the presence of AME in the line of sight towards SNRs represents an opportunity of better understanding AME properties.

The first sign of an AME contribution to SNRs was 
found in 3C\,396, a finding supported by a 7$\sigma$ excess 
 at 33\,GHz over the extrapolation of 
synchrotron emission from low frequencies 
\citep[at 9\,arcmin beam resolution,][]{2007MNRAS.377L..69S}.
Those claims were recently refuted using data on the 
same object from the Parkes 64\,m telescope, where the 
microwave spectra can be well understood in terms of 
flatter ($\alpha=-0.36\pm0.02$) synchrotron emission  
\citep{2016MNRAS.459.4224C}, with observations in the 
5--30\,GHz range being key to disentangling the AME, 
free--free and synchrotron contributions.
The AME in SNRs has been studied mainly in studies
characterizing the integrated spectral energy 
distribution on a scale of 1\,deg, where WMAP, 
\textit{Planck} and QUIJOTE data are considered.  
The integrated spectrum of the region dominated by  W44 
revealed a significant AME component at 6$\sigma$ 
\citep[][]{2017MNRAS.464.4107G}; this was the first 
remnant studied with QUIJOTE-MFI data. For W49 and W51, 
the AME contribution was detected at 4.7$\sigma$ and 
4.0$\sigma$ respectively \citep[][]{mfi_widesurvey_w51}, 
revealing the first AME detection for W51. 
They also provided evidence of no AME contribution 
for IC443 suggested by \cite{2019MNRAS.482.3857L} 
using observations on scales of few arc-minutes from the 
Sardinia Radio Telescope (SRT) at 1.6, 7.0, 21.4\,GHz and 
ancillary flux densities \citep[e.g. \textit{Planck} 
measurements at resolution $\lesssim$0.8\,deg,][]{2016A&A...586A.134P}.
In addition, \cite{2019MNRAS.482.3857L} ruled out the 
presence of AME for  W44, which was also supported by 
recent COMAP observations with similar beam resolutions 
\citep[around 4.5\,arcmin at 26--31\,GHz,][]{2022ApJ...933..187R},
having these higher resolutions than those of previous
studies with QUIJOTE-MFI, WMAP and \textit{Planck}.
There is still a controversy about the presence of AME 
towards SNRs, where interaction with the surrounding 
of remnants or the overlapping of different emissions in 
the line of sight are the topics under discussion.

\begin{figure*}
\begin{center}
\includegraphics[trim = 0cm 0cm 0cm 0cm,clip=true,width= 9. cm]{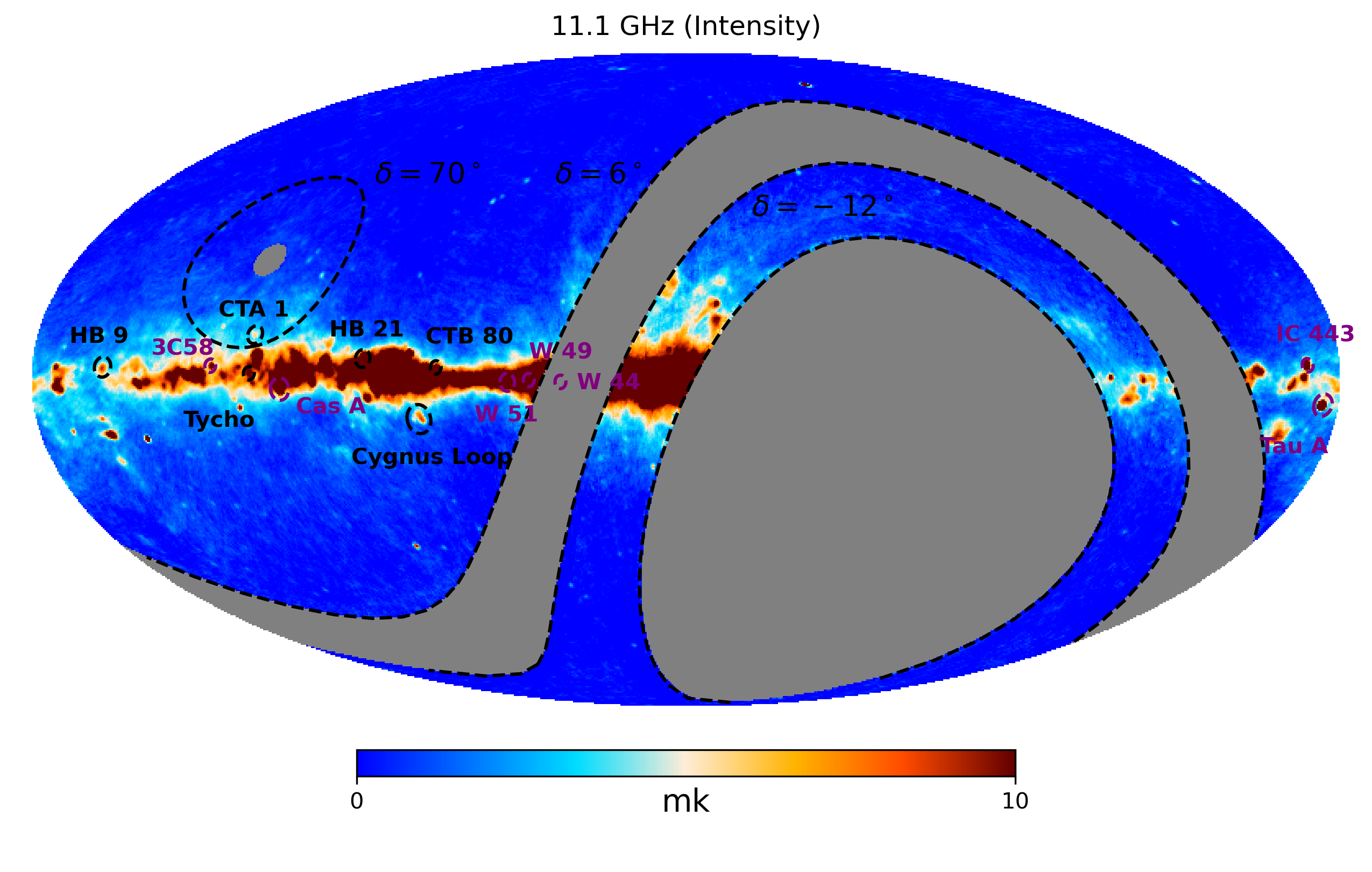}%
\includegraphics[trim = 0cm 0cm 0cm 0cm,clip=true,width= 9. cm]{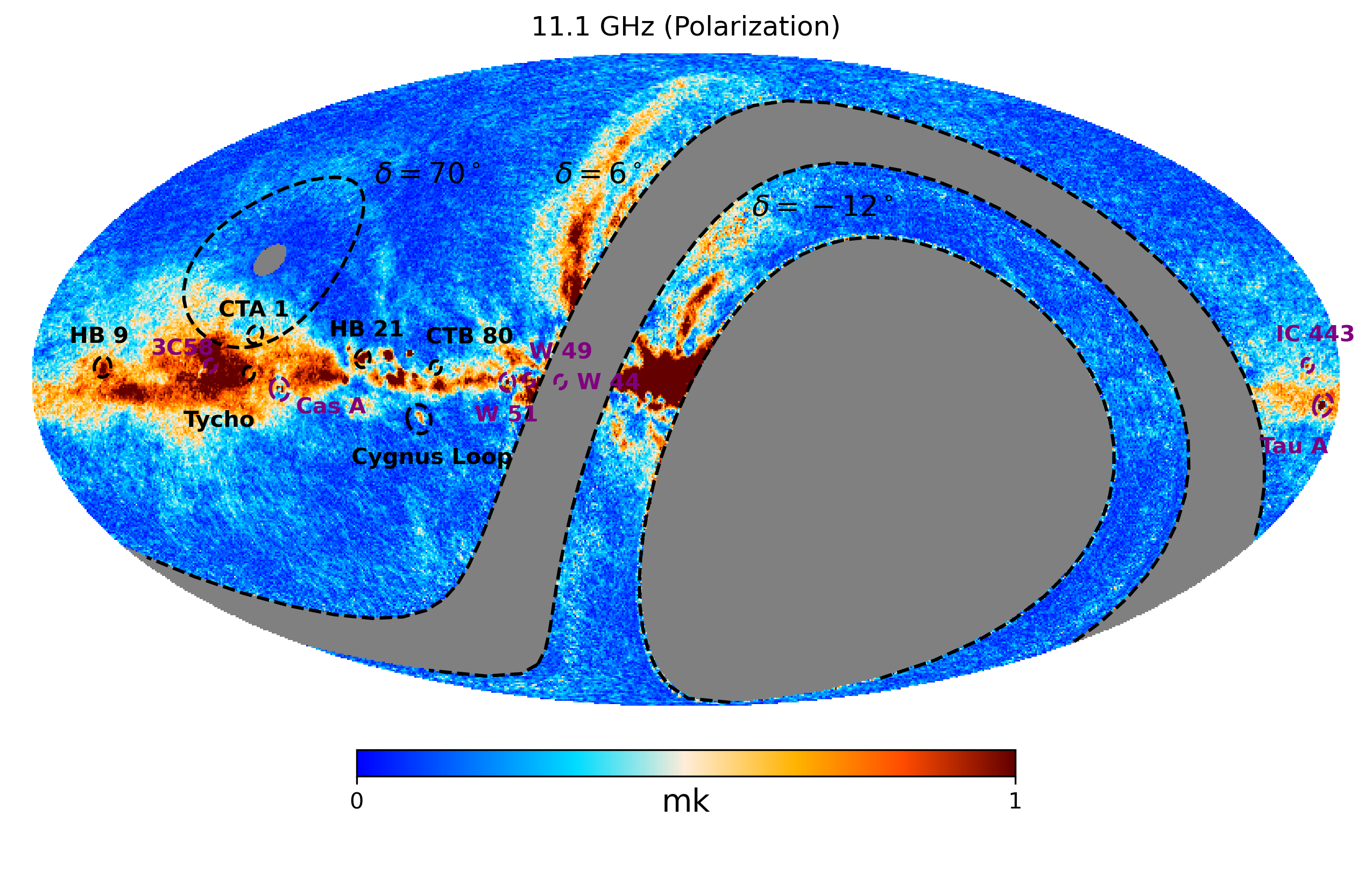}\\
\includegraphics[trim = 0cm 0cm 0cm 0cm,clip=true,width= 15 cm]{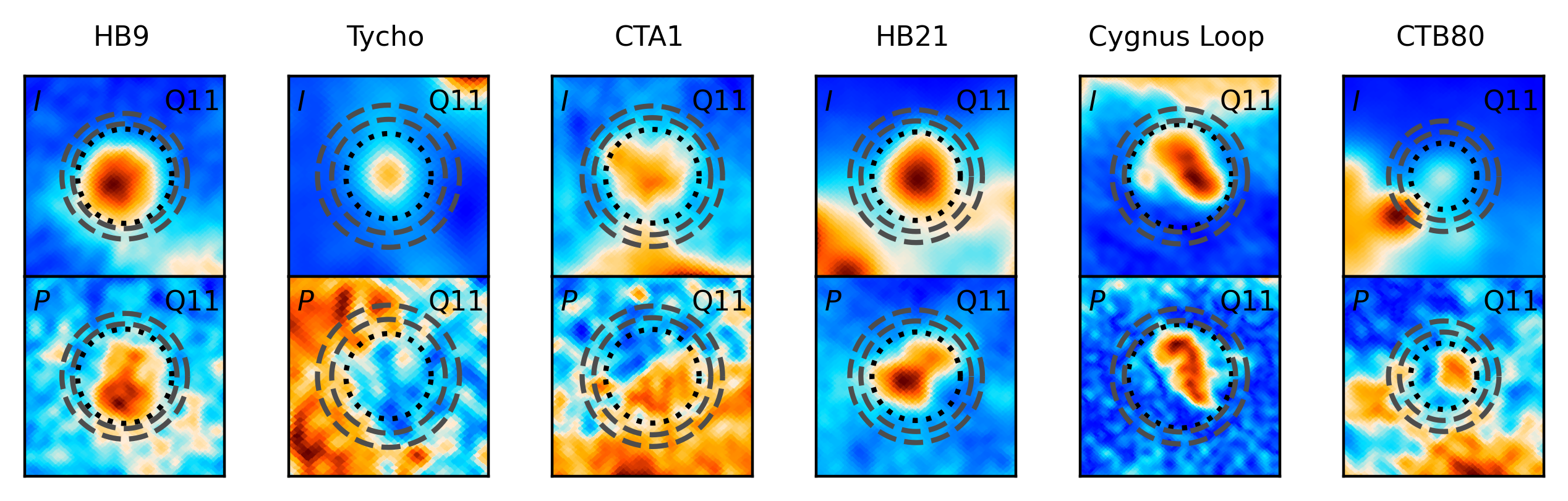}
\caption{Supernova Remnants studied using QUIJOTE-MFI 
data. Top left and right panels respectively show the 
intensity and polarized intensity ($P=\sqrt{Q^2+U^2}$, 
pixel-by-pixel) maps in the Galactic coordinate system, 
smoothed to 1\,deg, from the QUIJOTE-MFI wide survey at 
11\,GHz. 
The apertures and names identify the SNRs studied using the
 QUIJOTE-MFI facility (see Table~\ref{tab:description_SNRs}) 
in this study (black apertures, CTB\,80, Cygnus Loop, 
HB\,21, CTA\,1, Tycho and HB\,9) and in other publications 
(purple apertures,  W44, W49, W51, Cas\,A, 3C\,58, Tau\,A 
and IC443, see text). Bottom panels present the close 
view of intensity and polarized intensity maps of CTB\,80 
($6^\circ\times6^\circ$), Cygnus Loop 
($8.4^\circ\times8.4^\circ$), HB\,21 ($6^\circ\times6^\circ$), 
CTA\,1 ($5.7^\circ\times5.7^\circ$), Tycho 
($4.7^\circ\times4.7^\circ$) and HB\,9 ($6^\circ\times6^\circ$). 
Central positions and apertures are taken from 
Table~\ref{tab:phot_params}, which are used in the flux 
density estimates (Section~\ref{sec:flux_densities}).}
\label{ima:galacticplane_SNRsample}
\end{center}
\end{figure*}

In this work, part of a series of QUIJOTE-MFI scientific
papers, we characterize the intensity and polarized 
spectral energy distribution of six supernova remnants
using the QUIJOTE-MFI wide survey data. Observational 
properties of the synchrotron emission towards the remnants 
are presented, with emphasis on the spectral index behaviour 
and the integrated polarization degree. 
This study complements the analysis of SNRs in other 
QUIJOTE-MFI publications 
\citep{Genova15,mfi_widesurvey_w51,mfi_widesurvey_fan}, 
which collectively provide a sample of SNRs that can be 
used to explore the evolution of the spectral indices and 
the statistics of the polarization fraction, and 
investigate the presence of AME towards SNRs on scales of 1\,deg.

Section~\ref{sec:snr_sample} of this article gives a brief overview of 
the SNR sample of this study.
The QUIJOTE-MFI survey and ancillary data are described 
in Section~\ref{sec:data}. The methodology for building the 
intensity and polarized SED is presented in 
Section~\ref{sec:methodology}, which also shows the models 
and strategies used to perform the SED fits considering using a
Markov Chain Monte Carlo (MCMC) approach.
The main results and discussion about the characterization
 of each SNR are shown in Section~\ref{sec:results_discussion}, 
in addition to the statistical analysis of synchrotron 
properties and the possible AME contribution towards SNRs.
Our main conclusions and summary are listed 
in Section~\ref{sec:conclusions}.

%%%%%%%%%%%%%%%%%%%%%%%%%%%%%%%%%
%%---------- Section ----------%%
%%%%%%%%%%%%%%%%%%%%%%%%%%%%%%%%%
\section{Supernova remnants in the QUIJOTE-MFI survey}
\label{sec:snr_sample}

The QUIJOTE-MFI Wide Survey covers the full northern sky 
at four frequency bands between 10 and 20\,GHz \citep[see 
Section~\ref{sec:data_QUIJOTE} and ][]{mfi_widesurvey}, 
which allow us to access approximately half of the 
Galactic plane (see Figure~\ref{ima:galacticplane_SNRsample}).
In  Table~\ref{tab:description_SNRs}, we summarize the 
properties from literature of the SNRs studied here: 
CTB\,80, Cygnus Loop, HB\,21, CTA\,1, Tycho and 
HB\,9. For these remnants, the main radio and microwave 
properties are briefly described below (Section~\ref{sec:snr_sample_our}).
In addition, this sample is supplemented by the SNRs 
characterized by other QUIJOTE papers using MFI data, 
which are used in the statistical analysis of 
Section~\ref{sec:results_discussion}.
This supplementary group consists of  W44 \citep[the first 
remnant studied by QUIJOTE,][]{2017MNRAS.464.4107G}, 
IC443, W49 and W51 \citep[current QUIJOTE-MFI wide survey 
release combined with deeper observations towards these 
regions,][]{mfi_widesurvey_w51}.
We shall be using here some of the physical 
parameters derived in those publications for these SNRs.
The plerions or filled centre remnants, such as the Crab 
(Tau\,A), Cassiopeia\,A (Cas\,A) and 3C58, are not 
considered in the statistical analysis (see details 
in Section~\ref{sec:statistical_properties}). The intensity 
and polarization SED of 3C\,58 is characterized as part of 
the study of the Fan region \citep[][in prep.]{mfi_widesurvey_fan}. 
For Tau\,A and Cas\,A, there will be a paper devoted to 
describing in detail the spectral properties of these 
important sources \citep[][in prep.]{mfi_pipeline}, which 
are calibrators for the QUIJOTE-MFI data \citep[][]{mfi_widesurvey}.

%-----------------------
%%---------- Sub-section
%--> This Works (our sample)
\subsection{SNRs studied in this paper}
\label{sec:snr_sample_our}

\subsubsection*{CTB\,80}
This is an old-age SNR with a composite morphology and 
significant polarization level based on radio observations 
\citep[][]{1974A&A....32..375V,1981A&A....94..313A,1987ApJ...319L.103S}.
The structure of CTB\,80 covers approximately 
130$\times$100\,arcmin$^2$ and is located at Galactic coordinates 
$(l,b)= (68.8^\circ,+2.7^\circ)$. It shows a central nebula, 
with a bright source of flat radio spectrum, surrounded by 
an extended plateau and three extended arms with steeper 
spectral indices \citep[][]{1985A&A...145...50M,2003AJ....126.2114C}. 
Its pulsar wind nebula (PWN) is powered by the pulsar 
PSR\,B1951+32 \cite[discovered by][]{1988Natur.331...50K}, 
which has a period of 39.5\,ms, a dispersion measure (DM) of 
45\,cm$^{-3}$\,pc and rotation measure (RM) of 
$-182 \pm$ 8\,rad\,m$^{-2}$ \citep{2004MNRAS.353.1311H,2004ApJS..150..317W}.
The distance had been estimated to lie between 1.4 and up to 5\,kpc 
\citep[][]{1984ApJ...282..161B,2012MNRAS.423..718L,2018ApJS..238...35S},
which is compatible with the distance of 2\,kpc associated with 
PSR\,B1951+32 \citep[][]{1988Natur.331...50K}. This value is also
 used as a reference in the age estimation 
\citep[45 to 82\,kyr,][]{1990ApJ...364..178K,2002ApJ...567L.141M,2008ApJ...674..271Z,2021ApJ...914..103S}. 
We assume conservatively that the age is most probably $65 \pm 20$\,kyr. 
CTB\,80 has been detected in radio, optical, X-ray and 
gamma ray studies 
\citep[e.g.,][and references therein]{2001A&A...371..300M,2003AJ....126.2114C,2012MNRAS.423..718L,2021MNRAS.502..472A}.
The X-ray emission associated with the pulsar PSR\,B1951+32 was 
reported by \cite{1995ApJ...439..722S}, while the gamma ray 
emission analysis provides the cut-off energy, 
$E_{\rm c}=5$\,GeV and $\Gamma=1.75$  
\citep{1995ApJ...447L.109R,2010ApJ...720...26A}, which are common values for a PWN.
In addition, the observed extended GeV emission could 
suggest that the SNR shock interacts with the surrounding 
interstellar medium \citep{2021MNRAS.502..472A}.

The characterization of its radio spectrum is still not 
clear, owing to the spread of the integrated flux densities 
at frequencies lower than 3\,GHz and the steepening of the 
spectral index at $\gtrsim$\,10\,GHz \citep[][]{1983PASJ...35..437S,2016A&A...586A.134P}.
Several radio spectral indices $\alpha_{\rm radio}$ have 
been reported ranging from $-0.36\pm0.02$ up to $-0.45\pm0.03$ 
\citep[][]{2005A&A...440..171C,2006A&A...457.1081K,Gao2011A&A...529A.159}, 
which agree with spatial spectral variations observed 
across the SNR on arcmin scales \citep[][]{2005A&A...440..171C}.
polarized signal has been detected at radio frequencies reaching 
a level of up to 50\% at resolutions better than 10\,arcmin 
\citep[][]{1974A&A....32..375V,2006A&A...457.1081K,Gao2011A&A...529A.159}; 
e.g.\ 10--20\,\% (the plateau and central regions) and 30--40\% 
(towards arms) at 4750\,MHz \citep[][]{1985A&A...145...50M}.
Regarding average properties,\footnote{\label{fn:clarification_polarization}
At angular resolutions better than 1\,deg, the polarization 
intensity map can be computed to resolve the structure of the SNRs. 
In these cases, the average polarized intensity ($PI_{\nu}$) can 
be defined as the mean of the polarization measurements over the pixels 
towards the SNR. Consequently, the average polarization fraction 
also arises from the polarization fraction map estimated 
pixel-by-pixel. Nevertheless $\Pi_{\nu} = 100 * PI_{\nu} / S_{\nu}$ 
can be also consider as the polarization level towards the SNRs. 
These definitions depart from the integrated properties computed 
in this study (section~\ref{sec:metodology_sed_modelling}), 
causing differences in the interpretation of the properties of 
SNRs.} 
the average polarization fraction reaches values of 
$12^{+10}_{-6}$\,\% at 11\,cm \citep[][]{1974A&A....32..375V}, 
$7.6\pm0.7\%$ at 1420\,MHz \citep{2006A&A...457.1081K} and 
$13.2\pm2.0\%$ at 6\,cm \citep[][]{Gao2011A&A...529A.159}. 
These measurements suggest depolarization at lower frequencies, 
which is observed in high-resolution analyses 
\citep[][]{1985A&A...145...50M,2020RAA....20..186L}.
The rotation registers fluctuations of 35--85\,rad\,m$^{-2}$ 
\citep{1985A&A...145...50M,2021A&A...653A..62C}, which  favours the 
depth depolarization from the pulsar region over the beam 
depolarization \citep{2020RAA....20..186L}.

\begin{table*}
	\centering
	\caption{Summary of the properties of supernova 
remnant sampled from the literature. The name and Galactic 
coordinates are shown in columns 1 and 2 respectively. For 
the distance, age and the spectral index of intensity-integrated 
spectrum ($\alpha_{\rm radio}$), we present the available coverage 
range for each parameter, as well as the value considered in 
this paper (in parentheses). We adopt the spectral index definition 
such that the flux densities follow $S_{\nu} \propto \nu^{\alpha}$. 
The last column presents comments or remarks about the SNR. 
References are presented throughout Section~\ref{sec:snr_sample_our}.}
	\begin{tabular}{l c l c c c l}
		\hline
		Name & \multicolumn{2}{c}{Position} &  Distance & Age & $\alpha_{\rm radio}$ & Remarks or\\
         & \multicolumn{2}{c}{$l$, b [deg,deg]} & [kpc] & [kyr] & & comments \\
         \hline

%		\multicolumn{6}{l}{\bf SNRs sample (this work)} \\ [0.5 ex]
		CTB\,80 & \phantom{0}68.8& \phantom{0}+2.7\phantom{0} & 1.4 to 5 ($2\pm0.5$)& 45--82 ($65\pm20$) & $-$0.0 to $-$0.8 ($-0.45\pm0.03$)& Curved microwave spectrum\\
		Cyg. Loop & \phantom{0}74.2&  \phantom{0}$-$8.8\phantom{0}  & 0.4 to 1.4 ($0.74\pm0.03$)& 10--20 ($15\pm5$) &$-$0.40 to $-$0.53 ($-0.45\pm0.05$)& \\
		HB\,21 & \phantom{0}89.0&  \phantom{0}+4.7\phantom{0}& 0.7 to 3.0 ($1.74\pm0.5$)& 15--45 ($30\pm15$) & $-$0.27 to $-$0.41 ($-0.38\pm0.02$) & Curved microwave spectrum\\	
        CTA\,1 & 119.5&  +10.0 &1.1 to 1.7 ($1.4\pm0.3$) & 5--20 ($14\pm4$)& $-$0.5 to $-$0.7 ($-0.63\pm0.05$) &\\
		Tycho & 120.1&  \phantom{0}+1.0 & 2 to 5 ($2.5\pm0.8$) & 1572 A.D. & $-$0.65 to $-$0.58 ($-0.624\pm0.004$) & \\
		HB\,9 & 160.4&  \phantom{0}+2.8\phantom{0}  & 0.6 to 1.5 ($0.6\pm0.3$) & 4--20 ($6.6\pm1.5$) & $-$0.44 to $-$0.64 ($-0.59\pm0.02$) & Mixed-morphology type \\ [0.5 ex]
		\hline
	\end{tabular}
	\label{tab:description_SNRs}
\end{table*}

\subsubsection*{The Cygnus Loop}
This is a large and archetypal middle-aged shell-type 
SNR also know as the Veil Nebula or Network Nebula.
The Cygnus Loop is located at around $l=74.2^\circ$ and 
$b=-8.7^\circ$ in a region with low Galactic emission 
(far enough from the Galactic plane), and is well-studied 
across the whole electromagnetic spectrum 
\citep{1985A&A...145...50M,1986ApJ...303L..17H,1997ApJ...484..304L,1998ApJS..118..541L,2011ApJ...741...44K,2021MNRAS.500.5177L}. 
Its emission is moderate at radio--microwave frequencies 
and particularly intense in X-rays.
It is approximately $4^\circ\times3^\circ$ in apparent size 
and covers a nearly circular shell, which is composed of 
prominent radio--microwave regions such as NGC\,6960, 
NGC\,6992, NGC\,6995, the central filament and a southern 
shell 
\citep[referenced in the equatorial coordinate system,][]{2006A&A...447..937S,2021MNRAS.500.5177L}. 
This southern bubble-like structure was proposed to be 
another SNR based on its spectral index distribution and 
polarization properties \citep[][]{2002A&A...389L..61U}, 
however, this hypothesis has been refuted by X-ray observations 
\citep{2011ApJ...730...24K,2013ApJ...764...55L}. 
The Cygnus Loop is located at a distance of 735$\pm$25\,pc 
\citep{2018MNRAS.481.1786F,1958RvMP...30.1048M}, although 
distance estimates fall in the range 440--1400\,pc 
\citep[see][Table\,1 for a summary]{1958RvMP...30.1048M,2005AJ....129.2268B,2018MNRAS.475.3996F}. 
From X-ray and optical measurements, the estimated age spans the range 
10--20\,kyr, assuming that the Cygnus Loop is in an adiabatic 
phase and at the end of the Sedov stage 
\citep[][]{1984ApJ...278..615K,1986ApJ...300..675H,1994PASJ...46L.101M,1998ApJS..118..541L,2018MNRAS.475.3996F}.

% radio-microwave
In the radio range, Cygnus Loop is well described by 
observations from 400\,MHz up to 5\,GHz at resolutions of 
of 1--10\,arcmin 
\citep[see][]{1998ApJ...505..784L,2004A&A...426..909U,2006A&A...447..937S}.
In intensity, the integrated spectrum has a spectral index 
of $\alpha=-0.40\pm0.06$ for frequencies lower than 
4.9\,GHz \citep{2006A&A...447..937S,2004A&A...426..909U}.
When also considering higher frequencies, 
\cite{2021MNRAS.500.5177L} obtained $\alpha=-0.53\pm0.01$ 
and an amplitude of $164\pm5$\,Jy (normalised at 1\,GHz), 
using ancillary radio data 
\citep[][]{2004A&A...426..909U,2006A&A...447..937S} 
and measurements from Medicina and \textit{Planck} 
\citep[at 30\,GHz,][]{2016A&A...586A.134P}.
%polarization
Polarization varies across the Cygnus Loop, where some 
regions present levels of up to 35\,\% 
\citep[][]{1972AJ.....77..459K,1997AJ....114.2081L,2002A&A...389L..61U,2006A&A...447..937S}.
At 4.8\,GHz, NGC\,6992/5, NGC\,6960 and the central filament 
(in the northern shell) show levels of up to 30\,\%, and down 
to 20\,\% in the southern shell 
\citep{2006A&A...447..937S,1972AJ.....77..459K}.
In contrast, at lower frequencies there are major polarization 
differences between the southern and northern regions, 
the northern region being less intense \citep{2002A&A...389L..61U}.
At 1.42\,GHz, this reveals that the depolarization is 
significantly higher in northern areas than in the southern regions, 
where depolarization seems to be spatially correlated with 
X-ray emission \citep{1997AJ....114.2081L}.
Regarding rotation measure, differences remain between the 
north and south, with averages of $-$21\,rad\,m$^{-2}$ and 
$-$73\,rad\,m$^{-2}$ respectively 
\cite[using 6, 11 and 21\,cm,][]{2006A&A...447..937S}.

\subsubsection*{HB\,21}
This is a mixed-morphology and middle-aged SNR. 
HB\,21 consists of an irregular radio shell of around 
$2^\circ\times 1.5^\circ$ centred at $l=89.0^\circ$ and 
$b=4.7^\circ$, whose morphology reveals interaction with 
molecular clouds 
\citep{1974MNRAS.169...59H,1990A&A...237..189T,2006ApJ...637..283B} 
and the X-ray thermal emission is coming from its central area 
\citep{1996A&A...315..260L,1996A&AS..120C.339K}.
Observations of CO and infrared lines confirm its interaction 
with molecular clouds 
\citep{1991ApJ...382..204K,2009ApJ...693.1883S}, where HB\,21 
is associated with the Cyg\,OB7 complex located at $0.8\pm0.07$\,kpc 
\citep{1990A&A...237..189T,1996A&A...315..260L}.
\cite{2006ApJ...637..283B} recently suggested,
on the basis of a compendium from 
multiple analyses and datasets, a conservative 
higher distance value of $1.7\pm0.5$\,kpc.
Consequently, there is also a wide age range spanning the range
6--45\,kyr 
\citep[][]{1973AJ.....78..170K,2009ApJ...693.1883S,1999MNRAS.308..271F,2006ApJ...647..350L,2006ApJ...637..283B}. 
Here we adopt a conservative value of $30\pm15$\,kyr in order to remain consistent 
with that obtained from its expanding HI shell 
\citep[at 1.1\,kpc,][]{1973AJ.....78..170K}.

%--> radio--microwave
Concerning the radio--microwave properties, the intensity SED 
reveals a curved spectrum with a break frequency 
$\nu_{\rm b}=5.9\pm1.2$\,GHz, when a change of the spectral index 
of $\Delta \alpha=-0.5$ is assumed \citep[at 
$\lesssim$\,100\,GHz,][]{2013ApJ...779..179P}. The radio spectral 
index falls  between $-0.27$ to $-0.41$ in the range 
$\sim$0.03--\,5\,GHz 
\citep[e.g. see][and references therein]{1973A&A....26..237W,2003A&A...408..961R,2006A&A...457.1081K}, 
$\alpha_{\rm radio}=-0.36\pm0.03$ being  the adopted value 
\citep{Gao2011A&A...529A.159,2013ApJ...779..179P}.
%--> polarization
At radio frequencies, HB\,21 shows polarization level of up 
to 35\,\% in some areas 
\citep[][]{1973AJ.....78..170K,2006A&A...457.1081K,2010A&A...520A..80L,Gao2011A&A...529A.159}. 
At 1420\,MHz, polarized intensity is strongest in the east 
while it falls towards the western edge \citep[at $\sim$1\,arcmin 
scales,][]{2003A&A...408..961R,2010A&A...520A..80L}.
At 4.8\,GHz, \cite{Gao2011A&A...529A.159} found an average 
polarization intensity of $PI_{4.8}=11.3\pm1.1$\,Jy yielding a 
polarization fraction of $10.7\pm1.1$\,\%. The average polarization
 fraction seems to decrease at low frequencies 
\citep[$3.7\pm0.5$\,\% at 1.42\,GHz,][]{2006A&A...457.1081K}, 
although we also expect some differences caused by the 
resolutions of the maps used ($\sim$1 and 9.5\,arcmin 
at 1.4 and 4.8\,GHz respectively).

\subsubsection*{CTA\,1}
In our sample, this is the farthest SNR from the Galactic plane.
 It is located at $l=119.5^\circ$ and $b=10.2^\circ$ and was 
discovered by \cite{1960PASP...72..237H}.
CTA\,1 shows the typical morphology of a radio fragmentary shell,
 which is composed of strong limb-brightened arcs (mainly in the 
south and southwest), a central branch and a diffuse extended 
emission covering a circular region of radius $\sim$45\,arcmin. 
The breakout phenomenon is affecting the north-eastern part of the shell.
Observations of CTA\,1 span radio 
\citep{1975A&A....45..239C,1979A&A....74..361S,1981A&A...103..393S,1997A&A...324.1152P}, 
optical
\citep{1981ApJ...247..148F,1983ApJS...51..337F,2007A&A...461..991M,2013MNRAS.430.1354M}, 
X-ray 
\citep{1995ApJ...453..284S,1997ApJ...485..221S,2012MNRAS.426.2283L} 
and higher-energy bands 
\citep{1999A&A...348..868R,2013ApJ...764...38A,2003PhR...382..303T,2016ApJ...831...19L,2016MNRAS.459.3868M}.
\cite{1993AJ....105.1060P} 
obtained a kinematic distance of $1.4\pm0.3$\,kpc from H\textsc{i} observations.
Fermi-LAT data allowed the detection and characterization of 
its pulsar and PWN \citep{2008Sci...322.1218A,2012ApJ...744..146A}, 
yielding a period of $\sim$316\,ms, a spin-down power of 
$\sim$4.5$\times 10^{35}$\,erg\,s$^{-1}$ and a characteriztic 
age of 13.9\,kyr for the pulsar.
This age is in agreement with estimations based on radio and 
X-ray thermal emissions 
\citep[5--20\,kyrs,][]{1993AJ....105.1060P,1997ApJ...485..221S,2004ApJ...601.1045S}, 
where we adopted the $14\pm4$\,kyrs at 1.4\,kpc 
\citep[from Sedov expansion model,][]{2004ApJ...601.1045S} 
and accounting for an uncertainty of 25\%. 

%--> radio-microwave
At radio frequencies, the integrated spectrum of CTA\,1 is 
described by a power law with a spectral index $\alpha=-0.63\pm0.05$ 
in the 408--4800\,MHz range \citep[][]{2011A&A...527A..74S}, 
which is in agreement with previously reported spectral indices 
\cite[e.g.\ $-0.57\pm 0.006$,][]{1997A&A...324.1152P}. 
This spectral index from integrated SED is compatible with the 
spectral index distribution obtained at resolutions of around 
10\,arcmin, which has values around $-$0.5 (shell structure), 
$-$0.6 (central branch) and steepening in the breakout region 
\citep{1997A&A...324.1152P,2011A&A...527A..74S,1981A&A...103..393S}.
%--> polarization
Polarized emission has been reported from 1.4 to 4.8\,GHz 
\citep[][]{1981A&A...103..393S,1979A&A....74..361S,2011A&A...527A..74S}, 
where the average polarization fraction reaches levels of up 
to 40\% (at 4.8\,GHz) in the eastern shell 
\citep{2011A&A...527A..74S}, significantly decreasing towards 
the breakout region \citep{2011A&A...527A..74S}. 
The RM map shows variations between $\pm 40 {\rm rad\, m}^{-2}$ 
\citep[][]{1981A&A...103..393S,2011A&A...527A..74S},  in 
full agreement with the RMs obtained from extragalactic sources 
in the line of sight of CTA\,1 
\citep[][]{1992ApJ...386..170S,2009ApJ...702.1230T}.
These variations support the presence of a Faraday Screen 
(FS) in front of the remnant, driven by a regular magnetic 
field, between 2.7 and 1.9\,$\mu$G, in the direction opposite 
to the interstellar magnetic field.

\subsubsection*{Tycho}

Tycho is the remnant of a type Ia Supernova explosion, 
which was observed by Tycho Brahe in 1572 and has a long 
historical follow-up 
\citep[][and references therein]{1943ApJ....97..119B,1952Natur.170..364H,1970ApJ...160..199G,2011ApJ...730L..20A,2017hsn..book..117D} 
that includes the detection of optical scattered-light echoes 
\citep[][]{2008Natur.456..617K,2008ApJ...681L..81R}.
Tycho is catalogued as a shell-type SNR with a diameter 
of $\sim$8\,arcmin in radio bands, which expands at 
0.1126\%yr$^{-1}$ into the ISM, suggesting that it is still 
evolving towards the Sedov adiabatic phase \citep{1997ApJ...491..816R}.
Tycho is also called 3C\,10 and SN\,1572, and is located 
at $l=120.1^\circ$ and $b=1.4^\circ$.
Distance estimates range from 2 to 5\,kpc, and are computed 
from radio, optical, X-ray and gamma-ray observations 
\citep[see a summary in][]{2010ApJ...725..894H}, but the 
kinematic distance of 2.5\,kpc is widely used \citep{2011ApJ...729L..15T}.
%

%--> radio-microwave
At radio--microwave frequencies, the integrated spectrum is 
characterized by a spectral index of 
$\alpha = -0.624 \pm 0.004$ \citep[0.015--70\,GHz,][]{2021A&A...653A..62C}, 
which is compatible with previous estimates of about 
$\alpha = -0.58 \pm 0.02$ for a similar frequency range 
\citep[][]{2011A&A...536A..83S,2019MNRAS.482.3857L,2009MNRAS.396..365H}.
From higher-resolution observations, the spatial spectral 
variations fall in the $-0.4$ to $-0.8$ range 
\citep[][]{2000ApJ...529..453K,2019AJ....158..253A} and is 
steeper near the central region of the remnant. In addition, 
low-frequency free--free self-absorption has been reported for some 
regions of Tycho \citep{2019AJ....158..253A}, which is not 
observed in the integrated spectrum.
In contrast, some studies have reported a weak concave-up 
spectrum with a smooth spectral break of 
$\Delta\alpha\simeq 0.04$ at 1\,GHz \citep{1992ApJ...399L..75R,2021A&A...653A..62C}.
%--> polarization in radio-microwave
The polarization fraction seems to be weak at radio 
frequencies. At 1.42\,GHz, observations reveal that the 
polarization intensity consists of a faint arc structure, 
where the average percentage polarization is
 $2.5\pm 0.1$\%\ \citep[at $\sim$1\,arcmin, ][]{2006A&A...457.1081K}. 
\cite{2011A&A...536A..83S}, on the other hand, reported a 
lower value of 1\%, at 4.8\,GHz from a map with beam resolution 
of 9.5\,arcmin.

\begin{table}
\centering
\caption{Summary of the sky surveys considered in this 
study. The first column gives the name of the 
facility or survey. The second column is the notation that 
identifies the map used, which has a central frequency (GHz), 
original beam resolution (arcmin), calibration error (as a 
percentage) and Stokes maps available ($I$, $Q$ or $U$), as 
shown in the third, fourth, fifth and sixth columns
respectively. The main references are identified by the last column.}
\begin{threeparttable}
\begin{tabular}{l c c c c c c}
		\hline
		Name & ID & Freq. & FWHM & Cal. & Stokes & Ref.\\
		 &  & [\footnotesize{GHz}] & [arcmin] & [\%] & & \\\hline
		 Several facilities & Has & 0.408 & 51 & 10& $I$ & 1,2\\ [0.3 ex]
		 Dwingeloo\,25m & Dwi & 0.820& 72 & 10 & $I$  & 3\\ [0.3 ex]
		 DRAO\,25.6m & Wol & 1.410& 36 & 10 & $Q,U$  & 4\\ [0.3 ex]
		 Stockert\,25m \& & \multirow{2}{*}{Rei} & \multirow{2}{*}{1.420} & \multirow{2}{*}{36}  & \multirow{2}{*}{10}& \multirow{2}{*}{$I$} & \multirow{2}{*}{5,2}\\
	   Villa-Elisa\,30m & & & & & \\ [0.3 ex]
		 Urumqi\,25m & Urq & 4.8& 9.5 & 10 & $I,Q,U$ & 6\\ [0.3 ex]
		 QUIJOTE-MFI & Q11 & 11.1 & 55.4 & 5 & $I, Q, U$ & 7\\ [0.3 ex]
		 QUIJOTE-MFI & Q13 & 12.9 & 55.8 & 5 & $I, Q, U$ & 7\\ [0.3 ex]
		 QUIJOTE-MFI & Q17 & 16.8 & 39.0 & 5,6,6 & $I, Q, U$ & 7\\ [0.3 ex]
		 QUIJOTE-MFI & Q19 & 18.8 & 40.3 & 5,6,6 & $I, Q, U$ & 7\\ [0.3 ex]
		 WMAP\,9yr & WK & 22.8 & 50.7 & 3 & $I, Q, U$  & 8\\ [0.3 ex]
		 \textit{Planck}-LFI & P30 & 28.4 & 32.3 & 3 & $I, Q, U$  & 9\\ [0.3 ex]
		 WMAP\,9yr & WKa & 32.9 & 38.8 & 3 & $I, Q, U$  & 8\\ [0.3 ex]
		 WMAP\,9yr & WQ & 40.7 & 30.6 & 3 & $I, Q, U$  & 8\\ [0.3 ex]
		 \textit{Planck}-LFI & P44 & 44.1 & 27.0 & 3 & $I, Q, U$  & 9\\ [0.3 ex]
		 WMAP\,9yr & WV & 60.7 & 20.9 & 3 & $I, Q, U$  & 8\\ [0.3 ex]
		 \textit{Planck}-LFI & P70 & 70.4 & 13.2 & 3 & $I, Q, U$  & 9\\ [0.3 ex]
		 WMAP\,9yr & WW & 93.5 & 14.8 & 3 & $I, Q, U$  & 8\\ [0.3 ex]
		 \textit{Planck}-HFI & P100 & 100 & 9.7 & 3 & $I, Q, U$  & 9\\ [0.3 ex]
		 \textit{Planck}-HFI & P143 & 143 & 7.3 & 3 & $I, Q, U$  & 9\\ [0.3 ex]
		 \textit{Planck}-HFI & P217 & 217 & 5.0 & 3 & $I, Q, U$  & 9\\ [0.3 ex]
		 \textit{Planck}-HFI & P353 & 353 & 4.5 & 3 & $I, Q, U$ & 9\\ [0.3 ex]
		 \textit{Planck}-HFI & P545 & 545 & 4.8 & 6.1 & $I$ & 9\\ [0.3 ex]
		 \textit{Planck}-HFI & P857 & 857 & 4.6 & 6.4 & $I$ & 9\\ [0.3 ex]
		 COBE-DIRBE & D10 & 1249 & 37.1 & 11.6 & $I$ & 10 \\ [0.3 ex]
		 COBE-DIRBE & D9 & 2141 & 38.0 & 10.6 & $I$ & 10 \\ [0.3 ex]
		 COBE-DIRBE & D8 & 2997 & 38.6 & 13.5 & $I$ & 10 \\ [0.3 ex]
		\hline
\end{tabular} 
\label{tab:experiments}

\begin{tablenotes}
\small
      \item References. 1: \cite{Haslam1982}; 2: \cite{2003A&A...410..847P}; 3: \cite{Berkhuijsen1971}; 
4: \cite{2006A&A...448..411W}; 5: \cite{Reich1986}; 6: \cite{2007A&A...463..993S} 
and \cite{Gao2011A&A...529A.159}; 7: \cite{mfi_widesurvey}; 8: \cite{2013ApJS..208...20B}; 
9: \cite{2020A&A...641A...1P}; 10: \cite{1998ApJ...508...25H}.
\end{tablenotes}
\end{threeparttable}
\end{table}

\subsubsection*{HB\,9}
HB\,9 is a middle-aged supernova remnant with a mixed-morphology type.
In radio bands, shell-like shape of HB\,9 has a large extension 
of approximately $140\times 120$\,arcmin$^2$ with structures 
similar to fragmented shells or filaments 
\citep[][]{1973A&A....26..237W,1974AJ.....79.1253D,1976MNRAS.177..601C,1991AJ....101.1033L}.
The X-ray emission was discovered using HEAO--1\,A2 
soft X-ray data \citep[][]{1979MNRAS.189P..59T} and was later 
confirmed by X-ray missions such as GINGA and Suzaku 
\citep[][]{1993PASJ...45..545Y,2019MNRAS.489.4300S,2020PASJ...72...65S}. 
In particular, ROSAT observations allowed the detection of the compact 
thermal X-ray emission in the central part of HB\,9 
\citep[][]{1995A&A...293..853L}.
Other investigations included observations at various
frequencies from the optical 
\citep[][]{1975AZh....52...39L,1977A&AS...28..439D,1980A&A....84...26L} 
up to gamma-rays 
\citep[][]{2014MNRAS.444..860A,2019MNRAS.489.4300S,2020ApJS..247...33A}. 
For the latter, extended gamma-ray emission was detected in the 
line of sight to HB\,9 
\citep[Fermi-LAT data,][]{2014MNRAS.444..860A}, but was not 
detected in the TeV energy regime \citep[MAGIC observations,][]{2009arXiv0907.1009C}.
HB\,9 is also known as G160.9+2.6 (from early radio studies) 
although its central position is fixed at $l=160.4^\circ$ and $b=2.8^\circ$. 
The H\textsc{i} and $^{12}$CO ($J=1$--0) analysis provided a kinematic 
distance of $0.6\pm 0.3$\,kpc \citep[][]{2019MNRAS.489.4300S}, 
where the extinction distance of $0.54 \pm 0.10$\,kpc from stars in 
the line of sight \citep[GAIA survey,][]{2020ApJ...891..137Z} suggests
 a connection between interstellar gas and the remnant. These 
estimates are in agreement with previous higher values of 
$\lesssim$\,1.5\,kpc \citep[][]{2007A&A...461.1013L,1995A&A...293..853L}. 
These latter studies also provided age estimates  in the ranges
8--20\,kyr \citep[ROSAT data,][]{1995A&A...293..853L} and 
4--7\,kyr \citep[H\textsc{i} data,][]{2007A&A...461.1013L}, 
which establish HB\,9 as a middle-aged SNR with a Sedov age of 6.6\,kyr.
%

%--> radio-microwave
Regarding the radio properties, the integrated intensity SED is 
well described by a power law with a spectral index between $-0.47$ 
and $-0.64$ in the range from 0.01 to \,5\,GHz 
\citep[e.g. see][]{1982JApA....3..207D,2006A&A...457.1081K,2007A&A...461.1013L}; 
for example, \cite{Gao2011A&A...529A.159} found 
$\alpha=-0.59\pm0.02$ in the range 0.1--5\,GHz.
The spectral bend proposed by \cite{1973A&A....26..237W} at 1\,GHz 
was not confirmed by observations at higher frequencies 
\citep[][]{2003A&A...408..961R,Gao2011A&A...529A.159}.
%--> polarization
The observed linear polarization across HB\,9 implies a 
largely tangential orientation of the magnetic fields 
\citep[1.4 and 2.7\,GHz,][]{1983IAUS..101..377R,2004mim..proc..141F}.
At 1.42\,GHz, the average linear polarization fraction is 
$9.1\pm 0.1\%$, reaching a maximum of 59\% \citep{2006A&A...457.1081K}.
\cite{Gao2011A&A...529A.159} reported a 
similar polarization level of $15.6\pm1.6\%$ at 4.8\,GHz for the entire region.
In both cases, authors observed a spatial correlation between 
the highly polarized emission and the intensity signal coming 
from the outer filaments of the shell.

%%%%%%%%%%%%%%%%%%%%%%%%%%%%%%%%%
%%---------- Section ----------%%
%%%%%%%%%%%%%%%%%%%%%%%%%%%%%%%%%
\section{Data}
\label{sec:data}

In this study, we use QUIJOTE-MFI wide survey data together with
ancillary data from surveys such as WMAP, \textit{Planck} and 
DIRBE. We cover the radio up to the far-infrared range in intensity 
(0.4--3000\,GHz) while for polarization the range spans from 5 
to 353\,GHz. 
All maps are available in the \textsc{HEALPix}\footnote{HEALpix: 
\url{http://healpix.jpl.nasa.gov}.} pixelization scheme 
\citep[][]{healpix} and for different beam resolutions. 
For a coherent characterization of the spectral energy 
distribution of SNRs (Section~\ref{sec:flux_densities}), 
we use maps smoothed to a common angular resolution of 
1\,deg for an $N_{\rm side}=512$ pixelization scheme, which 
corresponds to a pixel size of $\sim$6.9\,arcmin and pixel 
solid angle $\Omega_{o}= 3.9947\times 10^{-6}$\,sr.
Table~\ref{tab:experiments} presents the main properties of 
the data used in our analyses, which are briefly described below.

%-----------------------
%%---------- Sub-section
\subsection{QUIJOTE-MFI wide survey}
\label{sec:data_QUIJOTE}

The Q-U-I JOint TEnerife CMB experiment  
\citep[QUIJOTE,][]{RubinoSPIE12} is a scientific 
collaboration devoted to characterizing the polarization 
of the CMB and astrophysical foregrounds in the range 
10--40\,GHz. The QUIJOTE telescopes operate from the Teide 
Observatory (Tenerife, Spain) at 2400\,m a.s.l. and cover 
most of the northern sky (above declination $-35^\circ$) 
at an angular resolution of $\sim$1\,deg. The QUIJOTE 
project consists of two telescopes \citep[of crossed-Dragone 
design,][]{QT1,QT2,QT2b} and three instruments 
\citep[][]{MFIstatus12,status2016SPIE,2022SPIE12190E..33H}.
The data used in this study come from the QUIJOTE-MFI wide survey, 
based on observations performed with the Multi-Frequency 
Instrument \citep[MFI,][]{MFIstatus12} between November 2012 
and October 2018 (almost 9000\,hr). 
The MFI consists of four polarimeters (called \textit{horns}) 
in the bands 10--14\,GHz (horns\,1 and 3) and 16–20 GHz 
(horns\,2 and 4). A detailed description of the QUIJOTE-MFI 
wide survey is presented in \cite{mfi_widesurvey} and \cite{mfi_pipeline}.

The data are publicly available on the QUIJOTE web 
page\footnote{\label{fn:QUIJOTE_webpage}QUIJOTE: \url{https://research.iac.es/proyecto/quijote}}. 
We consider the available intensity ($I$) and the 
linear polarization ($Q$ and $U$) maps smoothed to 
1\,deg (FWHM) which are centred at 11.1, 12.9, 16.8 
and 18.8\,GHz. Here, these are identified as Q11, Q13, 
Q17 and Q19 respectively. The Q11 and Q13 maps come from 
the  MFI horn\,3. However, the Q17 and Q19 maps are 
generated as a combination of data from two separate
horns (horn\,2  and horn\,4) in order to reduce the $1/f$ 
noise contribution \citep[see details in][]{mfi_widesurvey}.
The total calibration uncertainties are 5\% for intensity 
and polarization, except for polarized maps Q17 and Q19,
 which reach a level of 6\%.
Regarding the zero levels and colour corrections, their
 effects are discussed in Section~\ref{sec:data_systematic_effects}, 
and the noise correlations between bands and filtering scheme are
 discussed below. 

For each horn, the signal is processed to produce two 
maps at different frequencies (e.g.\ Q11 and Q13 are obtained 
from horn\,3), which share the same low-noise amplifier (LNA) configuration. 
These signals are therefore 
affected, to some degree, by a common $1/f$ noise component, 
which is known as the QUIJOTE band correlation, $\rho$.
The normalised correlation factors ($\rho$) are 
around 80\% in total intensity, and around 30\% in 
polarization (owing to smaller $1/f$ residuals), as discussed 
in detail in \citep{mfi_widesurvey}. In our analyses we 
account for this noise correlation by using the coefficients 
shown in Table~\ref{tab:band_correlation}.

One of the steps of the procedure to clean residuals 
radio frequency interference (RFI) involves the subtraction of 
the median of all pixels at the same declination. This 
template is called f($\delta$) or FDEC and is applied in both 
intensity and polarization 
\citep[see details in][]{mfi_widesurvey,mfi_pipeline}.
As FDEC filtering acts as a low pass filter, we 
expect a low impact on the local flux density estimate 
using the photometric approach, which is the strategy 
applied here (Section~\ref{sec:methodology}).
\cite{mfi_widesurvey} compares the polarized flux densities 
from maps with and without the FDEC, finding a negligible 
maximum difference of the two photometries of 0.4\,Jy and 
a standard deviation of the difference of 0.037\,Jy from 
different sky regions when the aperture size is 2 degrees. 
The results are lower for the case with apertures of 1 
degree (a maximum difference of 0.06\,Jy and a standard 
deviation of 0.007\,Jy). Consequently, we use the QUIJOTE-MFI 
maps with the FDEC filtering applied, which is the standard 
data product. No FDEC filtering is applied to the 
remaining ancillary maps.

\begin{table}
	\centering
	\caption{Band correlation measurements for the 
QUIJOTE-MFI maps. The second and third columns list
correlation factors, $\rho$, for the pair of bands 11--13\,GHz 
and 17--19\,GHz, for the intensity and polarization maps.}
	\begin{tabular}{l c c} % four columns, alignment for each
		\hline
		Case & Intensity & polarization \\
		\hline
		$\rho_{11,13}$ & 0.77 & 0.36 \\
		$\rho_{17,19}$ & 0.85 & 0.29 \\
		
		\hline
	\end{tabular}
	\label{tab:band_correlation}
\end{table}

%-----------------------
%%---------- Sub-section
\subsection{Surveys at radio frequencies}
\label{sec:data_lowfreq}

At low frequencies, we use data from three well-known 
radio surveys of sufficient resolution to carry out a 
comprehensive analysis in combination with the QUIJOTE-MFI maps.
These surveys have effective frequencies centred at 
408\,MHz \citep[Has map,][]{Haslam1982}, 820\,MHz 
\citep[Dwi map,][]{Berkhuijsen1971} observed with the 
Dwingeloo radio telescope, and 1420\,MHz 
\citep[Rei map,][]{Reich1986,Reich2001} from the 
Stockert 25 m telescope, whose original effective 
beam resolutions are 51, 72 and 34\,arcmin respectively.
In our analyses, the most relevant systematic effect 
is the calibration uncertainty.
In this regard, in the determination of flux densities 
of point sources it must be taken into account that the 
1.4\,GHz map is calibrated to the full-beam scale, so that
the calibration scale must be transferred to the 
main beam by using its efficiency.
A scale factor of 1.55, proposed by \cite{Reich1986}, is required to account for this 
effect  and has 
been widely used in previous studies 
\citep[e.g. see discussions in][and references therein]{2014A&A...565A.103P,2015MNRAS.448.3572I,2017MNRAS.464.4107G}.
In this context, for each map, we assume a conservative 
calibration error of 15\%, which will be added in 
quadrature to the flux density uncertainties.

%%%%
\subsubsection*{Galactic plane surveys}
\label{sec:data_urumqi}
At 4.8\,GHz, we use the Sino-German $\lambda$6\,cm 
survey\footnote{\url{http://zmtt.bao.ac.cn/6cm/} \label{foot:sinogerman}} 
of the Galactic plane (covering 
$10^\circ<l<230^\circ$ and $|b|<5^\circ$ ) produced with 
the 25\,m Urumqi telescope 
\citep{2007A&A...463..993S,2010A&A...515A..64G,2011A&A...527A..74S,2011A&A...529A..15X}.
The intensity and polarization maps are publicly available 
from the CADE\footnote{Centre d'Analyse de Donn\'ees 
Etendues \citep[CADE,][]{CADE}:  \url{http://cade.irap.omp.eu}\label{fn:cade}} 
webpage for the Urumqi 
dataset\footnote{\url{https://cade.irap.omp.eu/dokuwiki/doku.php?id=urumqi}} 
in \textsc{HEALPix} format ($N_{\rm side}=1024$) at 
beam resolution of 9.5\,arcmin.
Since we smooth the maps to 1\,deg (assuming the telescope
 beam to be Gaussian) the sky coverage of Urumqi is 
further restricted  where we extended the original mask 
by 0.5\,deg in order to avoid boundary effects in the 
smoothing process.
We can only apply the photometric method of Section~\ref{sec:flux_densities} 
to sources CTB\,80, Tycho 
and HB\,9. In contrast, HB\,21 is partially observed while the 
Cygnus Loop and CTA\,1 are outside of the observed region 
(at high latitudes).
\cite{2007A&A...463..993S} reported systematic errors from 
the calibration sources of around 4\% and 5\% for total 
intensity and polarized intensity respectively.
In the estimation of flux densities, they computed 
a typical error lower than 10\% in intensity and polarization, 
which was finally used as a global calibration uncertainty  
(10\%) for the study of large SNRs seen by Urumqi 
\citep{Gao2011A&A...529A.159,2011A&A...536A..83S}.
Neither can we adopt to the estimation of flux 
density uncertainties with random aperture photometry 
(Section~\ref{sec:flux_densities}) because of the 
limited sky area observed .
We therefore adopt a 10\% uncertainty as a global 
calibration error.

%%%%%
\subsection{Microwave and infrared surveys: WMAP, \textit{Planck} and DIRBE}
\label{sec:data_wmap_planck_dirbe}

The WMAP 
\citep[Wilkinson Microwave Anisotropy Probe,][]{2013ApJS..208...20B} 
and \textit{Planck} \citep[][]{2020A&A...641A...1P} missions 
provide $I$, $Q$ and $U$ maps that are publicly available 
in LAMBDA\footnote{\label{fn:lambda}Legacy Archive for
 Microwave Background Data Analysis (LAMBDA): 
\url{http://lambda.gsfc.nasa.gov/}} and in PLA\footnote{\label{fn_pla}
\textit{Planck} Legacy Archive (PLA): 
\url{http://lambda.gsfc.nasa.gov/}} web pages respectively.
The maps are in \textsc{HEALPix} format, where both 
the original and smoothed (1\,deg) versions are available.
For WMAP, we consider the nine-year smoothed maps with 
effective central frequencies of 22.8, 33.0, 40.6, 60.8 
and 93.5\,GHz. 
For \textit{Planck}, we use the maps from the PR3-2018 
data release. The nine bands cover from 28.4 to 857\,GHz 
in total intensity and from 28.4 to 353\,GHz in polarization 
(see effective central frequencies in Table~\ref{tab:experiments}).
For WMAP and \textit{Planck}, when working with intensity 
and polarized maps, we used a global calibration uncertainty 
of 3\% (22.8--353\,GHz), 6.1\% (545\,GHz) and 6.4\% (857\,GHz), 
which have been used previously in SED analyses that adopt
the aperture photometry strategy 
\citep[e.g. see][and references therein]{2014A&A...565A.103P,2017MNRAS.464.4107G,mfi_widesurvey_ame_srcs,mfi_widesurvey_galacticPlane,mfi_widesurvey_w51}.
In addition, we consider the CMB map recovered with the 
SMICA algorithm \citep[][]{2020A&A...641A...1P}, which 
is also available in the PLA web page.

%-----
% DIRBE data.
Finally, we consider the data product from the COBE-Diffuse 
Infrared Background Experiment \citep[DIRBE,][]{1998ApJ...508...25H}, 
which span the wavelength range from 1.25\,$\mu$m to 240 $\mu$m. We use the 
far-infrared maps at 240 $\mu$m (D10 at 1249\,GHz), 
140\,$\mu$m (D9 at 2141\,GHz) and 100\,$\mu$m (D8 at 2997\,GHz) 
available in LAMBDA$^{\ref{fn:lambda}}$, which are useful for 
modelling the spectrum of the thermal dust emission in intensity.
For these maps, we consider uncertainty levels of 11.6,10.6 
and 13.5\,\% respectively \citep[][]{1998ApJ...508...25H}.

%%%%%
\subsection{Colour corrections and additional remarks}
\label{sec:data_systematic_effects}

To correct our flux density estimates for the effect of finite 
bandwidth we apply a colour correction during the SED fitting 
(Section~\ref{sec:methodology}), which is estimated using the 
slope of the fitted spectrum at each frequency. 
In general, this redress is not negligible for experiments 
with broad bandpass receivers (QUIJOTE-MFI, WMAP, \textit{Planck} 
and DIRBE), and when the spectral shape of the sources of 
interest departs from the shape of the spectrum of their 
calibrator.\footnote{In general, QUIJOTE, WMAP, \textit{Planck} 
and DIRBE are calibrated using the CMB (mainly the dipole) or 
Galactic sources such as Tau\,A and Cas\,A (known as primary 
calibrators).}
Here, the colour correction coefficient, $cc_{\nu}(\theta)$ at 
frequency $\nu$, is defined as a multiplicative factor applied 
to the raw flux density and to its uncertainty. The estimation 
of $cc_{\nu}(\theta)$ depends on the non-ideal behaviour of 
detectors across the bandpass and the spectral shape of the 
source (identified by the parameter space $\theta$ of the 
fitted spectrum). 

This procedure requires the spectral integration over each 
bandpass for each model tested, which is computationally 
expensive and inefficient for our SED fit with the MCMC 
sampling approach (Section~\ref{sec:likelihood_analysis}). 
Therefore, as an alternative, we use the 
{\sc fastcc}\footnote{\label{fn:fastcc}{\sc fastcc} is 
available in IDL and Python language at 
\url{https://github.com/mpeel/fastcc}.} code 
\citep{2022RNAAS...6..252P}. {\sc fastcc} is optimized 
for several CMB experiments such as CBASS, QUIJOTE, 
WMAP, \textit{Planck} and DIRBE. It provides the colour 
correction factors by evaluating a previously modelled 
quadratic function, assuming a power law behaviour (with 
spectral index $\alpha$) across the bandpass.
Optionally, for DIRBE and \textit{Planck}-HFI bands, we 
use the dust temperature ($T_{\rm d}$) and emissivity 
index ( $\beta_{\rm d}$) as input when our spectrum is 
dominated by a thermal dust component 
\cite[see details in][]{2022RNAAS...6..252P}. This is 
available for dust temperatures between 10 and 40\,K and 
dust emissivity index between 1.0 and 4.0.

The uncertainties in the determination of the zero 
levels affect almost all the maps, although they are 
more significant in some surveys at the low frequencies. 
Nonetheless, the zero levels have a negligible effect 
on our flux density estimation, because the background 
correction applied by the photometric aperture method 
(Section~\ref{sec:methodology}) cancels them out. 
Therefore, we do not take into account the effects of 
zero levels.

Regarding polarization maps, we follow the COSMO 
convention used by WMAP, \textit{Planck} and QUIJOTE-MFI, 
which differs from the IAU convention by the negative 
sign in the Stokes $U$ maps \citep[i.e. 
$\mathrm{U} = -\mathrm{U}_{\rm IAU}$,][]{1996A&AS..117..161H}. 
Therefore, the Stokes $U$ map at 4.8\,GHz also includes the 
negative sign in accordance with the COSMO 
convention.

\begin{table}
	\centering
	\caption{Aperture photometry parameters selected 
to our SNRs. In addition, CTB\,80 and HB\,21 include a 
mask to avoid strong Galactic diffuse emission and point 
sources (see text for details).}
	\begin{tabular}{l c c c c} % four columns, alignment for each
		\hline
		%SNR & \multicolumn{2}{c}{Position} & Apertures & \multicolumn{3}{c}{Solid angle}\\
		 SNR & $l$ & b & $r_{\rm src}$, $r_{\rm int}$, $r_{\rm ext}$ & $\Omega_{\rm src}$ \\
		 & [deg] & [deg] & [arcmin] & [sr] \\
		\hline
%		\multicolumn{6}{l}{\bf Our SNRs candidates} \\ [0.5 ex]
CTB\,80 &68.8 & 2.7 & 60, 80, 100 & 9.51 $\times10^{-4}$\\
Cygnus L. & 73.9 & -8.8 & 130, 140, 170 & 4.50$\times10^{-3}$ \\
HB\,21 & 89.0 & 4.7 & 80, 100, 120 & 1.43$\times10^{-3}$ \\
CTA\,1 & 119.5 & 10.0 & 80, 100, 120 & 1.70$\times10^{-3}$ \\
Tycho & 120.1 & 1.4 & 60, 80, 100 & 9.51$\times10^{-4}$ \\
HB\,9 & 160.4 & 2.8 & 90, 100, 120 & 2.15$\times10^{-3}$ \\
		\hline
	\end{tabular}
	\label{tab:phot_params}
\end{table}

%%%%%%%%%%%%%%%%%%%%%%%%%%%%%%%%%
%%---------- Section ----------%%
%%%%%%%%%%%%%%%%%%%%%%%%%%%%%%%%%

\section{Methodology for SNR characterization}
\label{sec:methodology}

This section presents the methodologies used to obtain 
the observational properties of SNRs, based mainly 
on the description and fitting of the spectral energy 
distribution, both in intensity and in polarization. We 
start with the estimation of integrated flux densities in 
intensity and polarization, followed by their modelling, 
which includes the implementation of the Markov Chain 
Monte Carlo (MCMC) approach for sampling the posterior 
distribution in a Bayesian framework. In addition, 
strategies to perform the rotation measure fit and 
achieve constraints on the intensity amplitude of the 
AME are also described.

%-----------------------
%%---------- Sub-section
\subsection{Flux densities and uncertainties}
\label{sec:flux_densities}
The spectral energy distribution is determined by the 
flux densities observed for all frequency bands.
Since the QUIJOTE-MFI maps (and most of the low-frequency
 ancillary maps) have effective resolutions of 
$\lesssim$\,1\,deg, we are orientated towards the integrated
 properties of intensity and polarization signals.  
We therefore estimate the integrated flux densities in Stokes $I$, 
$Q$ and $U$ maps on angular scales of 1\,deg. 
For this purpose, we use the aperture photometry technique, 
which has the benefit of being robust against uncertainties in 
the beam or source shapes. In fact, this method is useful 
when the intensity and polarization spatial morphologies are different.
Aperture photometry is a widely used technique 
\citep[see implementations in][and references therein]{2011ApJ...729...25L,2012AdAst2012E..40R,2015MNRAS.452.4169G,mfi_widesurvey_galacticPlane},
 where we computed the flux density from the integrated 
flux (the averaged temperature) inside a circular aperture 
$r_{\rm src}$ defining the solid angle of the source 
($\Omega_{\rm src}$). The background emission in the line 
of sight is corrected by subtracting the median temperature in 
a circular corona (or ring) between radii $r_{\rm int}$ and 
$r_{\rm ext}$ with solid angle $\Omega_{\rm bkg}$. 
The flux density uncertainty associated with this measurement 
is computed as the propagation of the standard deviation from 
the corona, also taking into account the number of pixels in the 
source and  background regions \citep[see ][]{2015MNRAS.452.4169G}.

Table~\ref{tab:phot_params} shows the input 
parameters defining the source aperture and the background 
ring for each SNR. These parameters are the same for all the maps.
They are selected to cover the entire SNR emission in intensity 
and polarization across all bands and to reduce the effects 
of prominent background structures. For instance, the apertures
 of Cygnus Loop are off-centre to avoid contamination from
 the Galactic plane in the northern region of the remnant (see 
Figure~\ref{ima:cygnusloop}).
Although this method provides an efficient correction of the 
background emission, two SNRs are contaminated by strong point 
or extended sources. 
For this reason, we apply a mask for HB\,21 and CTB\,80 in 
all the maps to avoid emission from two sources lying in the 
background region.
The uncertainty estimator might also be biased owing to the 
high Galactic variation, large-scale correlated noise ($1/f$ 
noise) or residual CMB contribution, mainly in sources covering 
large regions.
Other alternatives, such as the selection of random control regions,
 are therefore explored to obtain a robust 
estimate of the errors.

\subsubsection{Random regions for uncertainty estimation}
\label{sec:uncertainties_controltests}

In general, our SNRs are relatively close to the Galactic 
plane in regions with diffuse Galactic emission that varies
 markedly with position (see 
Figures~\ref{ima:ctb80},$\ldots$,\ref{ima:tycho} in Appendix~\ref{appendix:fluxes_images_snr}).
Even the Cygnus Loop and CTA1, which are located at $|b| >$\,5$^\circ$, 
show non-negligible background diffuse emission. 
In this situation, estimation of final errors from the 
pixel-to-pixel scatter is not straightforward and needs 
knowledge of the exact correlation of the noise.
To have a more reliable assessment of the impact of background 
fluctuations on the noise, we have selected control regions in 
the neighbourhood of our SNR, taking care to 
\citep[following ][]{2011ApJ...729...25L,2015MNRAS.452.4169G,mfi_widesurvey_ame_srcs,mfi_widesurvey_w51} 
avoid zones with strong point or extended sources.
We apply the 
photometry method in the random control regions of each SNR
while taking into account the same apertures 
considered for that remnant (Table~\ref{tab:phot_params}). 
The uncertainty arising from the diffuse background fluctuations 
is then estimated through the standard deviation of the flux 
densities extracted from the random control regions. 
Overall, the background fluctuation contribution to the error 
bar is significant for the $I$ Stokes parameter with respect 
to the instrumental noise level, while its contribution is 
lower for Stokes $Q$ and $U$.
Note that this methodology cannot be applied to the 4.8\,GHz 
maps because these are restricted to the Galactic plane.

%%%% sub-section
\subsubsection{polarization measurements}
\label{sec:polarized_measurements}

The polarized SED is characterized via the integrated 
total linearly polarized intensities, $P_{\nu}$. First, 
we obtain the raw polarization intensity through the 
following expression:
\begin{equation}
    P_{\rm raw,\nu} = \sqrt{Q_{\nu}^2 + U_{\nu}^2},
\label{eq:pol_flux_integrated}
\end{equation}
where $Q_{\nu}$ and $U_{\nu}$ are the integrated flux 
densities from Stokes $Q$ and $U$ at frequency $\nu$.
However, this raw polarization is affected by the 
positive noise bias 
\citep[see][and references therein]{2006PASP..118.1340V,2012AdAst2012E..40R}. 
To obtain the true integrated polarization, $P_{\nu}$, 
a debiasing procedure is applied by using the Modified 
Asymptotic estimator \citep[MAS,][]{2014MNRAS.439.4048P}:
\begin{equation}
    P_{\nu} = P_{\rm raw}-b^2\frac{1-e^{-P_{\rm raw}^2/b^2}}{2P_{\rm raw}}
\end{equation}
where $b = \sqrt{(Q_\nu \cdot \sigma_{U_\nu})^2+(U_\nu \cdot \sigma_{Q_\nu})^2}/P_{\rm raw}$.
From the debiased polarized measurement, we can compute 
the integrated polarization fraction via 
$\Pi_\nu = 100\times P_{\nu}/I_{\nu}$. However, this 
calculation would be a biased estimate of the true 
polarization fraction, which becomes more pronounced when 
the signal-to-noise of the intensity estimate $I_\nu$ is 
low. We should then apply debiasing of the quantity 
$P_{\rm raw}/I_\nu$. To this aim we evaluate numerically, 
through Monte Carlo simulations, the probability density 
function of the true polarization fraction and integrate 
this curve to obtain the central value and confidence 
interval, as was done in \cite{2017MNRAS.464.4107G}.

In addition, the apparent orientation of the projected 
polarization electric vector is described by the polarization angle:
\begin{equation}
\label{eq:pol_angle}
\gamma_{\nu} = 0.5 \ \arctan(-U_{\nu}/Q_{\nu}),    
\end{equation}
where $\arctan(\ldots)$ represents the 4-quadrant inverse 
function providing angles between $-\pi$ and $\pi$, where 
the quadrant is assigned according to the sign of the 
$Q_{\nu}$ and $U_{\nu}$ flux densities. In expression~\ref{eq:pol_angle}, 
the negative sign appears to maintain the meaning of $\gamma$, 
since the COSMO convention differs from the IAU convention 
\citep[$U = -U_{\rm IAU}$,][]{1996A&AS..117..161H}. 

The flux densities from intensity and polarization are 
presented in Tables~\ref{tab:fluxes_ctb80}--\ref{tab:fluxes_hb9} (Appendix~\ref{appendix:fluxes_images_snr}), 
where uncertainties are computed by random 
control apertures analysis (section~\ref{sec:uncertainties_controltests}).
Overall, these measurements have not been colour corrected 
because this correction depends on the model considered 
(see section below).
Note that the polarization intensity ($P_{\nu}$) and the 
polarization fraction ($\Pi_\nu$) correspond to the debiased 
measurements.

\subsection{SED fitting procedure}
\label{sec:metodology_sed_modelling}

%%%%
\subsubsection{SED Models}
\label{sec:sed_models}

The spectral energy distributions of SNRs are dominated by 
synchrotron emission, which is the main component at low 
frequencies ($\le$\,100\,GHz).
However, in the microwave range other physical mechanisms 
can be present, such as free--free, AME, thermal dust emission 
and a residual emission from CMB anisotropies.
Some of these contributions arise as spurious emission in the 
line of sight (e.g.\ the CMB residual) while the rest can be 
associated with the SNR or their surrounding interstellar 
medium (such as free--free and thermal dust emissions), 
 the presence (or not) of AME in our SNR 
sample being 
of particular interest. The strategy used to constrain the intensity amplitude 
of AME is presented in Section~\ref{sec:ame_constraints}.

A preliminary analysis could determine whether the observed SEDs 
could be well fitted using a combination of synchrotron 
and thermal dust emission, where the former is associated 
with emission coming from the remnants and the latter 
is expected to be residual emission from the surrounding environment.
Visual inspection of the SED 
(Figures~\ref{fig:sed_CTB80}--\ref{fig:sed_HB9}) shows that 
the intensity signal appears to be dominated by synchrotron 
and thermal dust (low contribution) components, while the 
polarized signal is well modelled by one component (a synchrotron 
emission).
In what follows, we describe the parametric representation 
of synchrotron and thermal dust emission, along with models 
proposed to fit the intensity and polarization SEDs in joint 
and separate analyses.

\textit{Synchrotron emission}. The synchrotron radiation 
reflects the power-law energy spectrum of the relativistic 
particles responsible of the observed emission. Therefore, as 
usual, we consider the standard power-law function (labelled 
PL) with spectral index $\alpha$ 
\citep{1970ranp.book.....P,1990tra..book.....R}:
\begin{equation}
S_{\rm syn}(\nu;A_{\rm s},\alpha) = A_{\rm syn} 
\left(\dfrac{\nu}{\nu_{\rm ref}} \right)^{\alpha},
\label{eq:model_powerlaw}
\end{equation}
where $A_{\rm syn}$ is the synchrotron amplitude at the 
reference frequency $\nu_{\rm ref}$.
This synchrotron parametric model is used in this paper, 
in both the intensity and polarization analyses, where the 
emission amplitude is normalised at $\nu_{\rm ref}$=\,11.1\,GHz in both cases.

The intensity SEDs of CTB\,80 and HB\,21 show evidence 
of a break in the synchrotron spectrum at low frequencies 
($\lessapprox5$\,GHz), the spectral index being significantly
 steeper at high frequency.
In these cases, there is no standard model for the SED 
characterization, a power law with an exponential 
cut-off function being one of the most frequently used models 
\citep[e.g.][]{2013ApJ...779..179P,2021MNRAS.500.5177L,mfi_widesurvey_w51}.
In our cases, the spectral change is noticeably abrupt 
and the spectra follow a power law below (radio frequencies) 
and above the break frequency, which is not so compatible 
with the gradual change of spectral indices described by the 
cut-off curve or Rolloff models (see 
Appendix~\ref{appendix:curved_models} and 
Figure~\ref{fig:sed_powerlaw_examples}).
We therefore consider a smooth broken power-law spectral model (SBPL):
\begin{equation}
\begin{aligned}
S_{\rm syn}^{\rm SBPL}(\nu;A_{\rm syn},\alpha_{\rm bb},\alpha_{\rm ab},\nu_{\rm b},m) 
= & \ A_{\rm syn} \left( \dfrac{\nu}{\nu_{\rm ref}} \right)^{\alpha_{\rm bb}} \\
 & \cdot \left( \dfrac{ 1 + \left(\nu / \nu_{\rm b}\right)^m}{ 1 + 
\left(\nu_{\rm ref} / \nu_{\rm b}\right)^m}  \right)^{\frac{\alpha_{\rm ab} - \alpha_{\rm bb}}{m}},
\end{aligned}
\label{eq:model_brokenpowerlaw_smooth}
\end{equation}
where $\alpha_{\rm bb}$ and $\alpha_{\rm ab}$ are respectively 
the spectral indices below and above of the break frequency 
$\nu_{\rm b}$.
The synchrotron amplitude is normalised at $\nu_{\rm ref}=\,11.1$\,GHz.
The $m$ parameter determines the smoothing level of the spectral break.
Since we have assumed $m>0$, the sign of 
$\Delta\alpha =\alpha_{\rm ab} - \alpha_{\rm bb}$ determines 
whether the spectrum is flatter ($\Delta\alpha>0$) or steeper 
($\Delta\alpha<0$) after the spectral break. 
As $m$ increases, the spectrum converges to the sharp shape
 of the Broken Power Law (BPL, equation~\ref{eq:model_brokenpowerlaw_sharp}) 
function (see Appendix~\ref{appendix:curved_models} and 
Figure~\ref{fig:sed_powerlaw_examples}).
In our implementation, the $m$ value is fixed in order to 
have the same number of free parameters as in the BPL model 
(i.e.\ $A_{\rm syn}$, $\alpha_{\rm bb}$, $\alpha_{\rm ab}$ 
and $\nu_{\rm b}$). 
For HB\,21 and CTB\,80, we adopted a smooth broken 
power-law (SBPL) function with $m=10$, because it provides close 
concordance between the SBPL and BPL spectral models, and is 
also in agreement with the smooth transition of the power 
laws observed for these remnants (see 
Appendix~\ref{appendix:curved_models}).
%

%%%----
\textit{Thermal dust contribution}. It can be fitted by a 
single-temperature modified blackbody function assuming 
an optically thin scenario:
\begin{equation}
S_{\rm dust}(\nu;\tau_{353}, T_{\rm d},\beta_{\rm d}) = 
\tau_{353} \ \left(\dfrac{\nu}{\rm 353\,GHz} \right)^{\beta_{\rm d}} \dfrac{2 h \nu^3}{c^2 
\left({\rm \scalebox{1.3}{e}}^{h \nu/k T_{\rm d}} - \scalebox{1.1}{1} \right) } \Omega_{\rm src},
\label{eq:model_thermaldust}
\end{equation}
where $\tau_{353}$ is the optical depth at 353\,GHz for 
the solid angle subtended by the aperture defining the source 
($\Omega_{\rm src}$). $\beta_{\rm d}$ and $T_{\rm d}$ are 
the dust emissivity and dust temperature respectively.

Our estimates provide upper limits on 
the polarized flux densities for $\nu \gtrsim$\,90\,GHz.
These results are expected because the intensity of thermal 
dust emission is detected with low signal-to-noise.
Therefore, this type of emission is not considered 
in the polarization SED modelling.
%

%%%----
\textit{Remarks on proposed models.} 
We have considered and implemented three fitting 
procedures based on intensity data only (\textsc{Int}), 
on polarization data only (\textsc{Pol}) and on a combination
 of the two (\textsc{IntPol}).
The model for the \textsc{Int} fit case, 
$m^{\rm int}(\theta)$, consists of synchrotron (in either the PL 
or the SBPL case) and thermal dust emissions, except for
 CTB\,80 and CTA\,1, where the thermal dust component is not 
included because measurements at frequencies higher than 
$\sim 50$\,GHz are noise dominated. Table~\ref{tab:models_SNRs2} 
shows the model and parameter space $\theta$ for each SNR.
As an example, the model that is fitted to the intensity 
SED of HB\,21 consists of the parameter space $\theta_{\rm Int,HB\,21}
=\{A_{\rm syn}, \alpha_{\rm bb}, \alpha_{\rm ab}, \nu_{\rm b}, \tau_{353},
 T_{\rm d}, \beta_{\rm d}\}$, while CTA\,1 is modelled
 with two parameters ($\theta_{\rm Int,CTA\,1}=\{A_{\rm syn}, \alpha_{\rm bb}\}$).
In the case of the \textsc{Pol} fit, the synchrotron 
emission is modelled with only the power-law function (PL) 
because the sampling of the polarized SED is limited. 
Since polarized measurements are available above 11.1\,GHz 
(or above 4.8\,GHz for two cases), we do not consider  
polarized spectral break behaviour for either CTB\,80 or HB\,21.
Therefore, the proposed model is a power law, $\theta_{\rm Pol}
=\{A_{\rm syn}^{\rm pol}, \alpha^{\rm pol}\}$, for each 
SNR (see bottom of Table~\ref{tab:models_SNRs2}). 

Concerning the simultaneous intensity and polarization 
fit (\textsc{IntPol}), we assume that synchrotron 
radiation has the same spectral behaviour in both intensity 
and polarization; e.g.\ $\alpha^{\rm int} = \alpha^{\rm pol}$ 
in the power-law function case. Thus, the polarized synchrotron 
SED is written in terms of the intensity SED as 
$S^{\rm pol}_{\rm syn} = \Pi_{\rm syn} \cdot S_{\rm syn}$, 
where $\Pi_{\rm syn}$ is the polarization fraction of the 
synchrotron emission, on the assumption that $\Pi_{\rm syn}$ 
is constant.
For instance, the model for HB\,21 consists of intensity 
($m^{\rm int}= S_{\rm syn} + S_{\rm dust}$) and polarized 
($m^{\rm pol}= \Pi_{\rm syn} \cdot S_{\rm syn}$) components 
whose parameter space is $\theta_{\rm IntPol,HB21}=
\{A_{\rm syn}, \alpha_{\rm bb}, \alpha_{\rm ab}, \nu_{\rm b},\Pi_{\rm syn}, 
\tau_{353}, T_{\rm d}, \beta_{\rm d}\}$.
For each SNR, Table~\ref{tab:models_SNRs2} presents the 
intensity (int) and polarized (pol) components for the 
\textsc{IntPol} case, as well as its space parameter $\theta_{\rm IntPol}$.

\begin{table*}
	\centering
	\caption{Models and parameter spaces considered 
for each SNR. Parametric models are presented in 
Section~\ref{sec:sed_models} and consist of $S_{\rm syn}$ 
(equation~\ref{eq:model_powerlaw}) and $S_{\rm syn}^{\rm SBPL}$ 
(equation~\ref{eq:model_brokenpowerlaw_smooth}) for 
synchrotron, $S_{\rm dust}$ (equation~\ref{eq:model_thermaldust}) 
for thermal dust, and $\gamma_{\rm PA}$ 
(equation~\ref{eq:model_polang}) for polarization angle.}
	\begin{tabular}{l || l} % four columns, alignment for each
		\hline
		{\sc Int}  & {\sc IntPol} \\
		%\hdashline
		\hline
    \multicolumn{2}{l}{\textbf{HB\,21}} \\ [0.5 ex]
    $S^{\rm SBPL}_{\rm syn}(A_{\rm syn}, \alpha_{\rm bb}, \alpha_{\rm ab}, \nu_{\rm b}, m=10)$ + $S_{\rm dust}(\tau_{353}, T_{\rm d}, \beta_{\rm d})$ & int: $S^{\rm SBPL}_{\rm syn}(A_{\rm syn}, \alpha_{\rm bb}, \alpha_{\rm ab}, \nu_{\rm b}, m=10)$ + $S_{\rm dust}(\tau_{353}, T_{\rm d}, \beta_{\rm d})$\\ [0.5 ex]
    $\theta_{\rm \textsc{Int}}$:\,$\{A_{\rm syn}, \alpha_{\rm bb}, \alpha_{\rm ab}, \nu_{\rm b},\tau_{353}, T_{\rm d}, \beta_{\rm d}\}$& pol: $\Pi_{\rm syn} \times S^{\rm SBPL}_{\rm syn}(A_{\rm syn}, \alpha_{\rm bb}, \alpha_{\rm ab}, \nu_{\rm b}, m=10)$\\[0.5 ex]
    &$\theta_{\rm \textsc{IntPol}}$:\,$\{A_{\rm syn}, \Pi_{\rm syn}, \alpha_{\rm bb}, \alpha_{\rm ab}, \nu_{\rm b},\tau_{353}, T_{\rm d}, \beta_{\rm d}\}$\\  [1.0 ex]
    \multicolumn{2}{l}{\textbf{CTB\,80}} \\ [0.5 ex]
    $S^{\rm SBPL}_{\rm syn}(A_{\rm syn}, \alpha_{\rm bb}, \alpha_{\rm ab}, \nu_{\rm b}, m=10)$& int: $S^{\rm SBPL}_{\rm syn}(A_{\rm syn}, \alpha_{\rm bb}, \alpha_{\rm ab}, \nu_{\rm b}, m=10)$\\ [0.5 ex]
    $\theta_{\rm \textsc{Int}}$:\,$\{A_{\rm syn}, \alpha_{\rm bb}, \alpha_{\rm ab}, \nu_{\rm b}\}$& pol: $\Pi_{\rm syn} \times S^{\rm SBPL}_{\rm syn}(A_{\rm syn}, \alpha_{\rm bb}, \alpha_{\rm ab}, \nu_{\rm b}, m=10)$\\ [0.5 ex]
    & $\theta_{\rm \textsc{IntPol}}$:\,$\{A_{\rm syn}, \Pi_{\rm syn}, \alpha_{\rm bb}, \alpha_{\rm ab}, \nu_{\rm b}\}$\\  [1.0 ex]

    \multicolumn{2}{l}{\textbf{Cygnus Loop, HB\,9 and Tycho}} \\ [0.5 ex]
    $S_{\rm syn}(A_{\rm syn}, \alpha)$ + $S_{\rm dust}(\tau_{353}, T_{\rm d}, \beta_{\rm d})$& int: $S_{\rm syn}(A_{\rm syn}, \alpha)$ + $S_{\rm dust}(\tau_{353}, T_{\rm d}, \beta_{\rm d})$\\ [0.5 ex]
    $\theta_{\rm \textsc{Int}}$:\,$\{A_{\rm syn}, \alpha, \tau_{353}, T_{\rm d}, \beta_{\rm d}\}$& pol: $\Pi_{\rm syn} \times S_{\rm PL}(A_{\rm syn}, \alpha)$\\ [0.5 ex]
    & {$\theta_{\rm \textsc{IntPol}}$:\,$\{A_{\rm syn}, \Pi_{\rm syn}, \alpha, \tau_{353}, T_{\rm d}, \beta_{\rm d}\}$}\\  [1.0 ex]
    \multicolumn{2}{l}{\textbf{CTA\,1}} \\ [0.5 ex]
    $S_{\rm syn}(A_{\rm syn}, \alpha)$ & int: $S_{\rm syn}(A_{\rm syn}, \alpha)$ \\
    $\theta_{\rm \textsc{Int}}$:\,$\{A_{\rm syn}, \alpha\}$& pol: $\Pi_{\rm syn} \times S_{\rm PL}(A_{\rm syn}, \alpha)$\\
    & {$\theta_{\rm \textsc{IntPol}}$:\,$\{A_{\rm syn}, \Pi_{\rm syn}, \alpha\}$}\\  [1.0 ex]
\hline
{\underline{\textsc{Pol}}} & {\underline{\textsc{Pol. Angle}}} \\ [0.5 ex]
$S_{\rm syn}(A_{\rm syn}^{\rm pol}, \alpha^{\rm pol})$& $\gamma_{\rm PA}({\rm RM},\gamma_0)$ \\ [2.0 ex]
$\theta_{\rm \textsc{Pol}}$:\,$\{A_{\rm syn}^{\rm pol}, \alpha^{\rm pol} \}$& $\theta_{\rm \textsc{PA}}$:\,$\{{\rm RM}, \gamma_0 \}$\\ [2.0 ex]
		\hline
	\end{tabular}
	\label{tab:models_SNRs2}
\end{table*}

\subsubsection{Likelihood analysis}
\label{sec:likelihood_analysis}

For each model, the parameter space $\theta$ is explored 
using a Markov Chain Monte Carlo (MCMC) analysis, which 
allows us to extract the parameters and uncertainties 
through marginalization of the posterior probability 
distribution function of each parameter.
We used the publicly available affine-invariant MCMC 
sampler algorithm called 
{\sc emcee}\footnote{\label{fn:emcee}\textsc{emcee} is 
available at \url{https://emcee.readthedocs.io/en/stable/\#}} 
\citep{2013PASP..125..306F}, where the model is fitted 
to the data by maximizing the corresponding logarithmic 
posterior distribution:

\begin{equation}
\ln\left( \mathcal{L}(\theta) \ \prod_j \mathcal{P}(\theta_j) \right)  
\propto -\dfrac{1}{2}  \scalebox{1.3}{$\chi$}^2(\theta)
+ \sum_{j} \ln\left(\mathcal{P}(\theta_j)\right),
\label{eq:likelihood_sed}
\end{equation}
where $\mathcal{L}(\theta)$ is the likelihood and 
$\mathcal{P}(\theta_j)$ corresponds to the prior distribution
 on the $j$-th parameter.
Concerning priors, we impose positive top-hat 
priors\footnote{Positive priors are uniform priors, $\mathcal{U}(a,b)$, 
in the positive domain of the explored parameter, i.e.\ 
$\mathcal{U}(0,\infty)$.} on amplitudes of our components, 
such as $A_{\rm syn}$, $A^{\rm pol}_{\rm syn}$, $\tau_{\rm 353}$ 
and $\Pi_{\rm syn}$. 
For the synchrotron curved model, we set $|\alpha_{\rm ab} 
- \alpha_{\rm bb}| \geq$\,0.1 in order to avoid strong degeneracy. 
For the thermal dust, the priors on $\beta_{\rm d}$ and 
$T_{\rm d}$ are a proxy of the range available in the 
\textsc{fastcc} (Section~\ref{sec:data_systematic_effects}), 
yielding uniform priors  $\mathcal{P}(\scalebox{0.9}{$\beta_{\rm d}$})
 = \mathcal{U}\rm{(1.0,4.0)}$ and  $\mathcal{P}(\scalebox{0.9}{$T_{\rm d}$})
 = \mathcal{U}\rm{(10\,K,40\,K)}$ respectively.

As we establish positive or uniform priors on parameters 
(Table~\ref{tab:models_SNRs2}), the maximization of posterior 
distribution, equation~\ref{eq:likelihood_sed}, is driven by 
the shape of the objective function $\chi^2$:
\begin{equation}
\chi^2 = \big(\textit{\bf D} - {\textbf m(\theta)} \big)^T {\textbf N}^{-1} \big({\textbf D}
 - {\textbf m(\theta)} \big),
\label{eq:chi2_sed}
\end{equation}
in which ${\bf D}$ and ${\bf m(\theta)}$ identify vectors 
with the raw flux densities and the model predictions for 
all frequencies respectively.
In this equation, ${\bf N}$ is the noise covariance matrix, 
whose inverse matrix (${\bf N}^{-1}$) must exist.
We can now perform the characterization of the posterior 
distribution (equation~\ref{eq:likelihood_sed}).
However, the colour correction effects and the correlation 
between QUIJOTE-MFI bands are also elements that need to be taken into account.

\textit{Colour correction}.
As mentioned in Section~\ref{sec:data_systematic_effects}, 
 we use \textsc{fastcc}$^{\ref{fn:fastcc}}$ to compute 
the colour correction coefficients,  $cc_{\nu}(\theta)$, 
which depend on the spectral shape of the SED. These are 
multiplicative factors applied to the flux densities and 
uncertainties of QUIJOTE, WMAP, \textit{Planck} and DIRBE data. 
After factoring these into equation~\ref{eq:chi2_sed}, they
 are included in the likelihood formalism as follows:
\begin{equation}
    \chi^2 = \left({\bf D} - 
\dfrac{{\bf m(\theta)}}{{\bf cc(\theta)}} \right)^T {\bf N}^{-1} \left({\bf D} - 
\dfrac{{\bf m(\theta)}}{{\bf cc(\theta)}} \right),
\label{eq:chi2_sed_cc}
\end{equation}
where ${\bf cc(\theta)}$ is the vector with colour 
correction factors for all frequencies.
Note that $cc_\nu(\theta)$=\,1 when the colour correction 
is not applied, as in the case of low-frequency narrow-band surveys 
(Section~\ref{sec:data_lowfreq}).
\textsc{fastcc} provides accurate colour correction coefficients 
in our sample of SNRs as their spectra are reasonably well 
approximated by a power law. In fact, the differences between 
coefficients derived from \textsc{fastcc} and from the exact 
integration on the bandpass are very small (lower than 1\,\%).

\textit{QUIJOTE-MFI bands correlation}.
For QUIJOTE-MFI measurements, this correlation between 
frequency bands leads to a non-diagonal noise matrix 
(Section~\ref{sec:data_QUIJOTE}). Therefore, the normalised 
correlation terms, $\rho_{i,j}$, from the pairs of bands 
(identified as $i$ and $j$) are non-negligible:
\begin{equation}
  \begin{pmatrix}
    \sigma_{i}^2& \rho_{i,j}\ \sigma_{i} \sigma_{j}\\
    \rho_{j,i}\ \sigma_{i} \sigma_{j}&\sigma_{j}^2,\\
  \end{pmatrix}
\end{equation}
where $\rho_{j,i}=\rho_{i,j}\neq 0$ is assumed and 
$\sigma_k$ is the flux density uncertainty measured 
in the $k$ band. 
As discussed in Section~\ref{sec:data_QUIJOTE}, 
we take into account two cases, $\rho_{11,13}$ at low 
frequency and $\rho_{17,19}$ at high frequency, which 
have different values for intensity and polarization 
(see Table~\ref{tab:band_correlation}). 
Ideally, the noise is uncorrelated (i.e.\ 
$\rho_{i,j}$\,=\,0\,$\forall \ i,j$) and the covariance
 noise matrix is diagonal; i.e.\ $({\bf N}^{-1})_{i,j} = 
\delta_{i,j}/\sigma_i^2$, so the objective function $\chi^2$ becomes:
\begin{equation}
\chi^2 = \sum_{i=1}^{n} \dfrac{ \left( d_i - 
m_i(\theta)/cc_i(\theta) \right)^2}{\sigma_{i}^2}, \ 
\end{equation}
which is the standard and simplest representation 
applied in SED fitting. 
In this equation, $d_i$ and $\sigma_i$ are the flux 
densities and uncertainties at the $i$-th frequency band. 

For the \textsc{Int} and \textsc{Pol} analysis, the 
objective functions follow equation~\ref{eq:chi2_sed} 
with the respective data and model;  i.e.\ we have 
$\chi^2_{\rm int}(\theta_{\rm Int})$ and 
$\chi^2_{\rm pol}(\theta_{\rm Pol})$ separately for 
the intensity (\textsc{Int}) and  the polarization 
(\textsc{Pol}) fits respectively.
The objective function for the 
\textsc{IntPol} fit, however, is written in terms of the 
intensity and polarization  objective functions as follows:
\begin{equation}
\chi^2(\theta_{\rm \textsc{IntPol}}) = \chi^2_{\rm int} 
(\theta_{\rm \textsc{IntPol}})+ \chi^2_{\rm pol}(\theta_{\rm \textsc{IntPol}}),
\end{equation}
which is minimized by the MCMC method, assuming that the 
intensity and polarized noises are uncorrelated.
Overall, the spectral energy distributions presented
throughout this paper are generated with colour-corrected 
flux densities coming from the {\sc IntPol} analysis (see 
Figures~\ref{fig:sed_CTB80}--\ref{fig:sed_HB9}).

\begin{table*}
	\centering
	\caption{Results obtained from the characterization 
of the spectral energy distribution in intensity and 
polarization for the six SNRs: HB\,21, CTB\,80 (both with 
curved spectra), the Cygnus Loop, CTA\,1, Tycho and HB\,9. 
Three SED model approaches are performed: 1) only 
intensity measurements (\textsc{Int}), 2) only polarization 
measurements (\textsc{Pol}) and 3) fitting simultaneous intensity
 and polarization measurements (\textsc{IntPol}). The last
 is considered as our reference SED analysis. The models 
and parameter spaces are detailed in Table~\ref{tab:models_SNRs2}. 
For the {\sc Pol} approach, $A_{\rm syn}$ and $\alpha$ 
represent $A_{\rm syn}^{\rm pol}$, and $\alpha^{\rm pol}$ 
in Table~\ref{tab:models_SNRs2}. Parameters are estimated 
from the 50th percentile of the marginalized posterior 
distribution functions, while their uncertainties are
 estimated from the 16th and 84th percentiles.}
	\label{tab:parameters_from_modelling}
\begin{tabular}{l c c c || c c c }
\hline
& \textsc{IntPol} & \textsc{Int} & \textsc{Pol} &\textsc{IntPol} & \textsc{Int} & \textsc{Pol}\\
\hline
& \multicolumn{3}{c}{\textbf{CTB\,80}} & \multicolumn{3}{c}{\textbf{HB\,21}}\\
$A_{\rm syn}$ [Jy] & 21.0$^{+1.1}_{-1.0}$ & 20.9$^{+1.2}_{-1.1}$ & 0.8$\pm$ 0.1 & 58.1$^{+2.5}_{-2.2}$ & 56.1$^{+2.2}_{-2.1}$ & 3.2$\pm$ 0.2\\ [0.6 ex]  
$\Pi_{\rm syn}$ [\%] & 3.7$\pm$ 0.3 & - & - & 5.0$\pm$ 0.2 & - & -\\ [0.6 ex]  
$\alpha_{\rm bb}$ & -0.24$^{+0.07}_{-0.06}$ & -0.27$^{+0.06}_{-0.04}$ & - & -0.34$^{+0.04}_{-0.03}$ & -0.33$\pm$ 0.04 & -\\ [0.6 ex]  
$\alpha_{\rm ab}$ & -0.60$^{+0.04}_{-0.05}$ & -0.70$^{+0.10}_{-0.12}$ & -0.52$\pm$ 0.07 & -0.86$^{+0.04}_{-0.05}$ & -0.80$^{+0.04}_{-0.05}$ & -1.02$\pm$ 0.08\\ [0.6 ex]  
$\nu_{\rm b}$ [GHz] & 2.0$^{+1.2}_{-0.5}$ & 3.7$^{+1.2}_{-1.7}$ & - & 5.0$^{+1.2}_{-1.0}$ & 4.0$^{+1.1}_{-0.8}$ & -\\ [0.6 ex]  
$\tau_{\rm 353}$ [10$^{-7}$] & - & - & - & 143.7$^{+69.8}_{-52.5}$ & 146.0$^{+69.7}_{-53.2}$ & -\\ [0.6 ex]  
$T_{\rm d}$ [K] & - & - & - & 12.5$^{+2.6}_{-1.7}$ & 12.2$^{+2.4}_{-1.6}$ & -\\ [0.6 ex]  
$\beta_{\rm d}$ & - & - & - & $2.0\pm0.5$ & $2.2\pm0.5$ & -\\ [0.6 ex]  
$\chi^2_{\rm dof}$ & 1.83 & 2.24 & 0.70 & 0.63 & 0.47 & 0.54\\ [0.6 ex]  

\hline

& \multicolumn{3}{c}{\textbf{Cygnus Loop}} & \multicolumn{3}{c}{\textbf{CTA\,1}}\\[1.0 ex]
$A_{\rm syn}$ [Jy] & 44.6$^{+1.0}_{-1.1}$ & 44.6$\pm$ 1.0 & 2.9$^{+0.6}_{-0.5}$ & 8.9$\pm$ 0.6 & 9.0$\pm$ 0.6 & 1.0$\pm$ 0.2\\ [0.6 ex]  
$\Pi_{\rm syn}$ [\%] & 6.0$\pm$ 0.4 & - & - & 9.4$^{+1.2}_{-1.1}$ & - & -\\ [0.6 ex]  
$\alpha$ & -0.47$\pm$ 0.02 & -0.47$\pm$ 0.02 & -0.59$\pm$ 0.23 & -0.50$\pm$ 0.03 & -0.50$\pm$ 0.03 & -0.84$^{+0.29}_{-0.27}$\\ [0.6 ex]  
$\tau_{\rm 353}$ [10$^{-7}$] & 27.3$^{+9.3}_{-7.1}$ & 27.3$^{+9.2}_{-7.2}$ & - & - & - & -\\ [0.6 ex]  
$T_{\rm d}$ [K] & 15.8$^{+2.1}_{-2.0}$ & 15.8$\pm$ 2.1 & - & - & - & -\\ [0.6 ex]  
$\beta_{\rm d}$ & 1.8$^{+0.4}_{-0.3}$ & 1.8$^{+0.4}_{-0.3}$ & - & - & - & -\\ [0.6 ex]  
$\chi^2_{\rm dof}$ & 0.63 & 0.59 & 0.75 & 0.39 & 0.25 & 0.45\\ [0.6 ex]  

\hline

& \multicolumn{3}{c}{\textbf{Tycho}} & \multicolumn{3}{c}{\textbf{HB\,9}}\\[1.0 ex]
$A_{\rm syn}$ [Jy] & 12.9$\pm$ 0.6 & 12.9$\pm$ 0.6 & 0.1$^{+0.2}_{-0.1}$ & 20.6$\pm$ 0.7 & 20.7$\pm$ 0.7 & 1.4$\pm$ 0.1\\ [0.6 ex]  
$\Pi_{\rm syn}$ [\%] & 0.0$^{+1.5}_{-1.6}$ & - & - & 6.9$\pm$ 0.5 & - & -\\ [0.6 ex]  
$\alpha$ & -0.60$\pm$ 0.02 & -0.60$\pm$ 0.02 & -1.57$^{+1.29}_{-1.35}$ & -0.51$\pm$ 0.02 & -0.50$\pm$ 0.02 & -0.60$\pm$ 0.08\\ [0.6 ex]  
$\tau_{\rm 353}$ [10$^{-7}$] & 119.3$^{+28.3}_{-24.8}$ & 119.6$^{+28.3}_{-24.8}$ & - & 82.0$^{+22.2}_{-16.9}$ & 82.1$^{+23.0}_{-16.9}$ & -\\ [0.6 ex]  
$T_{\rm d}$ [K] & 14.7$^{+1.4}_{-1.3}$ & 14.6$^{+1.4}_{-1.3}$ & - & 14.9$^{+1.6}_{-1.7}$ & 14.8$^{+1.6}_{-1.7}$ & -\\ [0.6 ex]  
$\beta_{\rm d}$ & 2.2$^{+0.4}_{-0.3}$ & 2.2$^{+0.4}_{-0.3}$ & - & $1.5 \pm 0.3$ & $1.5 \pm 0.3$ & -\\ [0.6 ex]  
$\chi^2_{\rm dof}$ & 0.45 & 0.69 & 0.02 & 0.44 & 0.23 & 0.76\\ [0.6 ex]   
\hline
\end{tabular}
\end{table*}

%%%%%%%%%%%%%%%%%%%%%%%%%%%%%%%%%%%%%%%%%
\subsubsection{Estimation of parameters}

The characterization of the posterior distributions is performed by 
running {\sc emcee} with 32 chains, although the iteration 
steps and the \emph{burn-in} sample (first steps excluded 
of each chain) depend on the number of paremters of each 
case. For instance, the {\sc IntPol} analysis of HB21 (eight 
parameters) involves 30\,000 iteration steps, with a 
\emph{burn-in} sample of 3\,500 steps.
In order to obtain the parameters from the SED modeling, we 
marginalize the posterior distribution function over each 
parameter $\theta_i$ to yield the PDF($\theta_{i}$). The 
value of the parameter $\theta_{i}$ is then computed as the 
median (the 50th percentile) of its PDF($\theta_{i}$), 
while its associated uncertainty is obtained from the 16th 
and 84th percentiles.
Table~\ref{tab:parameters_from_modelling} reports the 
parameters and uncertainties obtained from the posterior 
distributions for the \textsc{Int}, \textsc{Pol} and 
\textsc{IntPol} fits of our SNR sample.

\subsection{Modelling of the polarization angle} 
%\clearpage
%\newpage

The multi-frequency characterization of the polarization 
angle provides information about the rotation measure. 
The RM is proportional to the product of the line-of-sight 
integral of the electron density and the parallel component 
of the magnetic field.
In our case, we can recover a RM value across the whole 
SNR, which is obtained from modelling the multi-frequency 
polarization angles using the model:
\begin{equation}
\gamma_{\textsc{pa}}(\nu;{\rm RM},\gamma_0) = 
\gamma_0 + {\rm RM} \ \left(\dfrac{c}{{\rm \nu}} \right)^2,
\label{eq:model_polang}
\end{equation}
where $c$ is the speed of light in vacuo and $\gamma_0$ 
corresponds to the intrinsic polarization angle (i.e.\  
the observed angle for $\nu \longmapsto \infty$).
The fitting procedure follows the MCMC approach described 
in Section~\ref{sec:likelihood_analysis}, and we build the 
likelihood via equation~\ref{eq:chi2_sed}. In this case, the 
colour correction factors have no effect, and, for simplicity, 
we do not consider the QUIJOTE-MFI band correlation.

%-----------------------
%%---------- Sub-section
\subsection{Constraint on the AME intensity} 
\label{sec:ame_constraints}

The intensity spectral analysis provide no evidence 
of the AME component towards any of the six SNRs studied here.
This is also supported by the low signal-to-noise 
detection of the intensity thermal dust emission.
In addition, the inclusion of an AME component does not 
improve the modelling of the intensity spectra, and it 
sometime worsens the exploration of parameters with the 
MCMC procedure. 

Notwithstanding, we are interested in constraining the 
intensity amplitude of AME for the six remnants.
For this purpose, we have included an AME contribution in our 
SED fit and have decided to impose strong priors on all components.
We consider the parametric description of the AME component 
\citep[][]{2014ApJ...781..113S,mfi_widesurvey_ame_srcs,mfi_widesurvey_galacticPlane} 
as follows:
\begin{equation}
   S_{\rm AME}(\nu;A_{\textsc{ame}}, \nu_{\textsc{ame}}, 
W_{\textsc{ame}}) = A_{\textsc{ame}} \  
\scalebox{1.3}{\rm e}^{-0.5 \bigl( \ln\left(\nu / \nu_{\textsc{ame}}\right) / W_{\textsc{ame}} \bigr)^2},
\label{eq:model_ame}
\end{equation}
$A_{\textsc{ame}}$ being the AME amplitude at the peak 
frequency $\nu_{\textsc{ame}}$, and $W_{\textsc{ame}}$ 
the width of the parabola in log--log space. 
This component is added to the SED fit of intensity 
(\textsc{Int}), where the parameters and uncertainties 
obtained in the case without AME (Table~\ref{tab:parameters_from_modelling}) 
are considered as Gaussian 
priors.\footnote{\label{fn:prior} A Gaussian prior follows 
the functional shape of a normal Gaussian distribution, 
$\mathcal{N}(\overline{x},\sigma_x)$, described by the mean 
and standard deviation of the $x$-parameter.} For example, 
we use a prior $\mathcal{P}(\scalebox{0.9}{$\alpha$}) 
= \mathcal{N} \rm{(-0.5,0.02)}$ for the synchrotron 
spectral index of HB\,9.
For estimating the  constraints on $A_{\textsc{ame}}$, 
we also impose the priors  
$\mathcal{P}(\scalebox{0.9}{$\nu_{\textsc{ame}}$}) = 
\mathcal{N} \rm{(21\,GHz,2\,GHz)}$ and $\mathcal{P}(\scalebox{0.9}{$W_{\textsc{ame}}$})
= \mathcal{N}(0.7,0.2)$ for the AME component. These 
values are consistent with a recent study of AME in the 
Galactic plane \citep[using QUIJOTE-MFI data at scales of 
1\,deg,][]{mfi_widesurvey_galacticPlane} and also span 
the AME parameters obtained for the SNRs W44, W41 and W51 
\citep[][]{2017MNRAS.464.4107G,mfi_widesurvey_w51}.
The upper limits on the AME amplitude are presented in 
Table~\ref{tab:upperlimit_ame} and their impact on the 
presence of AME on SNRs is discussed in Section~\ref{sec:ame_in_SNR}.

%%%%%%%%%%%%%%%%%%%%%%%%%%%%%%%%%
%%---------- Section ----------%%
%%%%%%%%%%%%%%%%%%%%%%%%%%%%%%%%%
%
\section{Results and discussion}
\label{sec:results_discussion}

In this section we discuss in detail the integrated 
SED in intensity and in polarization for our sample of SNRs. 
The obtained properties presented in 
Table~\ref{tab:parameters_from_modelling} correspond 
to modelling the SEDs with the three proposed approaches called 
\textsc{Int} (only intensity data), \textsc{Pol} (only 
polarization data) and \textsc{IntPol} (simultaneously fitting the
intensity and polarization data). We use the results 
of the \textsc{IntPol} fit as a reference.
In addition, we discussthe  statistical properties of the SNRs, 
such as the contribution of the AME mechanism in the SNR 
scenario on a scale of 1\,deg.
In our statistical studies 
(section~\ref{sec:statistical_properties}), we combine our 
results for these six objects with those obtained in other 
QUIJOTE publications for other SNRs.

%-----------------------
%%---------- Sub-section
\subsection{SED of CTB\,80}
\label{sec:snr_ctb80}

\begin{figure}
\begin{center}
\includegraphics[trim = 0cm 0cm 0cm 0.cm,clip=true,width= 8.5 cm]{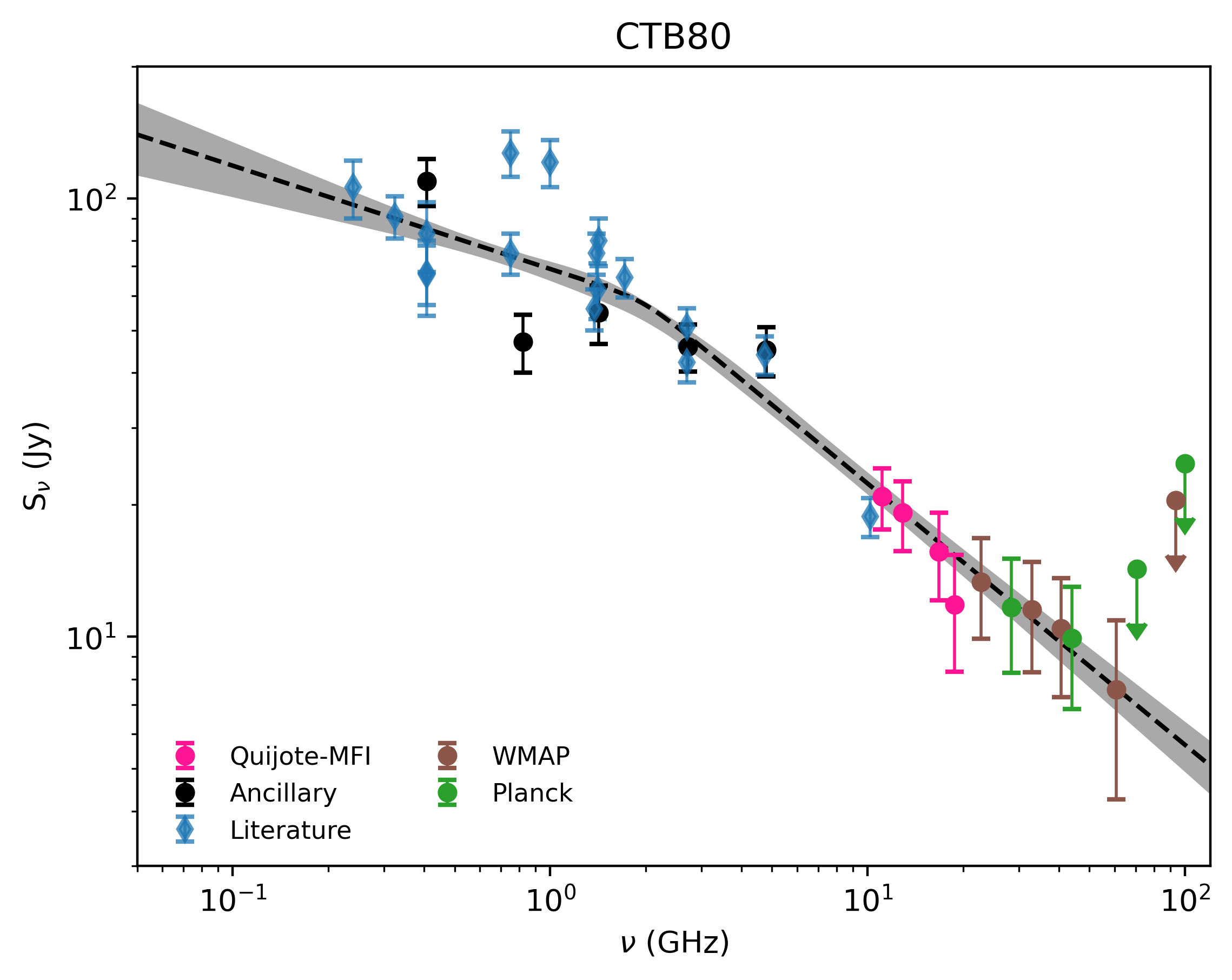}
\includegraphics[trim = 0cm 0cm 0cm .6cm,clip=true,width= 8.5 cm]{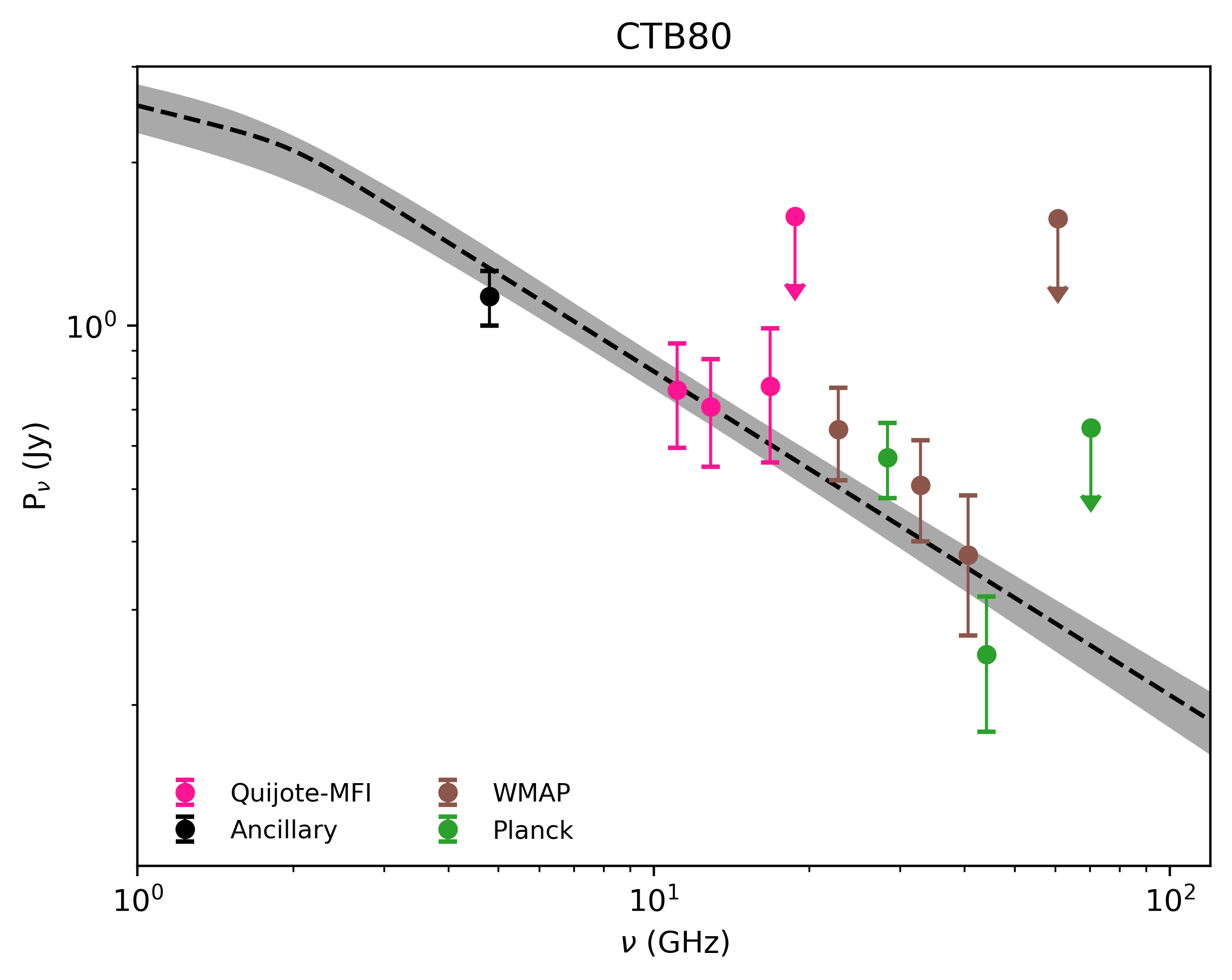}
\caption{Spectral energy distribution of CTB\,80 in 
intensity (top panel) and polarization (bottom panel). 
The integrated flux densities coming from different data 
sets are identified with colours. The blue diamonds are low
 frequency intensity flux densities from the literature 
(see Appendix~\ref{appendix:ctb80_hb21}). The dashed
 line corresponds to the SED model obtained from
 \textsc{IntPol} analysis (Table~\ref{tab:parameters_from_modelling}), 
while grey shading represents its uncertainties.}
\label{fig:sed_CTB80}
\end{center}
\end{figure}

The complex morphology of CTB\,80 at radio frequencies 
(central nebula, plateau and arms) is diluted into the 
1\,deg scale resolution, yielding only an extended structure 
(in intensity signal) well covered by an aperture of radius 
of $\sim$1\,deg (see Figure~\ref{ima:ctb80} and 
Table~\ref{tab:phot_params}).
CTB\,80 is located on the north-east edge of the Cygnus 
area, where extended diffuse emission covers 
the field of view. In addition, a source with a strong intensity 
signal lies partly in the external ring that we use for 
background subtraction. This source is masked with a circular 
aperture of radius 0.9\,deg centred at $l=\,70.2$\,deg and 
$b=\,1.55$\,deg (see Figure~\ref{ima:ctb80}).
Table~\ref{tab:fluxes_ctb80} lists the flux densities 
of CTB\,80.
In intensity, the spectrum is dominated by synchrotron 
emission with a spectral break at frequencies lower than the
QUIJOTE bands 
\citep[first evidence from][]{1983PASJ...35..437S,Gao2011A&A...529A.159,2005A&A...440..171C}. 
Spurious thermal dust emission is negligible; in 
fact, for $\nu \gtrsim$\,70\,GHz only upper limits (on 
flux densities) are computed from Table~\ref{tab:fluxes_ctb80}.
We also consider literature measurements in order  better to
characterize the curved shape of the intensity SED, although 
this procedure must be treated carefully because of the wide 
range of systematic effects of each measurement (see 
Appendix~\ref{appendix:ctb80_hb21}).
We do not take into account measurements below 200\,MHz 
in order to avoid the possible low frequency spectral 
turnover generated by the free--free thermal self-absorption 
\citep[proposed by][]{2005A&A...440..171C}, which is still 
debated on the basis of the inconsistency with the low level 
of dispersion reported for PSR\,B1951+32 
\citep{Gao2011A&A...529A.159,2004MNRAS.353.1311H}.
Low-frequency flux densities are listed in 
Appendix~\ref{appendix:ctb80_hb21} and identified by blue 
diamonds in Figure~\ref{fig:sed_CTB80}. Nevertheless, the data 
dispersion is high at low frequencies, which is also observed 
in our measurements.
At 0.408 and 0.82\,GHz, we measure $109\pm13$ and 
$47.1\pm7.1$\,Jy. These are the highest and lowest measurements 
in the ranges 0.3--0.5\,GHz and 0.7--1.0\,GHz respectively.
Regarding the map at 0.8\,GHz, one explanation may be 
connected to the fact that the apertures are optimized for a 
beam of 1\,deg while the map has a beam resolution of 1.2\,deg. 
At 11.1\,GHz, we obtain a raw flux density of $21.5\pm3.4$\,Jy, 
which is in agreement with the $18.8\pm1.3$\,Jy at 
10.2\,GHz reported by \cite{1983PASJ...35..437S}. Despite 
the high dispersion of the data, we decided to use 
measurements from the literature in order to be 
consistent with the analysis performed for the other SNR 
with a curved spectrum (HB\,21, Section~\ref{sec:snr_hb21}).
In polarization, it is not possible to trace an eventual 
break, since the polarized spectrum lacks data at low 
frequency. We then consider a single power-law synchrotron 
component for the \textsc{Pol} fit.
Finally, we use frequencies only up to 100 and 60.7\,GHz
 bands for intensity and polarization respectively. 
Figure~\ref{fig:sed_CTB80} shows the intensity and 
polarization SEDs, including upper limits, of CTB\,80 for 
the {\sc IntPol} analysis.

%--- SED: intensity 
In the intensity SED description, the {\sc IntPol} 
and {\sc Int} analyses provide good agreement for the 
synchrotron amplitude at 11.1\,GHz (21$\pm$1\,Jy) and 
the spectral index before the break ($-0.24\pm0.07$ and
 $-0.27\pm0.06$).
When these are compared with previously reported values 
such as $-0.36\pm0.02$ \citep[][]{2005A&A...440..171C}, 
$-0.45\pm0.03$ \citep[][]{2006A&A...457.1081K} and 
$-0.38\pm0.13$ \citep[using flux densities at 6, 11 and 
21\,cm from][]{Gao2011A&A...529A.159}, we found that the 
largest discrepancy is at 2.6\,$\sigma$. %, where we have 
This discrepancy is understood to be due to the wide dispersion 
of flux densities at low frequencies and to some 
studies using data at frequencies lower than 200\,MHz.
In that case, when all the flux densities at low 
frequencies are taken into account (including $\lesssim 
200$\,MHz), we found that the \textsc{IntPol} and 
\textsc{Int} results were in full agreement. However, we 
recover a flatter $\alpha_{\rm bb}$ and lower $\nu_{\rm b}$; 
e.g. $-0.10\pm 0.06$ and $1.5\pm 0.3$\,GHz for the 
\textsc{IntPol} analysis, although it is performed with a 
similar $\chi_{\rm dof}^2$ of 1.86.
This $\alpha_{\rm bb}$ around $-0.1$ therefore disagrees 
significantly from previous studies \citep{2005A&A...440..171C,2006A&A...457.1081K}.
Note that, our results provide $A_{\rm syn}\simeq$\,21\,Jy 
and $\alpha_{\rm ab}\simeq -0.6$ that remain stable in spite of
the model and the low-frequency data set considered, 
which is to be expected because these two parameters characterize 
the microwave SED (the QUIJOTE-MFI frequencies and higher).

The \textsc{Int} fit yields an asymmetric marginalized PDF 
for $\nu_{\rm b}$ and $\alpha_{\rm ab}$ (and also higher 
uncertainties for $\nu_{\rm b}$ and $\alpha_{\rm ab}$) 
with respect to the \textsc{IntPol} case.
The PDF of $\nu_{\rm b}$ shows a bimodal distribution 
peaking at $\sim 2$ and $\sim 4$\,GHz.
This also affects the $\alpha_{\rm ab}$ characterization,
 whose marginalized PDF yields $-0.70^{+0.10}_{-0.12}$, 
reaching maximum at around $-0.62$ (which agrees 
with $\alpha_{\rm ab}=-0.60\pm0.05$ from the \textsc{IntPol}).
Even so, the $\alpha_{\rm ab}$ and $\nu_{\rm b}$ are 
consistent at 1\,$\sigma$ between the {\sc IntPol} 
and {\sc Int} analyses.
Although our estimates of the spectral index in the
 microwave range, $-0.70^{+0.10}_{-0.12}$ and $-0.60\pm0.05$, 
are not compatible with the approximate spectral index 
of $-0.8$ inferred at 30 and 44\,GHz \citep{2016A&A...586A.134P}, 
our approach is the first SED fit to leave $\nu_{\rm b}$ 
and $\alpha_{\rm ab}$ as free parameters.

We confirm the curved shape of CTB\,80 inferred by 
\cite{1983PASJ...35..437S} and \cite{2016A&A...586A.134P}, 
where our models show the spectral break to occur
between 1.5 and 4\,GHz (most probably at $2^{+1.2}_{-0.5}$\,GHz) 
with a radio spectral index between $-0.1$ and 
$-0.3$.\footnote{For this range, the particle spectral index 
$\Gamma_{\rm p}$ falls in the range of 1.2 and 1.6, 
which is in agreement with the value of 1.72 used to model the 
Leptonic contribution of the non-thermal emission at high 
energy \citep{2021MNRAS.502..472A}. Note that they took into account
the radio spectral index $\alpha=-0.36\pm0.02$ reported by 
\cite{2005A&A...440..171C}.} and a microwave spectral index 
of $\alpha_{\rm ab}=-0.6\pm0.5$.
The spectral break $\Delta \alpha$ estimates are 
$-0.34\pm0.07$ and $-0.43\pm0.12$ for {\sc IntPol} 
and {\sc Int} respectively.
The former value departs from the prediction, $\Delta \alpha=-0.5$, 
of synchrotron losses in a homogeneous source with a
continuous injection of electrons \citep{2009ApJ...703..662R}, 
while the latter, $-0.43\pm0.12$, could be ascribed to 
the mechanism expected for a pulsar wind nebula \citep{2009ApJ...703..662R}. % 

%--- SED: polarization
Regarding polarization properties, we recover a 
polarization fraction $\Pi_{\rm syn}=3.7\pm0.3\%$ and 
a spectral index $\alpha_{\rm ab}=-0.60^{+0.04}_{-0.05}$ 
for the {\sc IntPol} analysis.
For comparison, the polarized spectrum is described 
by a single power law with spectral index 
$\alpha^{\rm pol}=-0.52\pm0.08$ and amplitude $A_{\rm syn}^{\rm pol}=0.8\pm0.1$\,Jy ({\sc Pol} fit).
This polarized spectral index, it will be recalled, is expected 
to be equivalent to $\alpha_{\rm ab}$ from the 
\textsc{IntPol} fit. We find that $\alpha^{\rm pol}$ 
is flatter than the other two cases using intensity data,
 with differences of 0.08 (at 1$\sigma$) and $-0.18$ 
(at 1.4$\sigma$) with respect to {\sc Int} and {\sc IntPol} 
respectively. 
This flattening could suggest depolarization between 
Urumqi band and microwave frequency bands, or that the 
intensity and polarization spectral indices are different.
However, the polarized SED from \textsc{IntPol} 
(Figure~\ref{fig:sed_CTB80}) shows a suitable model of 
the SED. Also, we observe a slight dip between the WMAP/\textit{Planck} 
(23--33\,GHz) and QUIJOTE-MFI (11--13\,GHz) bands. However, 
the uncertainties in the measured flux densities prevent 
more accurate modelling.
As a consistency test, the polarization fraction 
$\Pi_{\rm syn}^{\rm pol,int}=\,3.8\pm0.4\%$, computed from 
the intensity and polarization amplitudes of the {\sc Int} 
and {\sc Pol} analyses, is in agreement with the 
$\Pi_{\rm syn}=3.7\pm0.3\%$ for the {\sc IntPol} case.
This reveals that the polarization fraction at 11.1\,GHz is
 a reliable estimate, regardless of the model. 
In addition, $\Pi_{\rm syn}$ is unaffected by the different 
intensity low frequency data set considered in previous 
intensity analyses. 

The polarization degree $\Pi_{\rm syn}=3.7\pm0.3\%$ 
appears to be in disagreement with the high level 
reported in CTB\,80, where the averaged polarization 
fractions reported are $13.2\pm2.0\%$, $12^{+10}_{-6}\%$ 
and $7.6\pm0.7\%$ at 4.7, 2.73 and 1.42\,GHz respectively  
\citep[][]{1985A&A...145...50M,2005A&A...440..171C,Gao2011A&A...529A.159}.
However, literature measurements obtained using maps
 with higher resolutions and the polarization levels are
 provided at pixel scales. In contrast, we integrate $Q$ and 
$U$ in our aperture and derive the polarized intensity from 
those integrated quantities (Eq.~\ref{eq:pol_flux_integrated}), 
so our results are more prone to depolarization effects. 
A straightforward comparison is therefore not possible.

%-----------------------
%%---------- Sub-section
\subsection{SED of the Cygnus Loop}
\label{sec:snr_cygnusloop}

At an angular resolution of 1\,deg, the Cygnus Loop 
appears as a complex region with well-defined edges 
of diffuse plateau emission. Two extended structures 
can be clearly distinguished (see the intensity map 
in Figure~\ref{ima:cygnusloop}).
The smallest one is associated with the NGC\,6992/5 
regions (west region), while the main component (eastern 
region) predominantly encompasses the southern shell, 
NGC\,6960, the central filament and the remaining structures 
in the north (regions mentioned in Section~\ref{sec:snr_sample_our}).
These two regions are separated by an unpolarized canal, 
which is more noticeable at 22.8 and 28.4\,GHz (WK and 
P30 bands), as seen in the $P$ map in Figure~\ref{ima:cygnusloop}. 
In addition, the polarization angle distribution shows 
differences up to 45$^\circ$ between the NGC\,6992/5 regions 
and the rest of the remnant. This morphology is compatible with 
observations at lower frequencies and higher angular resolution,
 where weak diffuse polarized emission is observed between 
these two areas of the SNR 
\citep[e.g. see polarization maps in][]{1997AJ....114.2081L,2002A&A...389L..61U,2006A&A...447..937S}.

\begin{figure}
\begin{center}
\includegraphics[trim = 0cm 0cm 0cm 0.cm,clip=true,width= 8.5 cm]{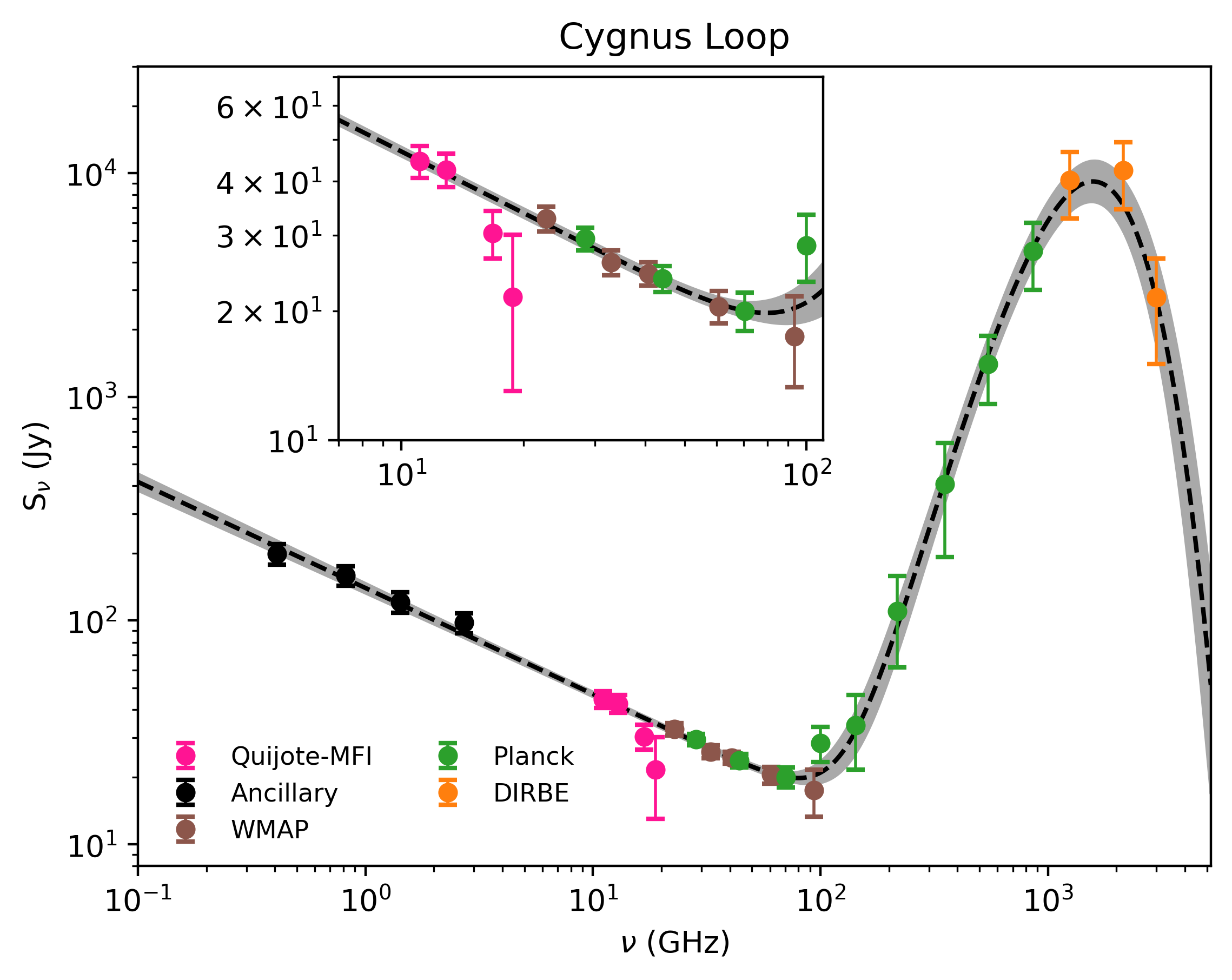}
\includegraphics[trim = 0cm 0cm 0cm 0.7cm,clip=true,width= 8.5 cm]{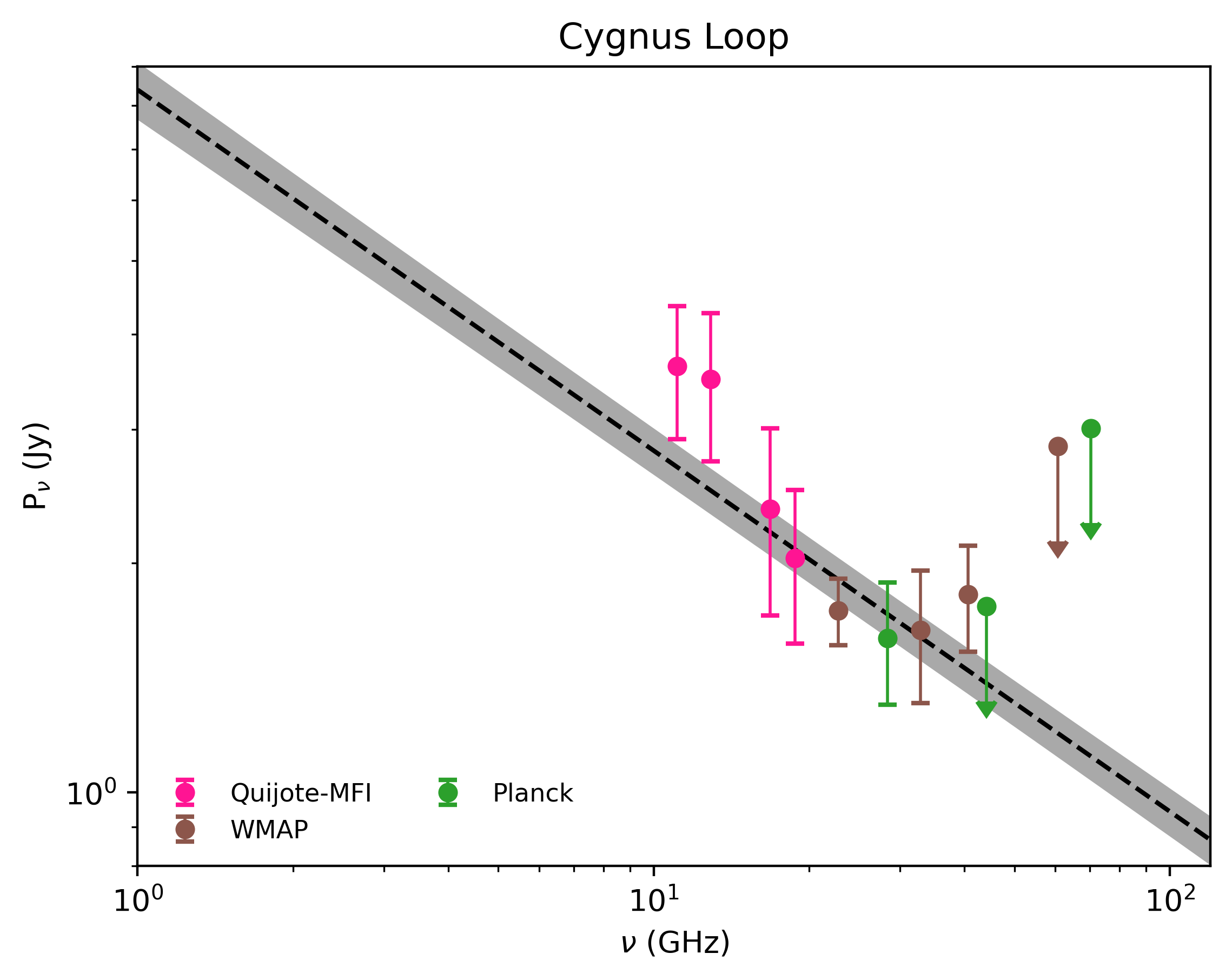}
\caption{Spectral energy distribution of Cygnus Loop 
in intensity (top panel) and in polarization (bottom panel). 
The integrated flux densities coming from different data 
sets are identified by colour. The dashed line corresponds 
to the SED model obtained from the \textsc{IntPol} analysis 
(Table~\ref{tab:parameters_from_modelling}), while grey 
shading represents its uncertainties.}
\label{fig:sed_CygnusLoop}
\end{center}
\end{figure}

Even though the Cygnus Loop is far enough from the Galactic
 plane, we observe Galactic diffuse emission in the northern
 part of our apertures, mainly affecting the intensity 
signal at high frequencies (see Figure~\ref{ima:cygnusloop}). 
Note that the central position of apertures is selected to 
reduce this contamination (Table~\ref{tab:phot_params}). Nonetheless, 
the intrinsic intensity emission of Cygnus Loop is negligible 
for frequencies $\gtrsim$100\,GHz. 
Its large size, however, requires a large aperture 
($r_{\rm src}$=\,130\,arcmin), where the CMB residuals become
 more important. Therefore, we subtract the SMICA-CMB map 
from the \textit{Planck} collaboration, which traces the CMB 
anisotropies (see Section~\ref{sec:data_wmap_planck_dirbe}). 
The raw (CMB-subtracted) flux densities  for the whole Cygnus 
Loop are listed in Table~\ref{tab:fluxes_cygnusloop}.
At 11.1\,GHz, we obtain an intensity flux density of 
46$\pm$4\,Jy, which is in agreement
with the value extrapolated (using $\alpha=-0.53$) from the 
flux density measured at 8.5\,GHz ($54\pm4$\,Jy) with the
 Medicina telescope \citep[][]{2021MNRAS.500.5177L}.
The intensity signal is well described by a power-law 
spectrum characteriztic of synchrotron emission (below 100\,GHz), 
while the observed thermal dust emission probably arises 
from Galactic background emission.
In polarization, we consider only one component associated 
with synchrotron emission and use measurements at 
frequencies lower than 80\,GHz. In Figure~\ref{fig:sed_CygnusLoop}, 
we show the intensity and polarization SEDs, including 
constraints, of the Cygnus Loop for the {\sc IntPol} case.

%-->Intensity
In intensity, we find a synchrotron amplitude at 
11.1\,GHz of $44.6\pm1.0$\,Jy and a spectral index of
$\alpha=-0.47\pm0.02$, since \textsc{Int} and 
\textsc{IntPol} yield the same parameter values, with 
the only difference being in the $\chi_{\rm dof}^2$ values (0.59 
and 0.62 respectively).
We also detect a weak thermal dust component associated 
with the background emission (with a $\tau_{353}$ 
significance of $3.8\sigma$), which is supported by the 
morphology observed at high frequencies (see, for example, the map 
at 217\,GHz in Figure~\ref{ima:cygnusloop}).
Our synchrotron spectral index falls in the range of $-0.40$ 
to $-0.53$ reported in previous studies 
\citep[e.g., see][]{2004A&A...426..909U,2006A&A...447..937S,2016A&A...586A.134P,2021MNRAS.500.5177L}, 
with differences of 2.7$\sigma$ in the worst case.
For instance, recently, \cite{2021MNRAS.500.5177L} obtained 
$\alpha =-0.53\pm0.01$ in the frequency range 0.022--30\,GHz.
In addition, \cite{2021MNRAS.500.5177L} modelled an 
exponential cut-off providing the spectral index 
$\alpha=-0.51\pm0.02$ with a cut-off frequency $\nu_{\rm c}=172\pm117$\,GHz, 
setting a weak restriction on the curvature of the 
spectrum, which becomes compatible with our power-law 
model for frequencies lower than 100\,GHz.
In another analysis, \cite{2006A&A...447..937S} reported a 
flatter spectral index ($\alpha=-0.40\pm0.06$) fitting only 
five bands in the range of 0.408--4.8\,GHz,  a frequency 
range that is comparable with our lower-frequency data set.
Our synchrotron model is therefore reliable and the low 
flux densities measured in the Q17 and Q19 bands do not
affect the fit since their uncertainties are conservative.
%

%-->polarization
The polarized emission is described by a single power law 
spectrum up to around 30\,GHz, and beyond that frequency 
our measurements are compatible with zero.
The \textsc{IntPol} characterization provides a polarization 
fraction of $\Pi_{\rm syn}=\,6.0\pm0.4$\,\%.
This polarized SED model predicts flux densities lower 
than observed in the Q11 and Q13 bands, although these are 
inside the 1\,$\sigma$ level when uncertainties are taken into account
(see bottom panel in Figure~\ref{fig:sed_CygnusLoop}). 
This effect is small because the \textsc{IntPol}, \textsc{Int} 
and \textsc{Pol} analyses provide coherent results. In fact, 
the $\Pi_{\rm syn}$ value agrees with the polarization 
fraction $\Pi_{\rm syn}^{\rm pol,int}=6.5\pm1.4$\%, where 
the latter is computed with amplitudes from the \textsc{Int} 
and \textsc{Pol} fits.
In contrast, the spectral index $\alpha^{\rm pol}=-0.59\pm0.23$ 
(from the \textsc{Pol} analysis) differs by 0.12 with respect
to the value found in the \textsc{IntPol} case, although 
they are compatible at 0.5$\sigma$.
As for CTB\,80, our integrated polarization fraction 
cannot be compared directly with values reported in the 
literature owing to the different methodologies and angular 
resolutions.
At 4.8\,GHz, the average polarization degree reaches levels 
of 30\% and 20\% around NGC\,6992/5 and the southern shell 
respectively \citep{2006A&A...447..937S,1972AJ.....77..459K}. 
In our $Q$ and $U$ maps at 1\,deg, the polarization is mainly 
observed in the eastern region (the main component) while it is
 weaker towards the NGC\,6992/5 region. For instance, in the Q11 
and Q13 bands, the emission from NGC\,6992/5 becomes 
indistinguishable from noise or background contamination (see the 
$P$ map in Figure~\ref{ima:cygnusloop}).

%-----------------------
%%---------- Sub-section
\subsection{SED of HB\,21}
\label{sec:snr_hb21}

\begin{figure}
\begin{center}
\includegraphics[trim = 0cm 0cm 0cm 0.cm,clip=true,width= 8.5 cm]{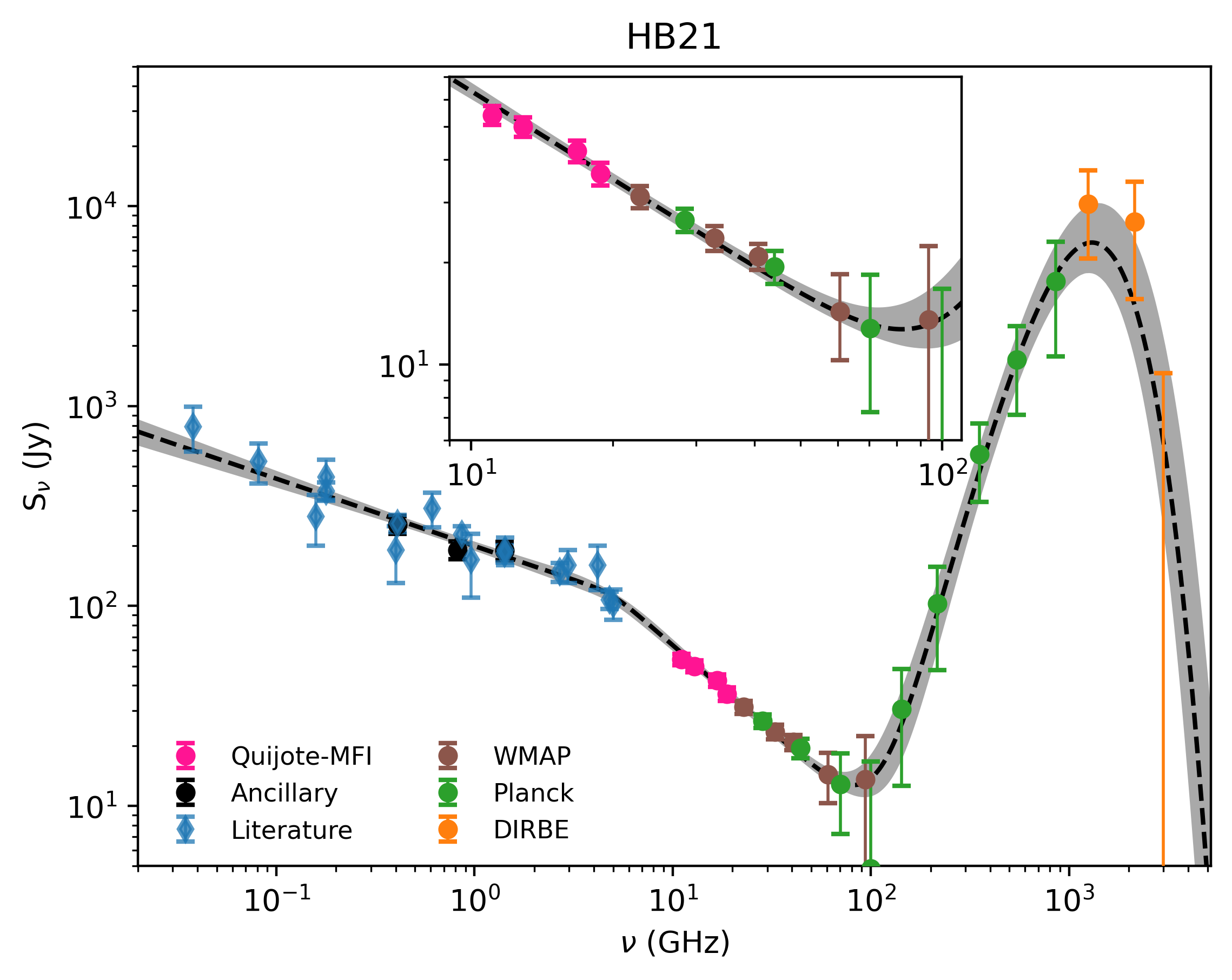}
\includegraphics[trim = 0cm 0cm 0cm 0.6cm,clip=true,width= 8.5 cm]{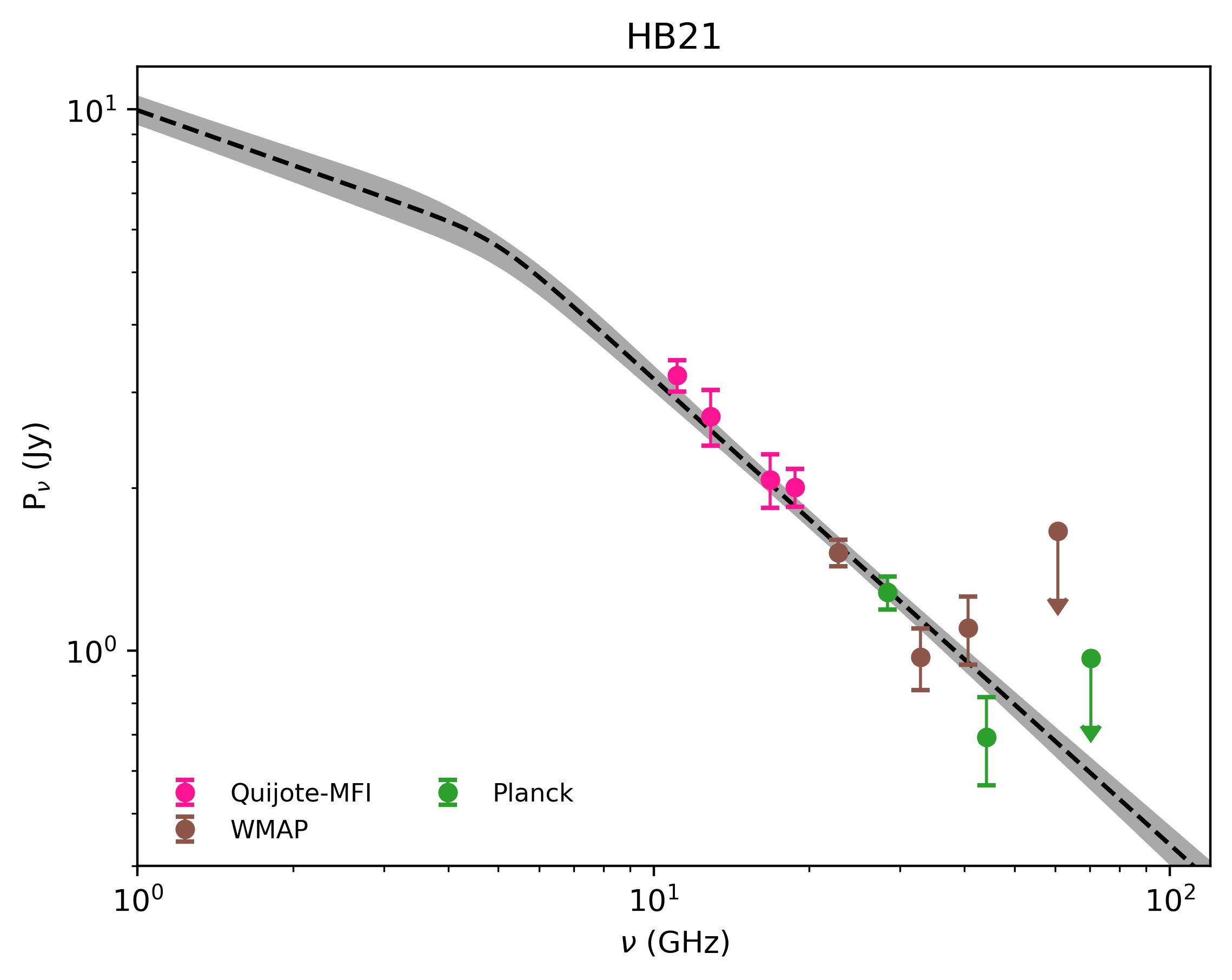}
\caption{Spectral energy distribution of HB\,21 in intensity 
(top panel) and polarization (bottom panel). The integrated
 flux densities coming from different data sets are identified
 by colour. The blue diamonds are low-frequency intensity 
flux densities from the literature (see Appendix~\ref{appendix:ctb80_hb21}). 
The dashed line corresponds to the SED model obtained from 
\textsc{IntPol} analysis (Table~\ref{tab:parameters_from_modelling}), 
while grey shading represents its uncertainties.}
\label{fig:sed_HB21}
\end{center}
\end{figure}

HB\,21 is located on the north-eastern edge of the Cygnus\,X 
area, and its radio emission appears to be superimposed on 
surrounding extended diffuse emission.
In the high-frequency range, where the synchrotron emission 
is below noise level, our default aperture lead to negative 
flux densities. This is a consequence of the thermal dust 
emission being brighter in the background ring than in the aperture.
In order to avoid this spurious emission, we use a mask to exclude 
part of the southern part of our background  annulus (see 
Figure~\ref{ima:hb21}). The flux densities extracted 
from the masked maps are then listed in Table~\ref{tab:fluxes_hb21}.
At 22.8\,GHz, we measure $32.6\pm2.5$\,Jy in intensity, 
which agrees with the $34\pm3$\,Jy value obtained by 
\cite{2013ApJ...779..179P} in the same band.
For the SED analyses, the intensity signal is described 
by a synchrotron component with a smooth broken spectral 
behaviour and thermal dust emission, while the polarization 
signal is well characterized by the synchrotron component 
only (see models in Table~\ref{tab:models_SNRs2}).
The thermal dust component is considered to be due to background emission.
Similarly to CTB\,80, fitting the synchrotron with a spectral 
break represents a challenge owing to our low number of measurements 
at low frequencies, which  strongly effects the estimation of $\nu_{\rm b}$.
We therefore use literature data available for $\nu \le$\,10\,GHz 
(see details in Appendix~\ref{appendix:ctb80_hb21}). 
In addition, the polarized measurements at high frequency, 
$\nu \gtrsim$\,100\,GHz, are noise dominated, so that we only 
consider polarized flux densities up to 70.4\,GHz (P70 band).
The polarization from the 60.7 and 70.4\,GHz frequencies are upper limits.
Figure~\ref{fig:sed_HB21} shows the modelled SEDs in 
intensity and in polarization from the {\sc IntPol} approach.

%--> Intensity
For the intensity signal, the {\sc IntPol} and \textsc{Int} 
approaches provide similar results for the synchrotron 
component (see Table~\ref{tab:parameters_from_modelling}), 
whose parameters agree within the uncertainties. 
We obtained amplitudes normalised at 11.1\,GHz of 58.1$^{+2.5}_{-2.2}$\,Jy 
and of 56.1$^{+2.2}_{-2.1}$\,Jy, and spectral indices 
below the spectral break of $-0.34^{+0.04}_{-0.03}$ and 
$-0.33\pm0.04$ for {\sc IntPol} and {\sc Int} respectively.
The largest differences, although still inside the 1-sigma 
level, are found for the spectral index after the break 
($\alpha_{\rm ab}$) and the break frequency ($\nu_{\rm b}$). 
The fit provides steeper $\alpha_{\rm ab}$ ($-0.86^{+0.04}_{-0.05}$)
 and higher $\nu_{\rm b}$ ($5.0^{+1.2}_{-1.0}$\,GHz) values 
when polarization and intensity signals are analysed together, 
in contrast to the intensity-only analysis 
($\alpha_{\rm ab}=-0.80^{+0.04}_{-0.05}$ and $\nu_{\rm b}=4.0^{+1.1}_{-0.8}$\,GHz).
Although the spectral break has already been reported in 
previous studies \citep[e.g,][]{2013ApJ...779..179P,2016A&A...586A.134P}, 
this is the first study in which the spectral indices and 
$\nu_{\rm b}$ are modelled as free parameters.
Using low-frequency and WMAP data, \cite{2013ApJ...779..179P} 
assumed a sudden spectral break of $\Delta \alpha=-0.5$ to 
recover a radio spectral index of $\alpha=-0.38\pm0.02$, 
(equivalent to our $\alpha_{\rm bb}$) and $\nu_{\rm b}=5.9\pm1.2$\,GHz.
Regarding the break frequency, our derived results are lower than those of 
\cite{2013ApJ...779..179P}, where {\sc IntPol} gives 
the closest value ($\nu_{\rm b}=5.0^{+1.2}_{-1.0}$\,GHz). 
For the radio spectral index, we obtained a steeper index 
($-0.34\pm0.04$). This provides a 
particle spectral index of $1.68\pm0.08$, which is also in 
agreement with the $\Gamma_{\rm p}=1.67\pm0.02$ reported by 
\citet{2013ApJ...779..179P}. 
In addition, we compute the spectral break, $\Delta \alpha = 
\alpha_{\rm ab} - \alpha_{\rm bb}$, of $-0.52\pm0.06$ and 
$-0.47\pm0.06$ for {\sc IntPol} and {\sc Int} respectively.
These values are in good agreement with the reference 
$\Delta \alpha=-0.5$ expected for middle-aged SNRs with 
synchrotron losses in a homogeneous source with continuous 
injection of electrons \citep{2009ApJ...703..662R}.

%--> polarization
Regarding the polarization properties, we recover a polarization 
fraction $\Pi_{\rm syn}=5.0\pm0.2\%$ and a spectral index 
$\alpha_{\rm ab}=-0.80^{+0.04}_{-0.05}$ in QUIJOTE-MFI 
(and higher) frequencies ({\sc IntPol} analysis).
For the {\sc Pol} case, the polarized spectrum is described 
by a single power law with spectral index $\alpha^{\rm pol}=-1.02\pm0.08$ 
and amplitude $A_{\rm 11.1}^{\rm pol}=3.2\pm0.2$\,Jy.
This spectral index of polarization is steeper than the 
other two cases using the intensity measurements ($\alpha_{\rm ab}$), 
where we have a difference of $-0.22$ (at 2.3$\sigma$) and 
$-0.16$ (at 1.7$\sigma$) for {\sc Int} and {\sc IntPol} respectively. 
Concerning the polarization level, the polarization 
fraction from amplitudes obtained in \textsc{Int} and 
{\sc Pol}, $\Pi_{\rm syn}^{\rm pol,int}=5.7\pm0.4\%$, 
is in agreement with 
$\Pi_{\rm syn}=5.0\pm0.2\,\%$ for the {\sc IntPol} case. 
Although a straightforward comparison is not possible, as 
mentioned for previous SNRs these polarization fractions 
are smaller than the average polarization fraction of 
$10.7\pm1.1$\% at 4.8\,GHz \citep{Gao2011A&A...529A.159} 
and $3.7\pm0.5$\% at 1.42\,GHz \citep{2006A&A...457.1081K} 
reported previously.

%-----------------------
%%---------- Sub-section

\subsection{SED of CTA\,1}
\label{sec:snr_cta1}
\begin{figure}
\begin{center}
\includegraphics[trim = 0cm 0cm 0cm 0.cm,clip=true,width= 8.5 cm]{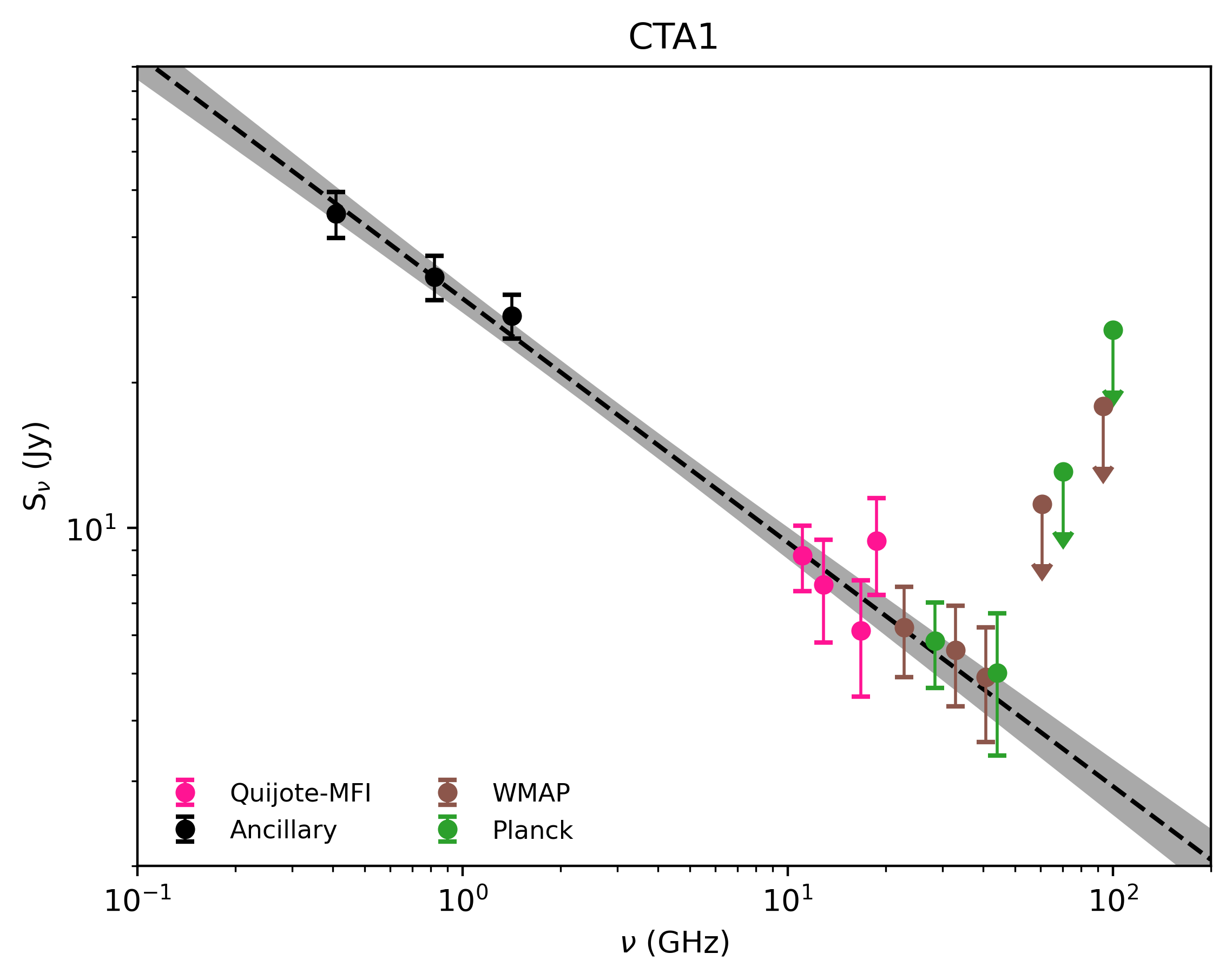}
\includegraphics[trim = 0cm 0cm 0cm 0.6cm,clip=true,width= 8.5 cm]{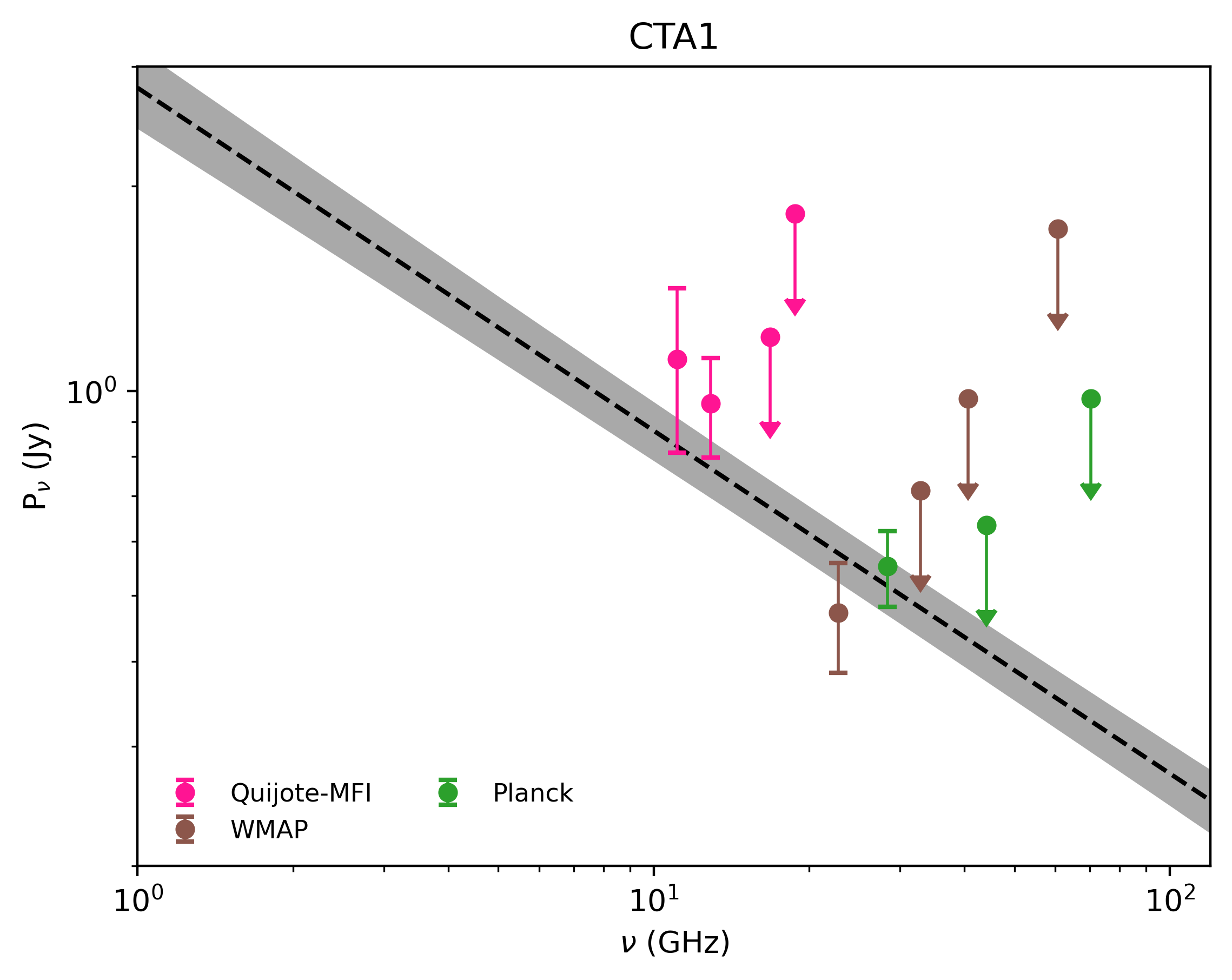}
\caption{Spectral energy distribution of CTA\,1 in intensity 
(top panel) and polarization (bottom panel). The integrated 
flux densities coming from different data sets are identified 
by colour. The dashed line corresponds to the SED model obtained and 
from \textsc{IntPol} analysis (Table~\ref{tab:parameters_from_modelling}), 
grey shading represents its uncertainties.}
\label{fig:sed_CTA1}
\end{center}
\end{figure}

At a resolution of 1\,deg, the intensity signal of 
CTA\,1 appears as an extended source where the emission 
from the shell, from the central branch and the 
breakout region, are diluted as a result of the coarse 
angular resolution (Figure~\ref{ima:cta1}).
There is a small diffuse region to the north-west that is 
associated with two extragalactic point sources and part 
of the diffuse emission coming from the western shell.
We also keep in mind the Faraday Screen (FS) reported by 
\cite{2011A&A...527A..74S}, which is located at 
$(l,b)=(121^{\circ},11^{\circ})$ with an approximated radius 
of 1.5$^\circ$. % (green dashed line in Figure~\ref{ima:cta1}).
Part of the FS covers the western shell, which could be 
associated with the negative spot in the $Q$ maps (mainly 
in Q11, WK and P30 bands).
For CTA\,1, we do not apply any mask despite the fact 
that our polarization measurements could contain the 
information coming from the FS. The integrated 
flux densities in Table~\ref{tab:fluxes_cta1} could therefore be 
affected by these effects.
In intensity, we obtained a flux density of 
$9.0\pm1.4$\,Jy at 11.1\,GHz. For frequencies higher than 
50\,GHz, measurements are noise dominated and upper limits 
are computed. In the modelling of the intensity SED we 
consider only a synchrotron component.
In polarization, we measure a polarized flux density of 
$11.4\pm0.31$\,Jy at 11.1\,GHz, which is in agreement with 
the extrapolation of $20.3\pm2.0$\,Jy, assuming a spectral 
index $\alpha_{\rm radio}=-0.63$ \citep[from $\lambda11$\,cm,][]{2011A&A...527A..74S}.
As for intensity analysis, we take into account a 
single power-law function for the synchrotron component only.
Figure~\ref{fig:sed_CTA1} shows the intensity and 
polarization SEDs, including upper limits, of CTA\,1 
for the \textsc{IntPol} analyses.

%-->Intensity
The intensity spectrum is described by a synchrotron 
component with amplitude of $8.9\pm0.6$\,Jy and spectral 
index of $-0.50\pm0.03$, as provided by both \textsc{Int} 
and \textsc{IntPol} analyses. This result is expected 
because the polarized SED consists only of four measurements, 
thus the \textsc{IntPol} fits is dominated by the intensity SED.  
This spectral index is $\sim$2.2$\sigma$ away from previous 
radio spectral index estimates, such as 
$\alpha=-0.63\pm0.05$ computed at frequencies lower than 
4.8\,GHz \citep[][]{2011A&A...527A..74S,1997A&A...324.1152P}. 
\cite{2011A&A...527A..74S} also found consistency between 
the integrated flux density spectral index and the spectral 
index computed from the TT plot strategy.
Therefore, the difference $\Delta \alpha=0.13$ between our 
results may be due to background contamination or to the 
worse beam resolution, which are not taken into account by our simple model.
Also, the spatial distribution of the spectral index 
towards CTA\,1 
\citep[at $\sim$10\,arcmin,][]{2011A&A...527A..74S,1997A&A...324.1152P,1981A&A...103..393S} 
shows a mean value of $-0.5$ in the shell region, a value 
that is closer to our derived spectral index. In addition, 
this spectral index map shows variations of up to 0.3, 
where the central branch has spectral indices of around $-0.6$ 
and the breakout region shows steeper values.

%-->polarization
In polarization the scarcity of high signal-to-noise 
measurements does not permit a reliable fit. 
Using four measurements and upper limits in the 10--90\,GHz 
range, we obtained an amplitude of 
$A_{\rm syn}^{\rm pol}=1.0\pm0.2$\,Jy and a spectral index 
of $\alpha_{\rm pol}=-0.8\pm0.3$ (\textsc{Pol} analysis).
Thanks to the high uncertainty of $\alpha_{\rm pol}$, 
this value is in agreement with the \textsc{Int} and 
\textsc{IntPol} analyses at just 1.4$\sigma$. 
From the \textsc{IntPol} analysis, the polarization 
fraction is $\Pi_{\rm syn}=9.4\pm1.2\%$, the 
highest $\Pi_{\rm syn}$ being computed for our SNR sample. 
We computed a polarization fraction 
$\Pi_{\rm syn}^{\rm pol,int}=11.1\pm2.4\%$ (from 
\textsc{Int} and \textsc{Pol} amplitudes) in full 
agreement with the $\Pi_{\rm syn}=9.4\pm1.2\%$ from 
the \textsc{IntPol} case.
From higher angular resolution maps, \cite{2011A&A...527A..74S} 
found average polarization fractions of about 26\%, 
14\% and 10\% for the eastern shell, the central 
branch and the southern shell respectively.

%-----------------------
%%---------- Sub-section
\begin{figure}
\begin{center}
\includegraphics[trim = 0cm 0cm 0cm 0.cm,clip=true,width= 8.5 cm]{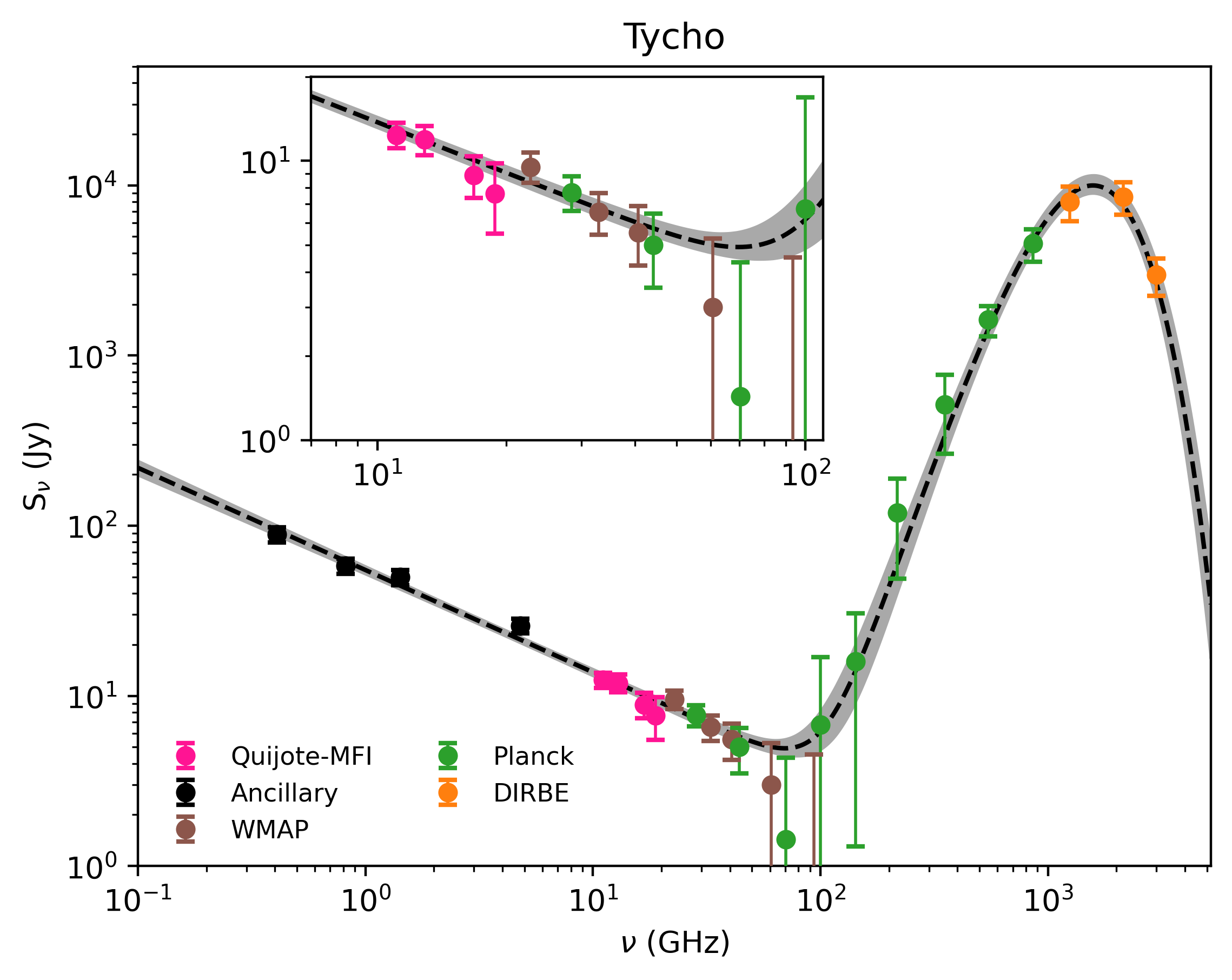}
\includegraphics[trim = 0cm 0cm 0cm 0.6cm,clip=true,width= 8.5 cm]{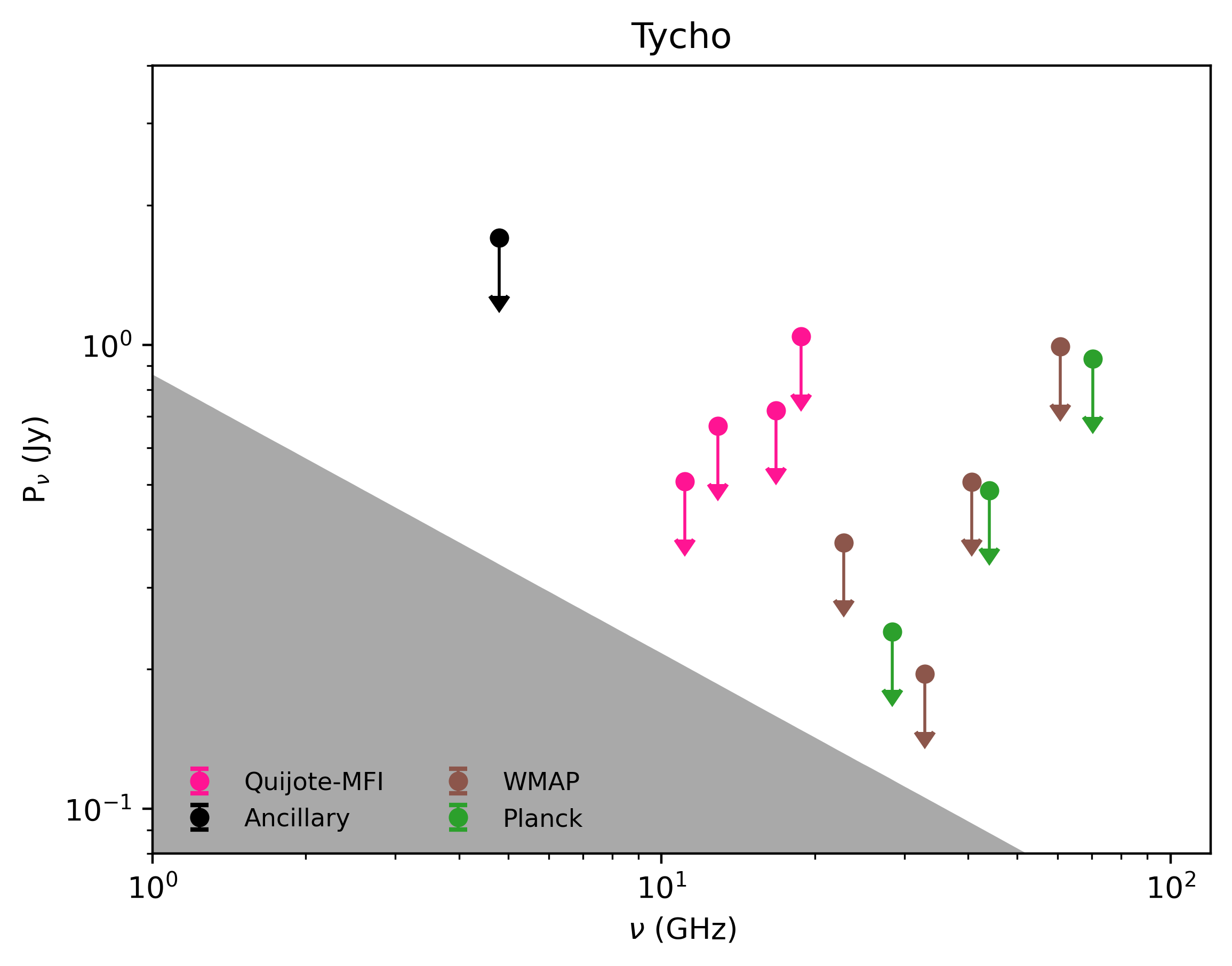}
\caption{Spectral energy distribution of Tycho in intensity 
(top panel) and in polarization (bottom panel). The 
integrated flux densities coming from different data sets 
are identified by colour. The dashed line corresponds to 
the SED model obtained from \textsc{IntPol} analysis 
(Table~\ref{tab:parameters_from_modelling}) and gray 
shading represents its uncertainties.}
\label{fig:sed_Tycho}
\end{center}
\end{figure}

\subsection{SED of Tycho}
\label{sec:snr_tycho}
In the QUIJOTE bands, the spatial structure of the total 
intensity emission of Tycho resembles that of a compact 
source, with low background contamination (see 
Figure~\ref{ima:tycho}). This is due to its small angular 
size \citep[$\sim 8$\,arcmin,][]{2019MNRAS.482.3857L} 
compared to our beam size.
In contrast, the polarization maps reveal negligible $Q$ 
and $U$ emission, which are dominated by noise fluctuations.
The raw integrated flux densities are presented in 
Table~\ref{tab:fluxes_tycho}.
In Q19 and WK bands, we measure flux densities in total 
intensity of $7.6\pm2.1$ and $9.9\pm1.2$\,Jy (not colour-corrected), 
which agree with the value of $8.8\pm0.9$\,Jy measured 
with the Sardinia Radio Telescope at 21.4\,GHz 
\citep[][]{2019MNRAS.482.3857L}. In contrast, our 
measurement of $25.9\pm2.6$\,Jy at 4.8\,GHz departs from 
the value of $20\pm2$\,Jy reported by \cite{2011A&A...536A..83S}.
In polarization, we quote upper limits on integrated 
polarized intensity for all frequencies. 

At 4.8\,GHz our measured flux density in total polarization, 
$1.4\pm0.2$\,Jy, is significantly higher than the value of 
$0.28\pm0.03$\,Jy obtained by \cite{2011A&A...536A..83S}. 
This difference could be due to spurious emission 
coming from the Faraday rotation of a source to the 
south-east inside the annulus (see map P in 
Figure~\ref{ima:tycho}). The source is only visible in 
the original Urumqi maps at 9.5\,arcmin and our angular 
resolution extends its contribution inside our aperture.
Therefore we only estimate an upper limit on the polarized 
flux density at 4.8 GHz.
The intensity and polarized SEDs are drawn in 
Figure~\ref{fig:sed_Tycho} using as a reference the 
\textsc{IntPol} analysis.

%-->Intensity: SED
The intensity spectrum of the remnant below 100\,GHz 
is well characterized by a spectral index of $-0.60\pm0.02$ 
and an amplitude of 12.9\,Jy at 11.1\,GHz (see 
Table~\ref{tab:parameters_from_modelling}). The thermal 
dust emission is coming from extended diffuse emission 
not associated with the SNR.
Our synchrotron component agrees with previous spectral 
indices of $\alpha=-0.624\pm0.004$ 
\citep[at 0.015--70\,GHz][]{2021A&A...653A..62C} and 
$\alpha=-0.58\pm0.02$ \citep[][]{2011A&A...536A..83S,2019MNRAS.482.3857L}.
Our data lack sufficient sensitivity and frequency coverage 
to be able to trace the weak concave-up change in the 
spectral index, $\Delta\alpha$\,$\simeq$\,0.04, at around 
1\,GHz reported by previous studies 
\citep{1992ApJ...399L..75R,2021A&A...653A..62C}.
Around 100\,GHz, the steepening of the spectrum could be 
understood in terms of a small negative residual CMB 
contribution. We rule out a spectral break of the synchrotron 
component because it is not expected in this frequency coverage.
In fact, the gamma-ray analysis suggests a spectral break 
at higher energies at around 100\,eV \citep{2011ApJ...730L..20A}.

%-->polarization: SED
As polarization measurements yield only upper limits for 
all frequency bands; the analyses taking into consideration the polarized 
SED just provides constraints on the polarized properties.
From the \textsc{IntPol}, we find an upper limit of 
$\Pi_{\rm syn}<$\,3.2\% (at 95\% of C.L.).
This is a conservative constraint with respect to the average 
percentage polarization of 1\% 
\cite[at 4.8\,GHz,][]{2011A&A...536A..83S} and 2.5\% 
\cite[at 1.42\,GHz,][]{2006A&A...457.1081K} reported previously.
At 4.8\,GHz, we obtain a constraint on the integrated 
polarization fraction of 7.4\% (at 95\% C.L.), and our most 
restrictive polarization levels are of the order of 4\% for 
the Q11 and WK bands. In this context, we do not have the 
sensitivity to reach such low polarization level observed 
at lower frequencies.

\subsection{SED of HB\,9}
\label{sec:snr_hb9}

\begin{figure}
\begin{center}
\includegraphics[trim = 0cm 0cm 0cm 0.cm,clip=true,width= 8.5 cm]{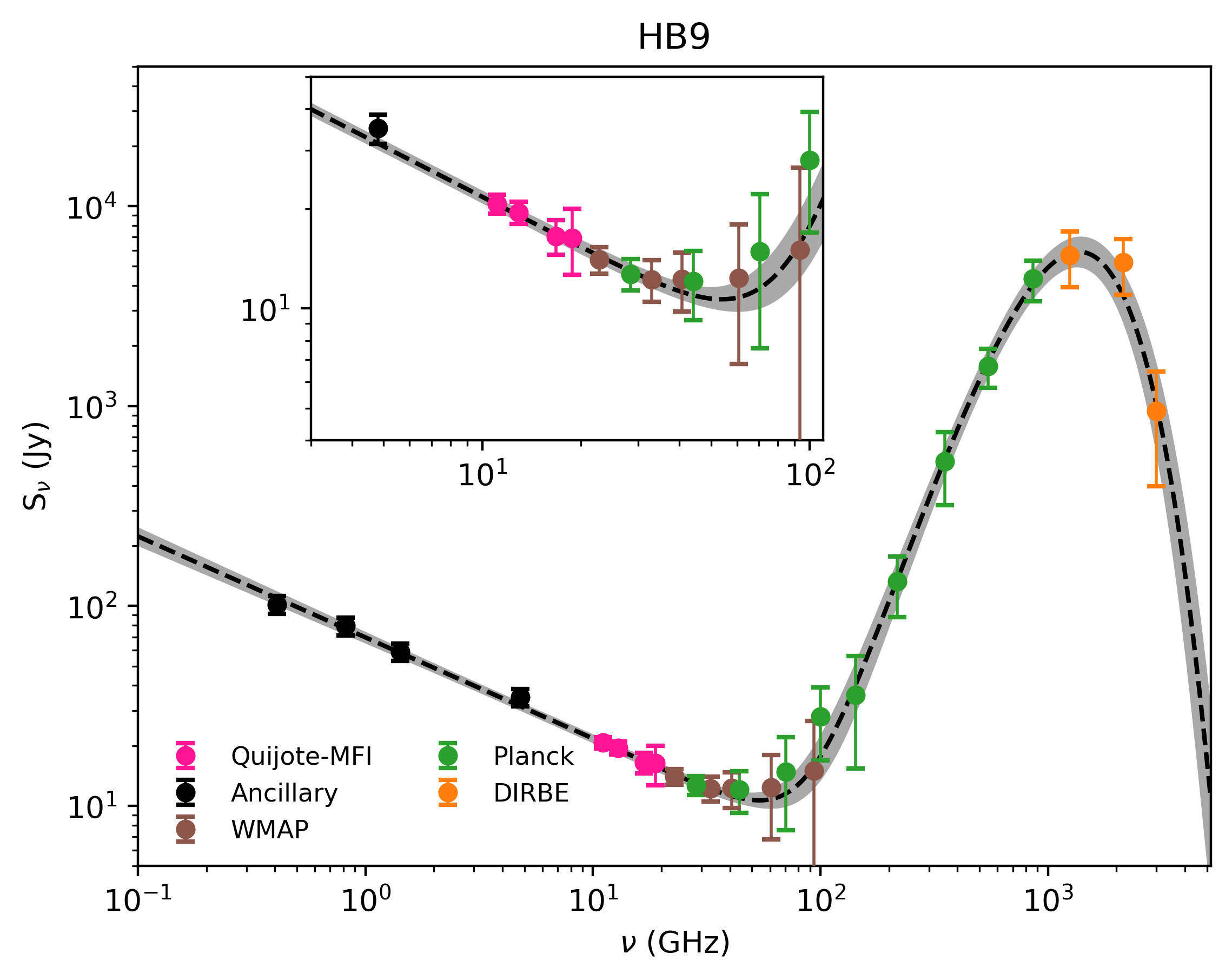}
\includegraphics[trim = 0cm 0cm 0cm 0.6cm,clip=true,width= 8.5 cm]{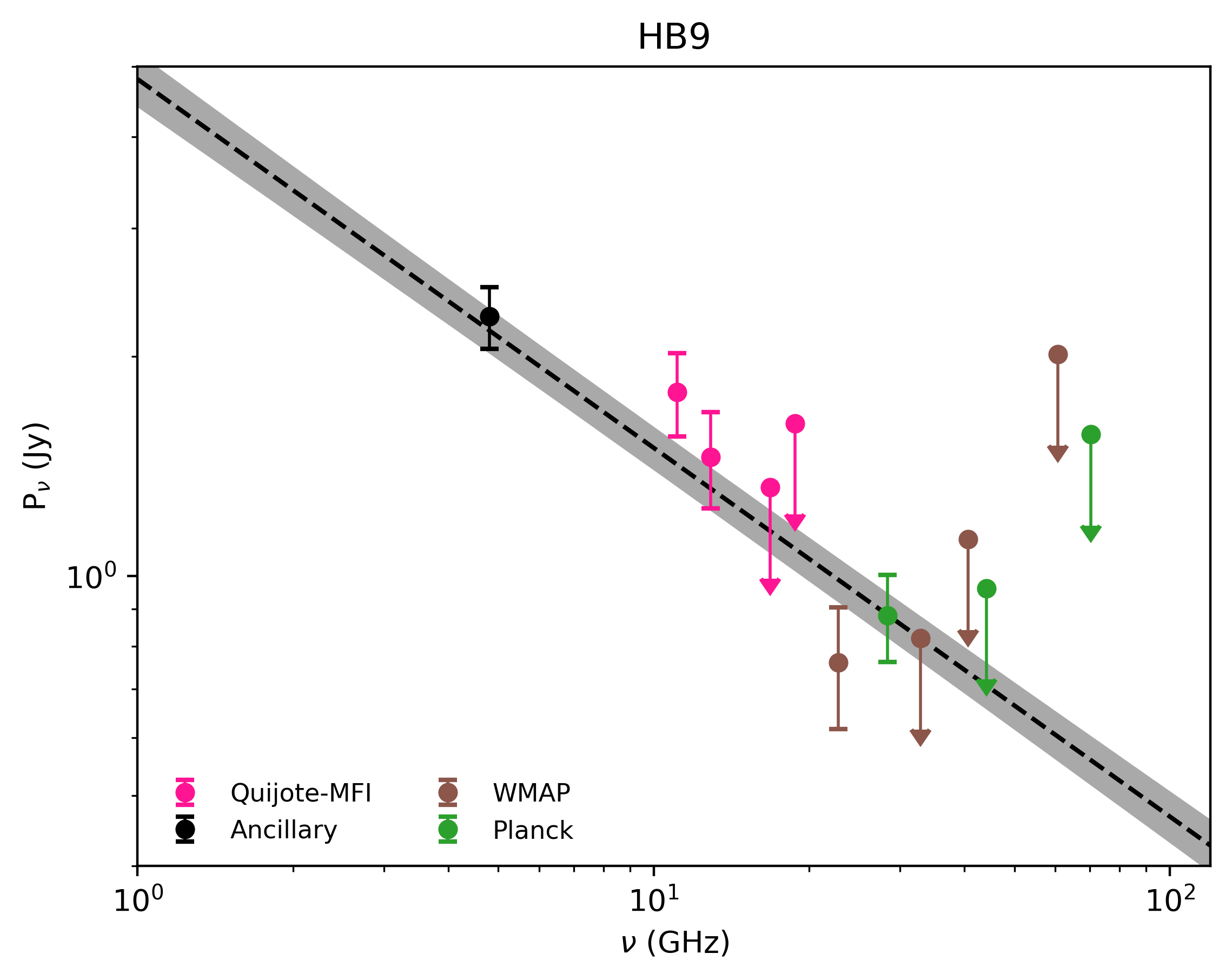}
\caption{Spectral energy distribution of HB\,9 in intensity 
(top panel) and in polarization (bottom panel). The integrated 
flux densities coming from different data sets are identified
by colour. The dashed line corresponds to the SED model obtained 
from \textsc{IntPol} analysis (Table~\ref{tab:parameters_from_modelling}) 
and grey shading represents its uncertainties.}
\label{fig:sed_HB9}
\end{center}
\end{figure}

On scales of 1\,deg, the shell structure of HB\,9 
dominates the total-intensity emission, where the peak 
emission is off-centre to the south-west (see Figure~\ref{ima:hb9}).
Synchrotron emission from HB\,9 is visible up to 70\,GHz, 
while a thermal dust component is captured inside the 
aperture although there is a weak spatial correlation between them.
In polarization, the smoothed $U$ maps, mainly in the 
Urumqi band, reveal two spots with positive (north-west) 
and negative (south-east) emission, and this pattern seems 
to prevail up to 40\,GHz. At higher frequencies, the polarized 
emission of HB\,9 becomes very weak (see $P$ map in Figure~\ref{ima:hb9}).
For HB\,9, the raw flux densities are presented in Table~\ref{tab:fluxes_hb9}.
In intensity, we measure a flux density of $35.0\pm 3.5$ 
at 4.8\,GHz that is consistent with the value of 
$33.9\pm 3.4$\,Jy reported by \cite{Gao2011A&A...529A.159}. 
At 11.1\,GHz, the QUIJOTE measurement is $21.2\pm1.4$\,Jy.
In polarization, upper limits are provided for 
frequencies higher than 30\,GHz. Therefore, we assumed 
that synchrotron emission drives the polarization spectrum of HB\,9.
The intensity and polarized SEDs are presented in Figure~\ref{fig:sed_HB9}
for the \textsc{IntPol} analysis.

%-->Intensity
For the intensity signal, the \textsc{IntPol} and \textsc{Int} 
fits provide similar amplitudes of about $21\pm1$\,Jy 
(at 11.1\,GHz) and spectral index of $-0.51\pm0.02$. Our 
fitted spectral index is flatter than that of 
$-0.59\pm0.02$ computed in the range 0.1--5\,GHz by 
\cite{Gao2011A&A...529A.159}, while it is steeper than 
the value of $-0.44\pm0.04$ obtained at frequencies below 
1\,GHz by \cite{1973A&A....26..237W}.
As pointed out by \cite{2003A&A...408..961R}, the differences
 in spectral index could be caused by the contribution from the 
HB\,9 plateau, which can be diluted on our maps due to our large beam.
These values are consistent with the spatial variation of the 
spectral index, $-0.4$ to $-0.9$, obtained with a TT-plot 
approach over fifty two regions across HB\,9 
\citep[408--1420\,MHz,][]{2007A&A...461.1013L}.
However, our spectral analysis shows no evidences for the 
spectral change at around 1\,GHz proposed by 
\cite{1973A&A....26..237W}. This result is consistent with 
the simple power law behaviour observed by previous authors
 \citep[see][]{2003A&A...408..961R,Gao2011A&A...529A.159}.

%-->polarization

We obtained a polarization 
fraction $\Pi_{\rm syn}=6.9\pm0.5$\,\% and a spectral 
index $\alpha=-0.51\pm0.02$ for the {\sc IntPol} analysis.
For the {\sc Pol} case, the polarized spectrum is described 
by a single power law with spectral index 
$\alpha^{\rm pol}=-0.60\pm0.08$ and amplitude 
$A_{\rm 11.1}^{\rm pol}=3.2\pm0.2$\,Jy.
This spectral index of polarization is steeper than that 
obtained with intensity data ($\alpha$ from \textsc{Int} 
and \textsc{IntPol} fits), although they are consistent when 
the uncertainties are taken into consideration.
Furthermore, $\alpha^{\rm pol}=-0.60\pm0.08$ is in agreement 
with the value of $\alpha=-0.59\pm0.02$ obtained with only 
intensity flux densities between 0.1 and 5\,GHz \citep{Gao2011A&A...529A.159}. 
Moreover, we computed a polarization fraction of 
$\Pi_{\rm syn}^{\rm pol,int}=6.8\pm0.7$\% (from \textsc{Int} 
and \textsc{Pol} amplitudes) in full agreement with the 
previous $\Pi_{\rm syn}=6.9\pm0.5$\% value from the 
\textsc{IntPol} case, which shows the stability of the estimation 
of the polarization fraction.
This value is only half the average polarization degree of 
$9.1\pm 0.1$\% and $15.6\pm 1.6$\% obtained with better angular 
resolutions at 1.42\,GHz by \cite{Gao2011A&A...529A.159} and 
at 4.8\,GHz by \cite{2006A&A...457.1081K} respectively.

%-----------------------
%%---------- Sub-section
%\subsection{Further statistical properties}
%\label{sec:statistical_properties}
\subsection{Overall description of the QUIJOTE SNR sample}
\label{sec:statistical_properties}

This section is devoted to exploring the properties of SNRs 
from a statistical point of view. To this end, the observational 
properties obtained in this study are complemented with previous 
QUIJOTE-MFI studies of SNRs.
A joint study of these sources is possible because the SED 
analyses and methodologies used in those previous studies are 
fully consistent with those used in this article 
(described in Section~\ref{sec:methodology}).

\begin{figure}
\begin{center}
\includegraphics[trim = 0cm 0cm 0cm .cm,clip=true,width= 8.55 cm]{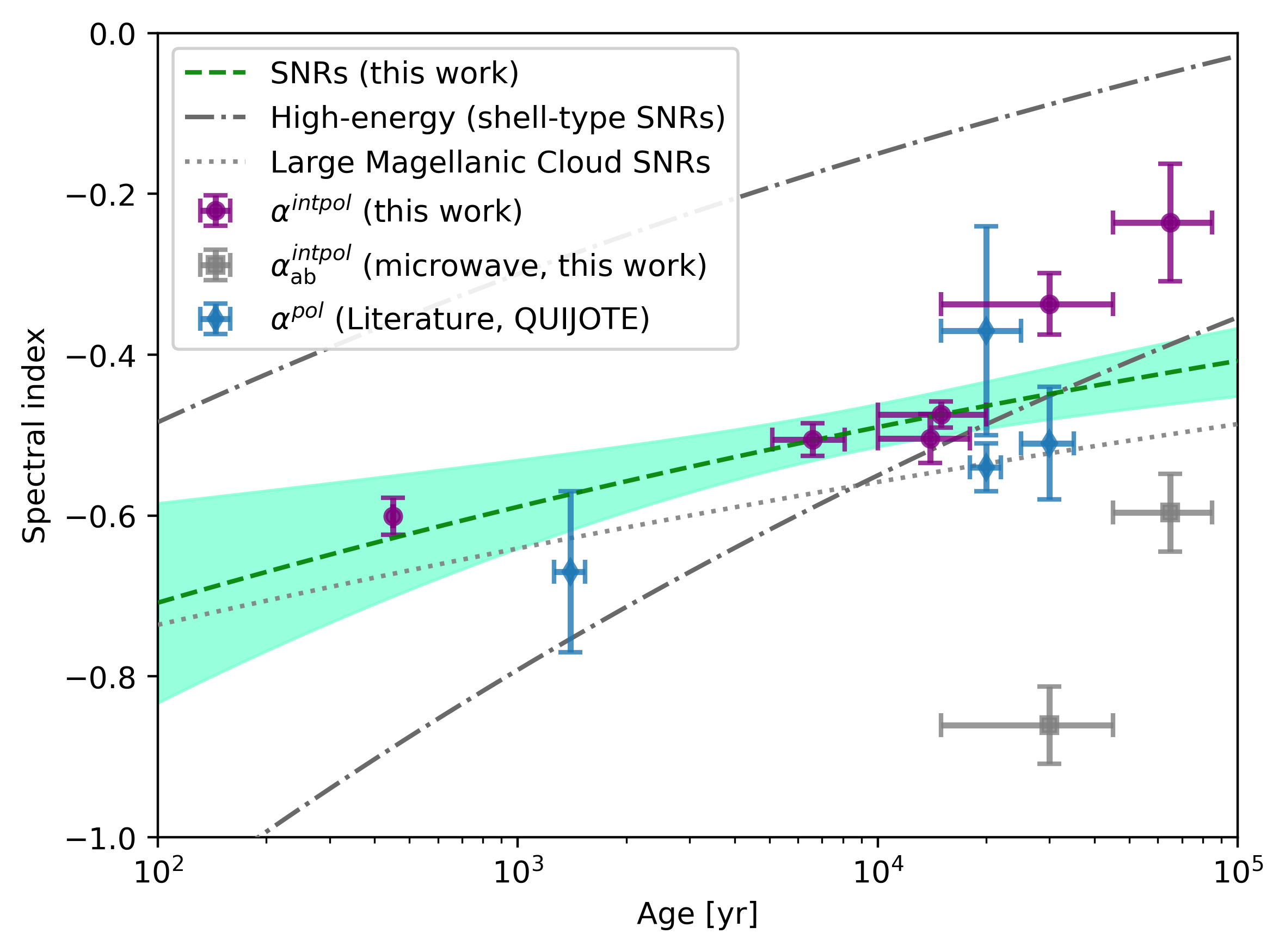}
\caption{The spectral index evolution for SNRs observed 
with QUIJOTE-MFI data. The spectral indices are 
obtained from SEDs built at 1 deg resolution (FWHM), 
which also span information from the intensity and 
polarization SED. The blue points correspond to W44, IC443,
 W49 and W51 from other QUIJOTE-MFI studies. The ages of SNRs 
are taken from Table~\ref{tab:description_SNRs}. The purple 
points identify the SNRs studied in this paper. For remnants 
with curved spectra (HB\,21 and CTB\,80), we display the 
spectral indices below (purple circles) and above (grey 
squares) the break frequency.
The evolution of the spectral index is modelled with a 
power-law function fitting the purple and blue points, 
which provides $a=\,-0.49\pm0.02$ and $b=\,-0.07\pm0.03$ 
(the green dashed line). In addition, we also drawn the 
relation between the spectral index and age for the Milky 
Way (the dashed and dotted lines, see text for details) and 
for the Large Magellanic Cloud 
\citep[the dotted line,][]{2017ApJS..230....2B}.
}
\label{fig:stat_sample_age_index}
\end{center}
\end{figure}

\subsubsection{SNR spectral indices and ages}\label{sec:statistical_indexevolution}

We explore the evolution of the polarized spectral 
index of SNRs studied with QUIJOTE-MFI on scales of 
1\,deg. This sample consists of our six SNRs and other 
sources such as W44 \citep{2017MNRAS.464.4107G}, IC443, W49 
and W51 \citep{mfi_widesurvey_w51}. 
For our six SNRs, we select $\alpha^{\rm intpol}$ which 
contains information from the intensity and polarized SEDs.
For  W44, IC443, W49 and W51, we use the polarized spectral 
index $\alpha^{\rm pol}$ derived in those references, since 
these did not simultaneously fit intensity and polarization signals.

On angular scales of 1\,deg, we obtain an average spectral 
index of $-0.48\pm0.16$ for the whole sample, where the
term 0.16 represents  the scatter of the measurements.
For evolved SNRs, i.e.\ older than 20\,kyr, we find flatter 
spectral indices covering from $-0.54$ to $-0.24$ 
(CTB\,80, HB\,21, W51,  W44 and IC443, see 
Figure~\ref{fig:stat_sample_age_index}), with an average 
spectral index of $-0.37\pm0.18$. In contrast, we recover a 
mean spectral index of $-0.55\pm0.08$ for younger remnants 
(<\,20\,kyr).
The spectral indices are significantly flatter for the older 
CTB\,80 ($\sim$\,65\,kyrs) and HB\,21 ($\sim$30\,kyrs), 
reaching indices of $-0.24$ and $-0.36$ respectively. In fact, 
CTB\,80 and HB\,21 exhibit a curved spectrum; however, here 
we use the radio spectral indices $\alpha^{\rm intpol}_{\rm bb}$ 
(purple points in Figure~\ref{fig:stat_sample_age_index}).
Therefore, we obtain flatter spectral indices for the 
older SNRs, which agrees with other Galactic studies, 
although the scatter of the spectral index distribution 
is still large 
\citep[]{1976MNRAS.174..267C,1962ApJ...135..661H,2019ApJ...874...50Z,2023arXiv230206593R}.

To find a relation between spectral index and age, we 
consider a power-law function following the same procedure 
implemented for the Large Magellanic Cloud 
SNRs~\citep[LMC,][]{2017ApJS..230....2B}:
\begin{equation}
\alpha(t;a,b) = a (t/t_{\rm ref})^{b},
\label{eq:alpha_time}
\end{equation}
where $a$ is the amplitude, $t$ is the age normalised 
to $t_{\rm ref}=10^4$\,yr and $b$ is the slope. 
The fit provides $a = -0.49\pm 0.02$ and $b = -0.07 \pm 0.03$. 
This evolution of the spectral index is drawn by dashed 
green line in Figure~\ref{fig:stat_sample_age_index}.
This fit underestimates the spectral indices of HB\,21 
and CTB\,80, with a significance of 2.0$\sigma$ and 
4.6$\sigma$ respectively. This could suggest a different 
trend for these older SNRs ($\gtrsim$\,30\,kyr) with 
curved spectra. Note that the W51 ($\sim$30\,kyr),  W44 and 
IC443 ($\sim$20\,kyr) remnants are inside the modelled 
tendency although the dispersion is higher than in younger 
SNRs (<\,20\,kyr).

Although our modelling is conservative and limited by our
 number of SNRs, it is in agreement with other Galactic constraints. 
The dash-dotted lines of Figure~\ref{fig:stat_sample_age_index}
 show that our spectral index tendency is contained within 
the evolutionary trend extrapolated from \cite{2019ApJ...874...50Z}. 
Actually, they obtained the correlation of the particle 
spectral index and the age of Galactic SNRs 
\citep[Figure\,3b of][]{2019ApJ...874...50Z}, from which we put
 upper and lower edges by power laws with slope $-0.09$ and 
amplitudes of 2.1 and 1.3 (at $t_{\rm ref}=10^4$\,yr) 
respectively. Then these edges are translated to the 
$\alpha_{\rm radio}(t)$ using the established relationship 
$\alpha_{\rm radio} = (1-\Gamma_{\rm p})/2$ to yield the 
previously mentioned dash-dotted lines of Figure~\ref{fig:stat_sample_age_index}.
The region bounded by these curves is wide and spans the 
flatter spectra of CTB\,80 and HB\,21 (for the radio spectral 
indices); although their microwave spectral indices fall 
outside this trend (grey squares in Figure~\ref{fig:stat_sample_age_index}). 

Regarding comparisons with extragalactic studies, 
\cite{2017ApJS..230....2B} obtained a slope 
$b_{\rm LMC}$=\,--0.06$\pm$0.02 and amplitude 
$a_{\rm LMC}$=\,--0.97$\pm$0.09 at 1\,yr for the LMC (see dotted 
line in Figure~\ref{fig:stat_sample_age_index}). They established 
the correlation of the radio spectral index and age using 
29 LMC SNRs, which includes multi-epoch observations of the 
younger remnant SN1987A. The latter was crucial to confirm 
that evolved remnants have flatter spectral indices than the younger ones. 
The model evolution of spectral indices for the LMC exhibits 
steeper spectral indices than our predictions from the Galaxy 
model ($b=-0.07\pm0.04$). Although the differences fall in the 
range 0.07 to 0.08 for ages $\gtrsim$\,15\,kyr, we found a 
consistency when uncertainties are considered.
As final remark, we emphasise that results for our Galaxy 
\citep[e.g. ][]{2019yCat.7284....0G,2019ApJ...874...50Z} and 
the LMC \citep{2017ApJS..230....2B} arise from the characterization 
of the intensity SED, while our results also span the polarization properties.

\begin{table}
	\centering
	\caption{Summary of the polarization properties considered for statistical analysis.}
	\begin{tabular}{l c c c c} % four columns, alignment for each
		\hline
		SNR & $d$ [kpc]& age [yrs] & $\Pi$ [\%]& $\delta_{\pi}$ [\%] \\
		\hline 
		\multicolumn{5}{c}{}\\[-1.5 ex]
Tycho & 2.50 & 450 & $\lesssim$3.2 & 0.0$\pm$2.2\\ 
W49 & 11.00 & 1400 & 11.0$\pm$6.3 & 15.4$\pm$8.8\\ 
HB\,9 & 0.60 & 6600 & 6.9$\pm$0.5 & 10.0$\pm$0.7\\ 
CTA\,1 & 1.40 & 14000 & 9.4$\pm$1.2 & 13.5$\pm$1.8\\ 
Cygnus Loop & 0.74 & 15000 & 6.0$\pm$0.4 & 8.7$\pm$0.6\\ 
IC443 & 1.90 & 20000 & 2.7$\pm$1.1 & 4.0$\pm$1.6\\
 W44 & 3.10 & 20000 & 15.2$\pm$1.8 & 21.8$\pm$2.5\\ 
W51 & 5.40 & 30000 & 5.5$\pm$2.5 & 8.0$\pm$3.6\\ 
HB\,21 & 1.74 & 30000 & 5.0$\pm$0.2 & 7.5$\pm$0.3\\ 
CTB\,80 & 2.00 & 65000 & 3.7$\pm$0.3 & 5.7$\pm$0.5\\  
		\hline
	\end{tabular}
	\label{tab:statistic_polarization_properties}
\end{table}

\subsubsection{Observed polarization fraction}
\label{sec:statistical_poldegreee}

We investigate the observed polarized fraction of SNRs 
studied with QUIJOTE-MFI on a scale of 1\,deg. We consider 
the same sample as in the previous section: our six 
remnants and sources W44 \citep{2017MNRAS.464.4107G}, 
IC443, W49 and W51 \citep{mfi_widesurvey_w51}. 
For our six SNRs, we select the $\Pi_{\rm syn}$ values yielded 
by the joint fit of the intensity and polarized SEDs (\textsc{IntPol} case).
For  W44, IC443, W49 and W51, we use the polarized 
fractions computed from the intensity and polarized amplitudes
 fitted respectively to the intensity and polarized SEDs. 
In addition, as the observational polarization degree is 
equivalent to a decreasing factor of the intrinsic 
polarization, i.e.\ $\Pi_{o}(\alpha_{\rm radio}) \cdot \delta_{\pi}$ 
\citep{1966MNRAS.133...67B,1979A&A....74..361S}, we also 
compute this $\delta_{\pi}$ as follows:
\begin{equation}
     \delta_{\pi}= \dfrac{\Pi_{\rm obs}}{\Pi_{o}(\alpha_{\rm radio})},
     \label{eq:delta_pi_obs}
\end{equation}
$\alpha_{\rm radio}$ being the spectral indices considered 
in the previous section ($\alpha$ or $\alpha_{\rm bb}$). 
Table~\ref{tab:statistic_polarization_properties} presents 
the $\Pi_{\rm obs}$ and $\delta_{\pi}$ values for each SNR.

On an angular scale of 1 deg, we found a mean and standard 
deviation of $7.2\pm4.9\%$, which agrees with previous studies 
that obtain polarization fractions lower than 20\% 
\citep[e.g. see,][]{2006A&A...457.1081K,Gao2011A&A...529A.159}. 
We also found lower polarization fraction levels 
($5.5 \pm 2.5$, $5.0\pm 0.2$, $3.7\pm 0.3$\%) for the older 
SNRs (W51, HB\,21 and CTB\,80), although some younger remnants 
also have weak polarization; for example, we compute an upper
 limit of 3.2\% at 95\% C.L. for Tycho.
Nevertheless, the depolarization effects change towards each 
region 
\citep[][]{1966ARA&A...4..245G,1966MNRAS.133...67B,1973SvA....16..774S}, 
resulting in unclear correlations between the different parameters 
obtained from the SED; for instance, the expected evolution of the 
intrinsic polarization fraction (equations~\ref{eq:intrinsic_pi_alpha}
 and~\ref{eq:alpha_time}).
Thus the observed low polarization fractions for our older SNRs does 
not constitute robust evidence of a correlation between the $\Pi_{\rm obs}$ 
and the age of SNRs. Even so, this tendency could be explored in 
future studies, which would require observations with better resolution 
to avoid the effects of our large beam. 

In the ideal case that $\delta_{\pi}$ is not dominated by the 
dilution due to our large beam, the observed low polarization 
levels trace the importance of a disordered and/or turbulent 
magnetic field component \citep{1966MNRAS.133...67B,1973SvA....16..774S} 
as $\delta_{\pi} = H_0^2/(H_0^2+H_r^2)$.\footnote{In this equation, 
$H_0$ is the regular magnetic field component and $H_r$ is a 
random isotropic magnetic field component (with Gaussian variance 2/3$H_r^2$).}
In our Galaxy, the average of 
$\langle H_r \rangle /\langle H_0 \rangle$ falls in the range 3 to 5 
\cite[][and references therein]{2001SSRv...99..243B}, yielding a 
$\delta_\pi$ between 10\% and 4\%.
For our SNR sample, we found a mean and standard deviation 
of $10.5\pm5.2\%$ for $\delta_{\pi}$ (see 
Table~\ref{tab:statistic_polarization_properties}), which 
agrees with the expected 10--4\% observed in our Galaxy.
In addition, the $\delta_{\pi}$ does not seem to depend on 
distance or position.

%-----------------------
%%---------- Sub-section
\subsubsection{The AME in SNRs}
\label{sec:ame_in_SNR}

As was pointed in previous sections, the presence of AME 
towards our SNRs is not favoured by the SED analysis.
In fact, the spectra are dominated by the synchrotron 
radiation in the 10--40\,GHz range. Moreover, the thermal 
dust contribution is low, so the AME contribution is 
also expected to be low, reaching values of the order of, or 
smaller than, our flux density uncertainties.
We can estimate the expected AME contribution at 22.8\,GHz 
considering the AME emissivity, 
$\epsilon^{\rm 22.8\,GHz}_{\rm AME,\tau_{353}}$, based on 
the ratio between the AME amplitude at 22.8\,GHz and the 
dust emissivity at 353\,GHz. We use the value 9\,K as 
reference for our Galaxy 
\citep[][and references therein]{2016A&A...594A..25P,2022MNRAS.513.5900H,mfi_widesurvey_galacticPlane}. 
For instance, the study of AME in the Galactic plane 
provided $9.84\pm3.57$\,K 
\citep[using QUIJOTE-MFI,][]{mfi_widesurvey_galacticPlane}, 
which is consistent with a full-sky analysis 
\citep[$8.3\pm0.8$\,K,][]{2016A&A...594A..25P}.
Table~\ref{tab:upperlimit_ame} presents the predicted AME 
amplitudes at 22.8\,GHz, $A_{\rm AME,pred}^{\rm 22.8GHz}$, 
considering the $\tau_{353}$ values from the \textsc{IntPol} case.
These are unavailable for CTB\,80 and CTA\,1  since the 
thermal dust component was not considered in the SEDs fit.

We also study the presence of AME following the  methodology 
proposed in Section~\ref{sec:ame_constraints}, where an AME 
component is considered in the SED fit.
Overall, we obtain AME amplitudes compatible with zero, 
thus we provide only upper limits on the AME amplitude (at 
95\% C.L) for our six SNRs (see  Table~\ref{tab:upperlimit_ame}).
These limits are dominated by the high uncertainties behind 
$A_{\textsc{ame}}$ and are driven mainly by the uncertainties 
of the low-frequency (microwave) part of the spectrum. Although 
these upper limits are not a strong constraint on AME amplitudes,
 they suggest that the AME contribution is negligible in the 
range 10--40\,GHz for our six remnants.
Moreover, this is also supported by the AME emissivity, 
$\epsilon^{\rm 22.8\,GHz}_{\rm AME,\tau_{353}}$, obtained from 
$A_{\rm AME}$ and $\tau_{353}$ provided by the SED fit. We 
compute upper limits on $\epsilon^{\rm 22.8\,GHz}_{\rm AME,\tau_{353}}$ 
(at 95\,C.L.) of between 5 and 13\,${\rm K}$. Tycho and HB\,9 
show, respectively, the stronger constraints of 6.8 and 4.9\,${\rm K}$ 
(at $2\sigma$), which are lower than the observed values in 
our Galaxy 
\citep[$8-11\,{\rm K}$,][]{2016A&A...594A..25P,2022MNRAS.513.5900H,mfi_widesurvey_galacticPlane}.

Other QUIJOTE studies also favour the non-detection of AME 
towards SNRs on a  scale of 1\,deg: $i)$ the main QUIJOTE-MFI 
calibrators, the Crab Nebulae and CasA 
\citep[][in prep.]{mfi_widesurvey,mfi_widesurvey_calibrators_TauA_CasA}, 
$ii)$ 3C\,58 \citep[the Fan region analysis,][ in prep.]{mfi_widesurvey_fan}, 
and $iii)$ the G010.19-00.32, G037.79-00.11 and G045.47+00.06 
catalogued as semi-significant AME sources 
\citep[][]{mfi_widesurvey_ame_srcs}. 
Therefore, sources W44 \citep{2017MNRAS.464.4107G}, W49 and 
W51 \citep{mfi_widesurvey_w51} appear to be the only SNRs with
 an AME contribution on a  scale of 1\,deg.
These three SNRs share two features: first, they interact 
with H\textsc{ii} or molecular clouds; and second, they are all 
located within the first quadrant of the Galactic plane.
However, these two characteriztics cannot be assumed as drivers 
of AME emission in SNRs.
Regarding morphology, CTA\,1 and HB\,21 are considered as 
remnants that interact with their surroundings, and both 
lack an AME component in their spectra.
In addition, we do not measure AME in SNRs catalogued as 
plerion types, e.g.\ compact source remnants w.r.t.\ our 
beam size, such as TauA, CasA and 3C58 
\citep{mfi_widesurvey,mfi_widesurvey_fan}, mainly owing to the 
lack of a thermal dust component in their spectra.
Concerning location, CTB\,80, the Cygnus Loop and HB\,21 are also in 
the first quadrant. In addition, SNRs located far from 
Galactic plane (the Cygnus Loop, CTA\,1 and
HB\,21) also show no evidence of an AME spectrum, or in 
cases that are relatively close to the Galactic plane in 
the second and third quadrants (such as Cas\,A, 3C58 and HB\,9).

Therefore, we point to an overlap of different mechanisms 
along the line of sight rather than to  a strong relationship 
between the AME and SNR scenarios.
For instance, on arcminute scales, \cite{2019MNRAS.482.3857L} 
used the Sardinia Radio Telescope (SRT, 1.55, 7.0 and 21.4\,GHz) 
to discount an AME spectrum for  W44 that had previously been reported by 
\cite{2017MNRAS.464.4107G} at 1\,deg scale. Recently, the COMAP 
maps of Galactic plane \citep{2022ApJ...933..187R}, covering 
the 26--31\,GHz range at 4.5\,arcmin, helped to confirm the SRT 
results, for which the spectrum is well modelled by a simple 
synchrotron mechanism with a spectral index of $-0.54\pm0.06$.
In this context, this observational panorama stresses the 
need for surveys with better angular resolutions in order 
to access the structure of W44, W49 and W51 in the range 10--40\,GHz. 

\begin{table}
	\centering
	\caption{Upper limits on the presence of AME towards our 
SNRs. The second and third rows show respectively the constraints 
on the amplitude of AME ($A_{\textsc{ame}}$ in units of Jy) and 
on the AME emissivity ($\epsilon^{\rm 22.8GHz}_{\rm AME,\tau_{353}}$ 
in units of ${\rm K}$) when an AME component is considered in 
the SED fit (see section~\ref{sec:ame_constraints}). These upper 
limits are computed at 95\% of confidence level (95\% C.L.). The 
last row shows the AME amplitude at 22.8\,GHz,
 $A_{\rm pred,AME}^{\rm 22.8GHz}$, obtained from the AME emissivity 
$\epsilon^{\rm 22.8GHz}_{\rm AME,\tau_{353}} = 9$ ${\rm K}$. }
	\begin{tabular}{c c c c c c c} % four columns, alignment for each
		\hline
		&CTB\,80&Cyg.\,L.&HB\,21&CTA\,1&Tycho&HB\,9\\
		\hline 
\multicolumn{7}{c}{}\\ [-1.5 ex]
\multicolumn{7}{c}{Upper limits at 95\% C.L.}\\ [0.5 ex]
$A_{\rm AME}$ &2.8&2.4&3.9&1.4&1.3&1.5\\ [0.5 ex]
$\epsilon^{\rm 22.8GHz}_{\rm AME,\tau_{353}}$ &--& 11.0& 12.3 &--& 6.8 & 4.9\\ [0.8 ex]
%[\%] &19.7&7.0&11.3&20.3&14.6&9.7\\ [0.2 ex]

\multicolumn{7}{c}{Predictions from the AME emissivity }\\ [0.5 ex]
$A_{\rm AME,pred}^{\rm 22.8GHz}$ &$-$&1.8&2.9&--&1.6&2.5\\ [0.8 ex]
		\hline
	\end{tabular}
	\label{tab:upperlimit_ame}
\end{table}

%%%%%%%%%%%%%%%%%%%%%%%%%%%%%%%%%
%%---------- Section ----------%%
%%%%%%%%%%%%%%%%%%%%%%%%%%%%%%%%%
\section{Conclusions}
\label{sec:conclusions}

We exploit the new QUIJOTE-MFI wide survey, at 10--20\,GHz, 
to study the intensity and polarization properties of radio--microwave 
emission towards the SNRs CTB\,80, the Cygnus Loop, HB\,21, CTA\,1, Tycho and HB\,9.
For each SNR we present updated modelling of the intensity 
and polarization SEDs, which also takes into account a simultaneous 
fit of intensity and polarization spectra, taking advantage of  
the low depolarization level expected in the  frequency range 
of QUIJOTE-MFI  (and at higher frequencies).
For all six sources, we obtain the properties of the synchrotron 
radiation on angular scales of 1\,deg, mainly associated with the 
remnant emission of each region (Table~\ref{tab:parameters_from_modelling}), 
and these properties are consistent with previous studies. However, 
the weak intensity signal of the thermal dust 
emission captured inside the apertures in some of them is 
considered as background coming from the Galaxy.

In intensity, we confirm the curved spectra of CTB\,80 and HB\,21,
 where a power law with an abrupt break is preferred over other 
models, such as a power law with an exponential cut-off. 
In addition, we compute constraints on the intensity amplitude of 
the Anomalous Microwave Emission that suggest a negligible AME 
contribution towards all six remnants. 
From the simultaneous intensity and polarization fit, we recover 
radio spectral indices as flat as $-0.24$, with a mean and scatter 
values of $-0.44\pm 0.12$. In addition, the polarization degree 
levels are lower than 10\% with a mean and scatter values of 
$6.1\pm 1.9\%$, while we set the upper limit of 
$\Pi_{\rm syn}\lesssim 3.2\%$ (at 95\% of confidential level) only for Tycho.

When combining our results with the measurements from other 
QUIJOTE-MFI studies of SNRs, we find that radio spectral indices 
are flatter for mature SNRs and are significantly flat for 
CTB\,80 ($-0.24^{+0.07}_{-0.06}$) and HB\,21 ($-0.34^{+0.04}_{-0.03}$).
We found the evolution of the spectral indices to be adequately described by a 
power-law function with an exponent $-0.07\pm 0.03$ and 
amplitude $-0.49\pm 0.02$, which is conservative and agrees with 
previous studies of our Galaxy and the Large Magellanic Cloud. 
Further observations at higher sensitivities and beam resolutions 
at lower frequencies in these and other remnants, will help to 
complete this study by increasing the number of SNRs and improving 
the SED modelling. Therefore, data product from CBASS at 5\,GHz 
\citep{2018MNRAS.480.3224J} and the upcoming updated version of 
the MFI instrument of QUIJOTE \citep[MFI2,][]{2022SPIE12190E..33H} 
will be valuable.

Regarding the AME component towards SNRs on angular scales of 
1\,deg, the data indicate the overlapping of components along the 
line of sight rather than a strong relationship between the AME 
and synchrotron radiation from the SNRs as the reason why 
only certain SNRs (W44, W49 and W51) show an AME contribution in 
their spectra.
Since they  interact with H\textsc{ii} or molecular clouds 
and are  embedded inside the Galactic plane in the first quadrant, 
further observations at higher angular resolution of  W44, W49 
and W51 will be useful to understanding the nature of the anomalous 
emission observed on angular scale of 1\,deg. This is one of the 
goals of the TFGI instrument of QUIJOTE \citep[30 and 
41\,GHz,][]{status2016SPIE}, which is currently operating at
Teide Observatory.

%%%%%%%%%%%%%%%%%%%%%%%%%%%%%%%%%
%%---------- Section ----------%%
%%%%%%%%%%%%%%%%%%%%%%%%%%%%%%%%%

\section*{Acknowledgements}
%-----------------------------
%--> Acknowledgments from the authors
CHLC appreciates the knowledge, professional training and affection 
received from Rodolfo Barb\'a, who has become the most important 
Supernova of my life. Now I can find you among the stars (RIP, 2021-12-07). 
We thank Terry Mahoney (Scientific Editorial Service of the IAC) for proofreading this manuscript.
%-----------------------------
%--> QUIJOTE acknowledgements
We thank the staff of the Teide Observatory for invaluable assistance in the commissioning and operation of QUIJOTE.
The {\it QUIJOTE} experiment is being developed by the Instituto de Astrofisica de Canarias (IAC),
the Instituto de Fisica de Cantabria (IFCA), and the Universities of Cantabria, Manchester and Cambridge.
Partial financial support was provided by the Spanish Ministry of Science and Innovation 
under the projects AYA2007-68058-C03-01, AYA2007-68058-C03-02, AYA2010-21766-C03-01, AYA2010-21766-C03-02,
AYA2014-60438-P,  ESP2015-70646-C2-1-R, AYA2017-84185-P, ESP2017-83921-C2-1-R, PID2019-110610RB-C21, PID2020-120514GB-I00, PID2019-110614GB-C21,
IACA13-3E-2336, IACA15-BE-3707, EQC2018-004918-P, the Severo Ochoa Programs SEV-2015-0548 and CEX2019-000920-S, the
Maria de Maeztu Program MDM-2017-0765 and by the Consolider-Ingenio project CSD2010-00064 (EPI: Exploring
the Physics of Inflation).
We acknowledge support from the ACIISI, Consejeria de Economia, Conocimiento y Empleo del Gobierno de Canarias and the European Regional Development Fund (ERDF) under grant with reference ProID2020010108.
This project has received funding from the European Union's Horizon 2020 research and innovation program under
grant agreement number 687312 (RADIOFOREGROUNDS).

%-----------------------------
%--> Others acknowledgements
We acknowledge the use of data provided by the Centre d'Analyse 
de Donn\'ees Etendues (CADE), a service of IRAP-UPS/CNRS 
\citep[http://cade.irap.omp.eu,][]{CADE}. This research has made use 
of the SIMBAD database, operated at CDS, Strasbourg, France~\citep{simbad}.
Some of the results in this paper have been derived using the 
healpy and \textsc{HEALPix} package~\citep{healpix,healpix2}. 
We have also used {\tt scipy}~\citep{scipy}, 
{\tt emcee}~\citep{2013PASP..125..306F}, {\tt numpy}~\citep{numpy}, 
{\tt matplotlib}~\citep{matplotlib}, {\tt corner}~\citep{corner} 
and {\tt astropy}~\citep{astropy1, astropy2} python packages.

%%%%%%%%%%%%%%%%%%%%%%%%%%%%%%%%%%%%%%%%%%%%%%%%%%
\section*{Data Availability}

The QUIJOTE-MFI wide survey maps are available in the QUIJOTE collaboration webpage$^{\ref{fn:QUIJOTE_webpage}}$ (\url{https://research.iac.es/proyecto/quijote}).
The ancillary maps used in this work are publicly available in databases such as
LAMBDA$^{\ref{fn:lambda}}$, PLA$^{\ref{fn_pla}}$ and
CADE$^{\ref{fn:cade}}$, as explained in Sect.~\ref{sec:data}.

%The main data used in this work, the QUIJOTE-MFI Wide Survey data, are available in the QUIJOTE collaboration webpage$^{\ref{fn:QUIJOTE_webpage}}$ (\url{https://research.iac.es/proyecto/quijote}).
%
%As was presented in section~\ref{sec:data}, the ancillary
%data are publicly available in databases such as LAMBDA$^{\ref{fn:lambda}}$,
%PLA$^{\ref{fn_pla}}$ and CADE$^{\ref{fn:cade}}$.

%The inclusion of a Data Availability Statement is a requirement for articles published in MNRAS. Data Availability Statements provide a standardised format for readers to understand the availability of data underlying the research results described in the article. The statement may refer to original data generated in the course of the study or to third-party data analysed in the article. The statement should describe and provide means of access, where possible, by linking to the data or providing the required accession numbers for the relevant databases or DOIs.


\bibliographystyle{mnras}
\bibliography{bibtex_SNR_quijote} % if your bibtex file is called example.bib

%\begin{thebibliography}{99}
%\end{thebibliography}

%%%%%%%%%%%%%%%%%%%%%%%%%%%%%%%%%%%%%%%%%%%%%%%%%%

%%%%%%%%%%%%%%%%% APPENDICES %%%%%%%%%%%%%%%%%%%%%

\clearpage \pagebreak
\setcounter{page}{1}
\section*{ONLINE MATERIAL}
\appendix

%-----------------------
%%---------- Sub-section
\section{Flux densities and images of our SNRs sample}
\label{appendix:fluxes_images_snr}

\begin{figure*}
\begin{center}
\includegraphics[trim = 0.38cm 0.25cm 0.25cm 0.15cm,clip=true,height= 2.75 cm]{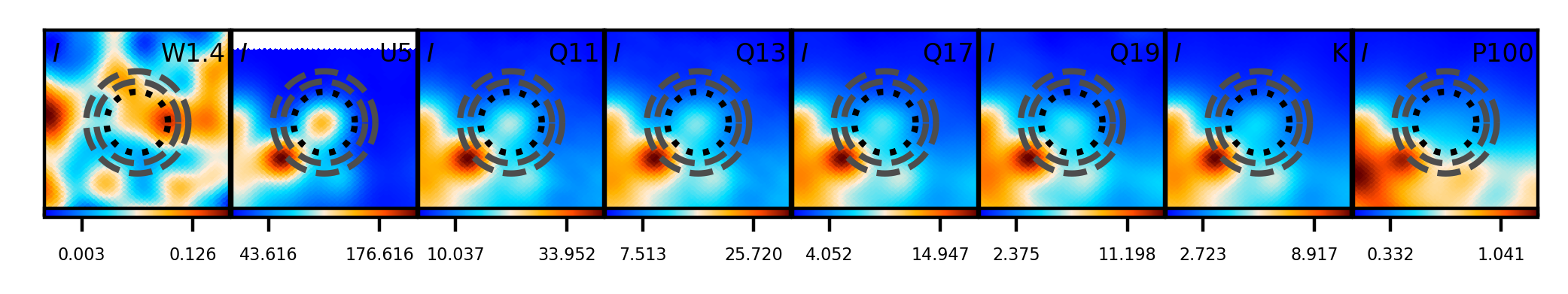}\\
\includegraphics[trim = 0.38cm 0.25cm 0.25cm 0.15cm,clip=true,height= 2.75 cm]{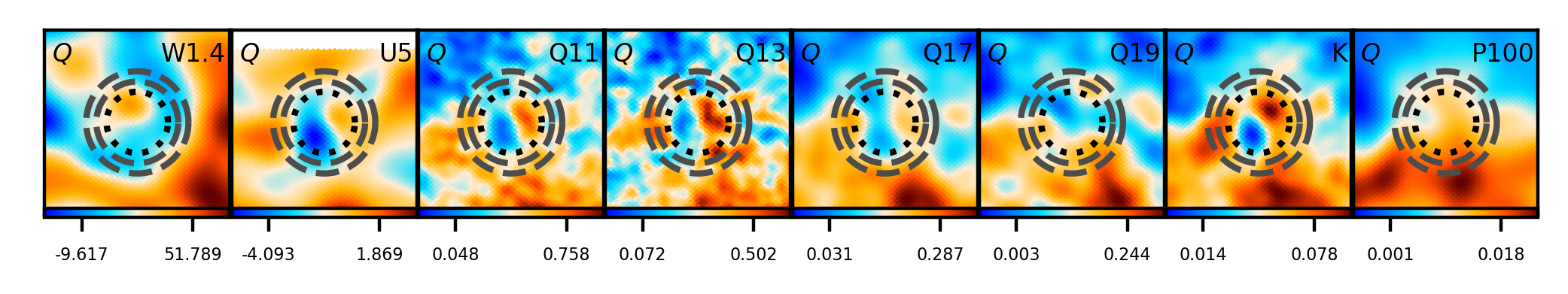}\\
\includegraphics[trim = 0.38cm 0.25cm 0.25cm 0.15cm,clip=true,height= 2.75 cm]{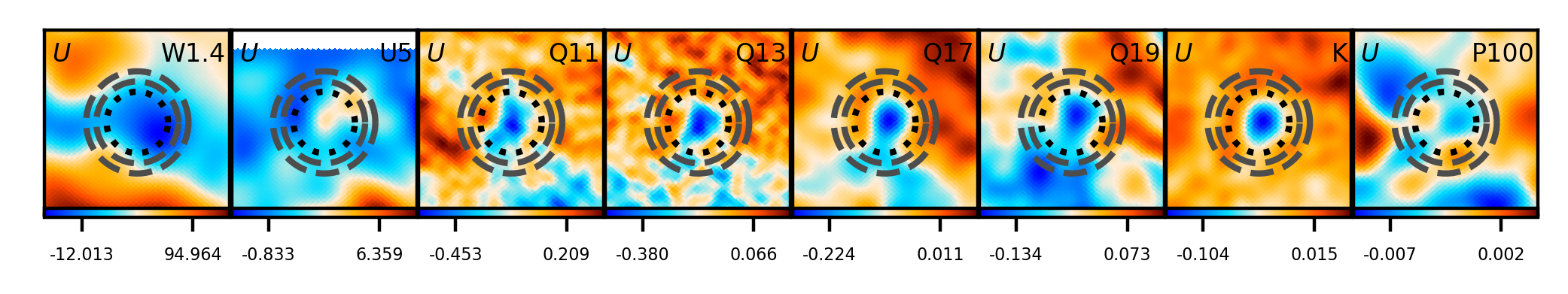}\\
\includegraphics[trim = 0.38cm 0.25cm 0.25cm 0.15cm,clip=true,height= 2.75 cm]{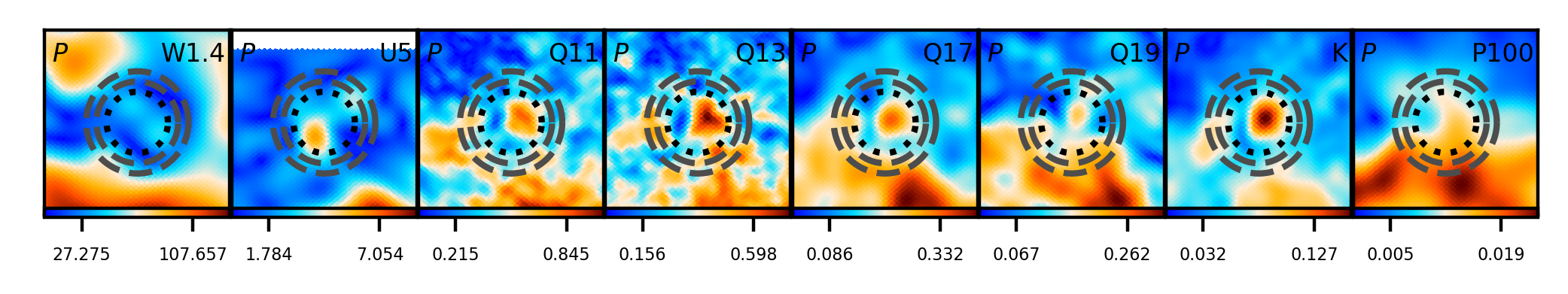}\\
\caption{Intensity and polarization maps of CTB\,80 
from the Urumqi, QUIJOTE, WMAP and \textit{Planck} 
data. Maps are smoothed to an angular resolution of 1\,deg, and centred on $l,b = 68.8^\circ,2.7^\circ$ with a field of view of $6.0^\circ \times 6.0^\circ$ (Galactic coordinate system).}
\label{ima:ctb80}
\end{center}
\end{figure*}

\begin{figure*}
\begin{center}
\includegraphics[trim = 0.38cm 0.25cm 0.25cm 0.15cm,clip=true,height= 2.75 cm]{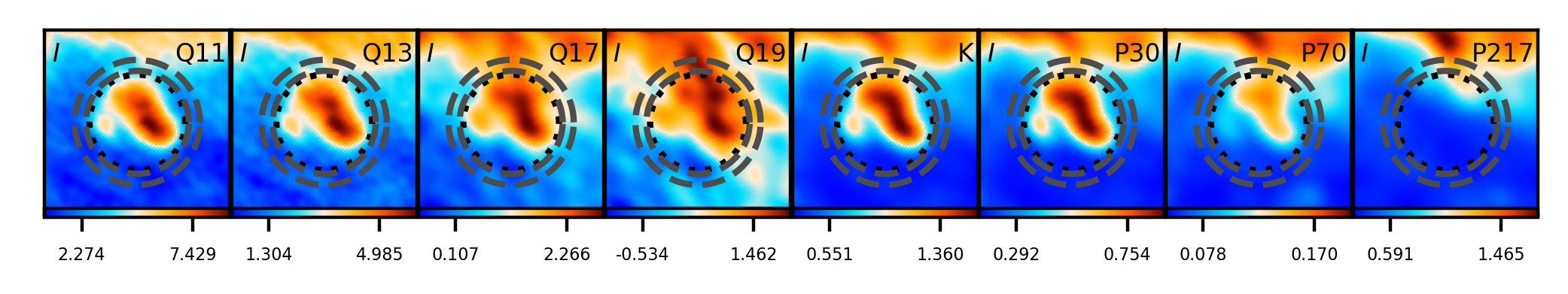}\\
\includegraphics[trim = 0.38cm 0.25cm 0.25cm 0.15cm,clip=true,height= 2.75 cm]{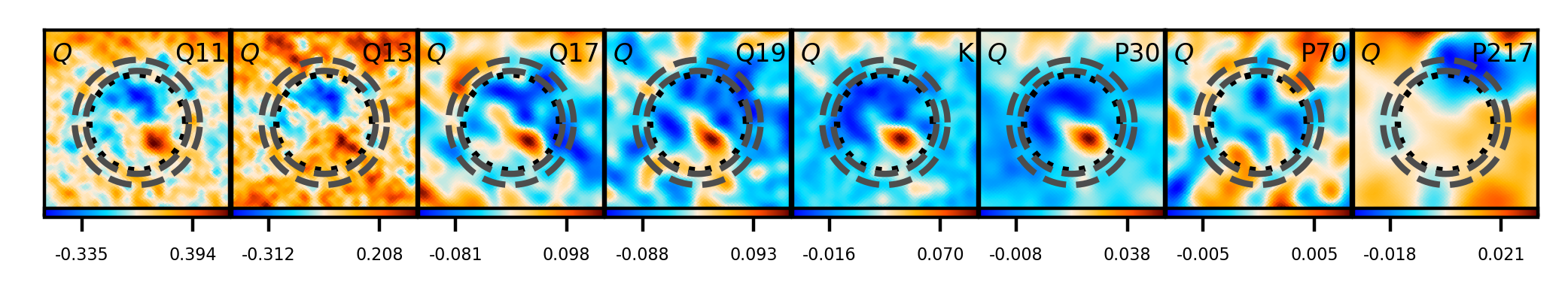}\\
\includegraphics[trim = 0.38cm 0.25cm 0.25cm 0.15cm,clip=true,height= 2.75 cm]{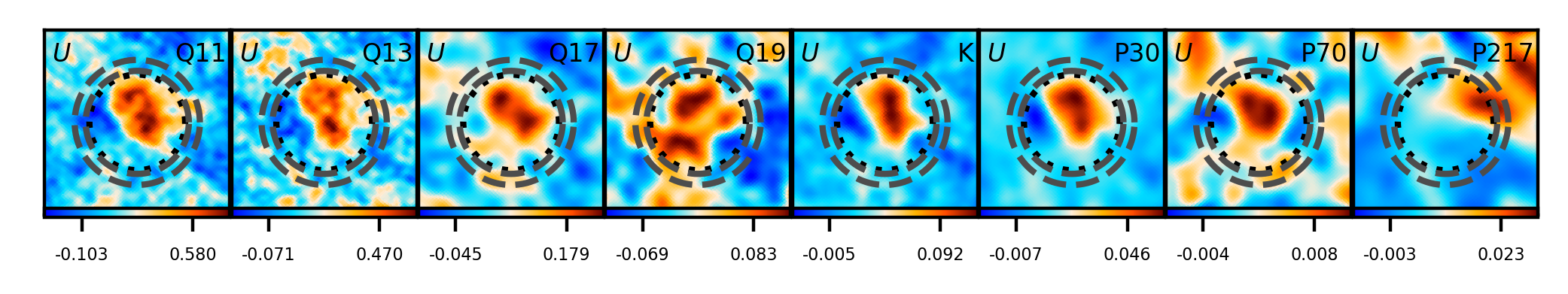}\\
\includegraphics[trim = 0.38cm 0.25cm 0.25cm 0.15cm,clip=true,height= 2.75 cm]{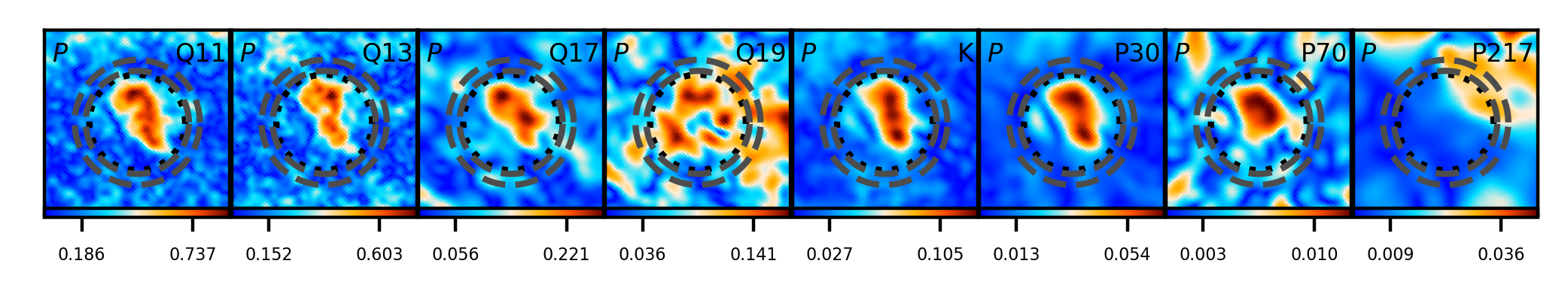}\\
\caption{Intensity and polarization maps of Cygnus Loop from the QUIJOTE, WMAP and \textit{Planck} data. Maps are smoothed to an angular resolution of 1\,deg, and centred on $l,b = 73.9^\circ,-8.8^\circ$ with a field of view of $8.4^\circ \times 8.4^\circ$ (Galactic coordinate system).
.}
\label{ima:cygnusloop}
\end{center}
\end{figure*}

\begin{figure*}
\begin{center}
\includegraphics[trim = 0.38cm 0.25cm 0.25cm 0.15cm,clip=true,height= 2.75 cm]{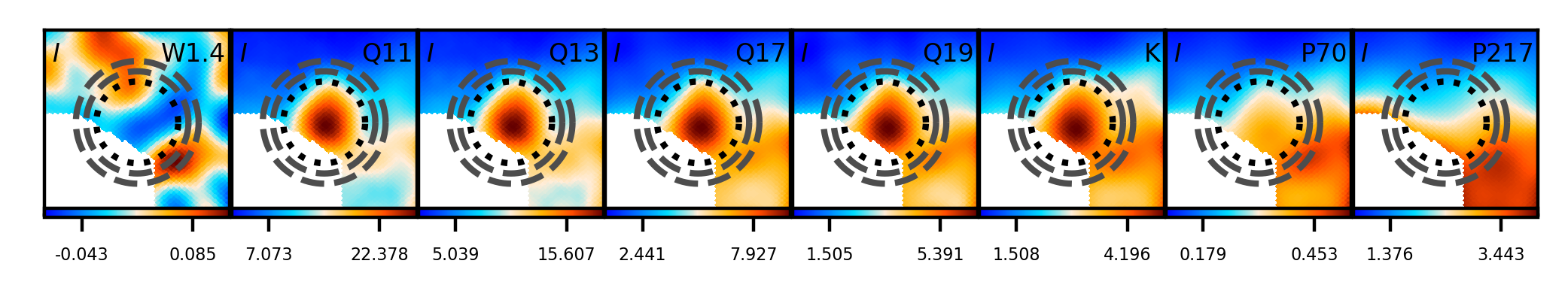}\\
\includegraphics[trim = 0.38cm 0.25cm 0.25cm 0.15cm,clip=true,height= 2.75 cm]{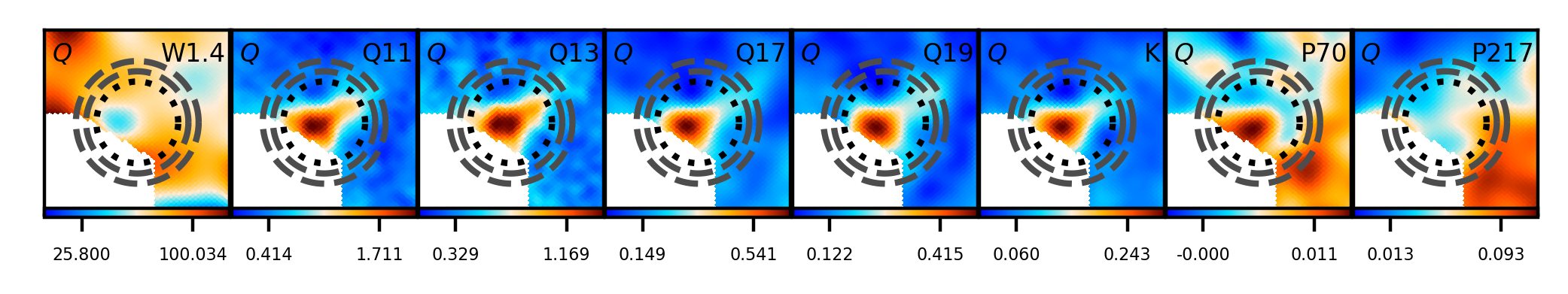}\\
\includegraphics[trim = 0.38cm 0.25cm 0.25cm 0.15cm,clip=true,height= 2.75 cm]{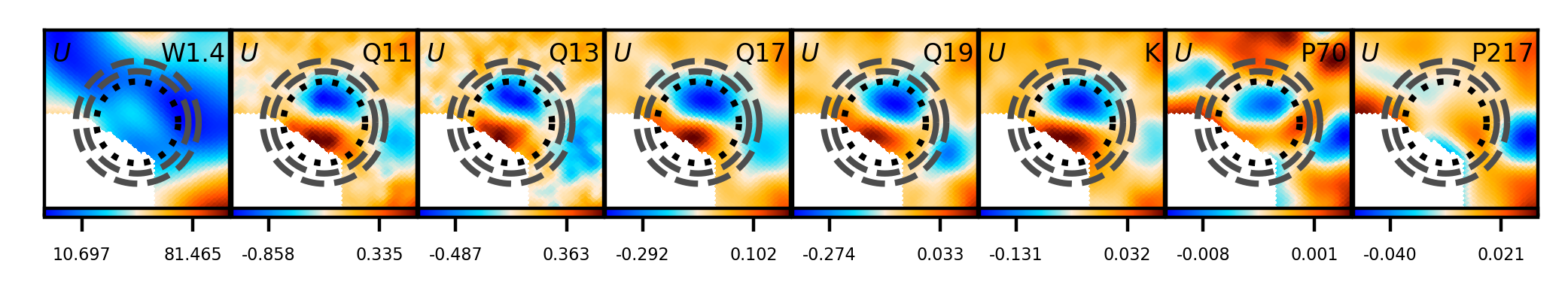}\\
\includegraphics[trim = 0.38cm 0.25cm 0.25cm 0.15cm,clip=true,height= 2.75 cm]{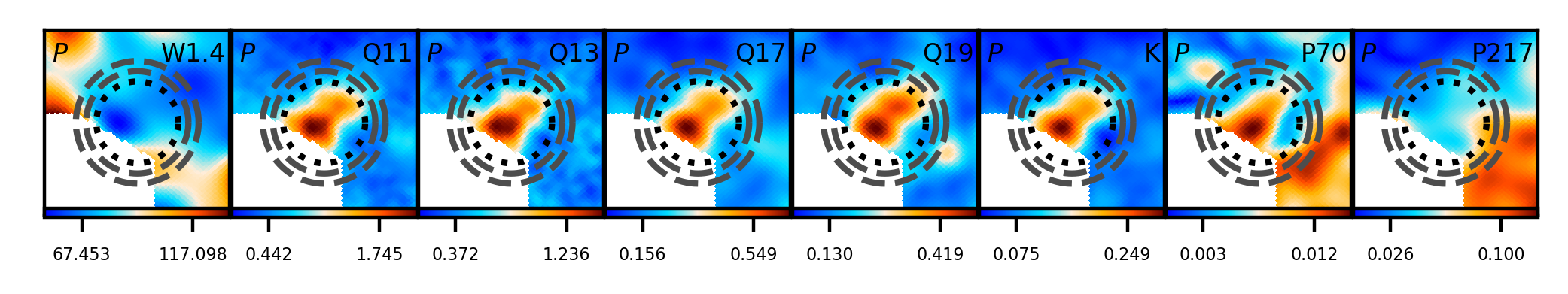}\\
\caption{Intensity and polarization maps of HB\,21 
from the QUIJOTE, WMAP and \textit{Planck} data. Maps 
are smoothed to an angular resolution of 1\,deg, and 
centred on $l,b = 89.0^\circ,4.7^\circ$ with a field 
of view of $6.0^\circ \times 6.0^\circ$ (Galactic coordinate system).}
\label{ima:hb21}
\end{center}
\end{figure*}

\begin{figure*}
\begin{center}
\includegraphics[trim = 0.38cm 0.25cm 0.25cm 0.15cm,clip=true,height= 2.75 cm]{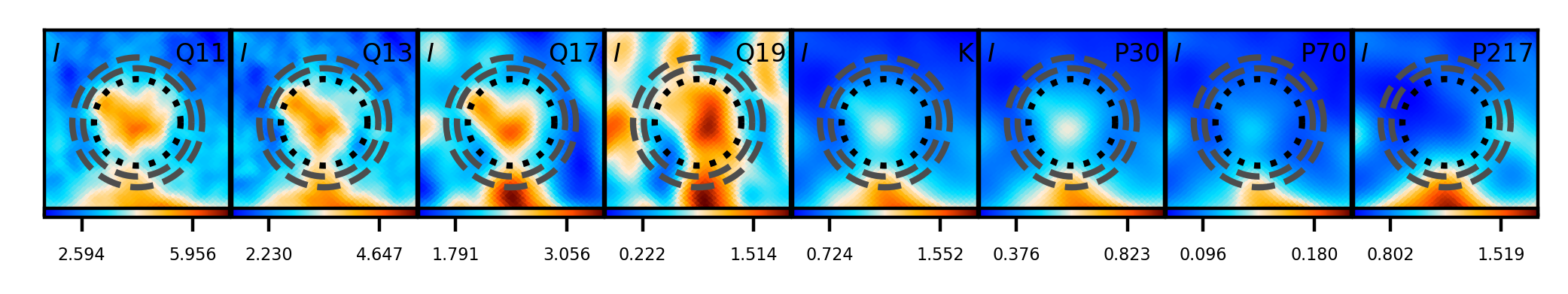}\\
\includegraphics[trim = 0.38cm 0.25cm 0.25cm 0.15cm,clip=true,height= 2.75 cm]{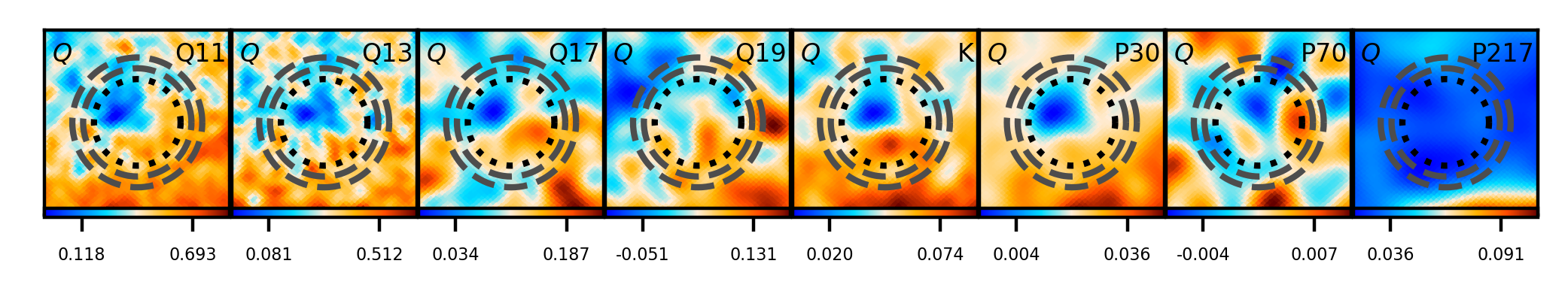}\\
\includegraphics[trim = 0.38cm 0.25cm 0.25cm 0.15cm,clip=true,height= 2.75 cm]{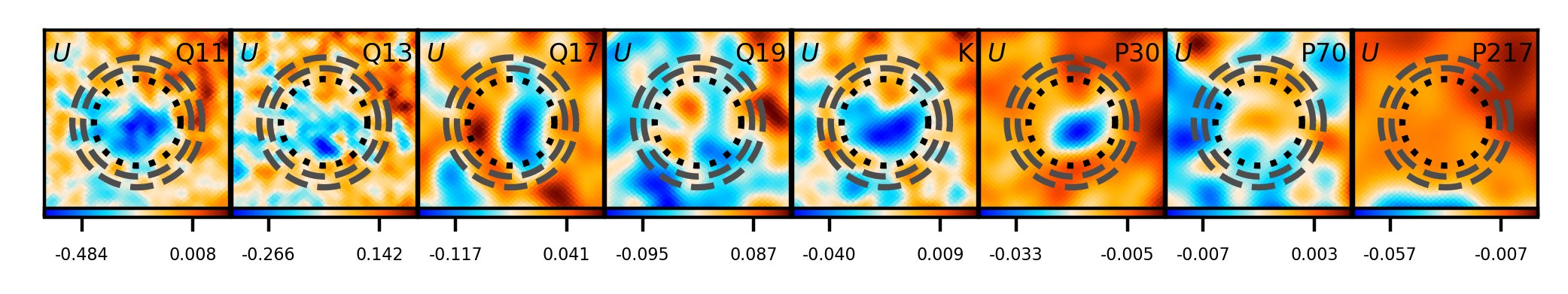}\\
\includegraphics[trim = 0.38cm 0.25cm 0.25cm 0.15cm,clip=true,height= 2.75 cm]{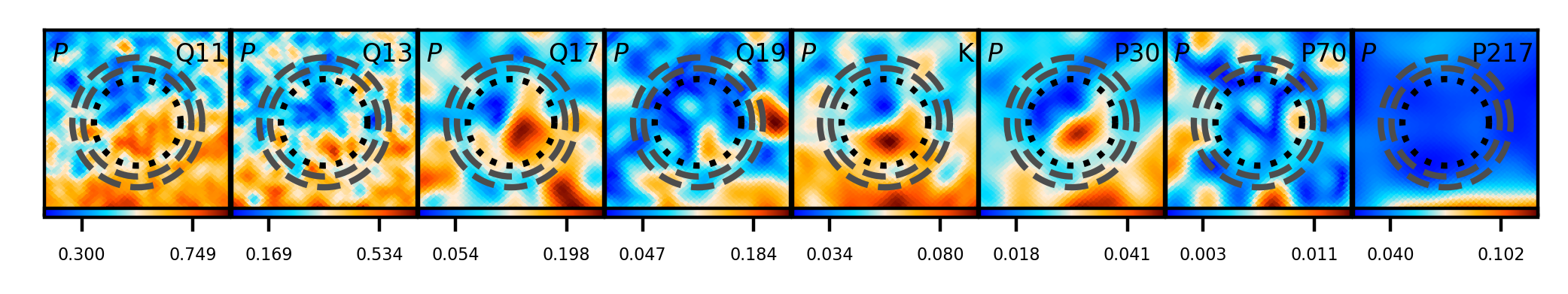}\\
\caption{Intensity and polarization maps of CTA\,1 
from the QUIJOTE, WMAP and \textit{Planck} data. Maps 
are smoothed to an angular resolution of 1\,deg, and 
centred on $l,b = 119.5^\circ,10.0^\circ$ with a field 
of view of $5.7^\circ \times 5.7^\circ$ (Galactic coordinate system).}
\label{ima:cta1}
\end{center}
\end{figure*}

\begin{figure*}
\begin{center}
\includegraphics[trim = 0.38cm 0.25cm 0.25cm 0.15cm,clip=true,height= 2.75 cm]{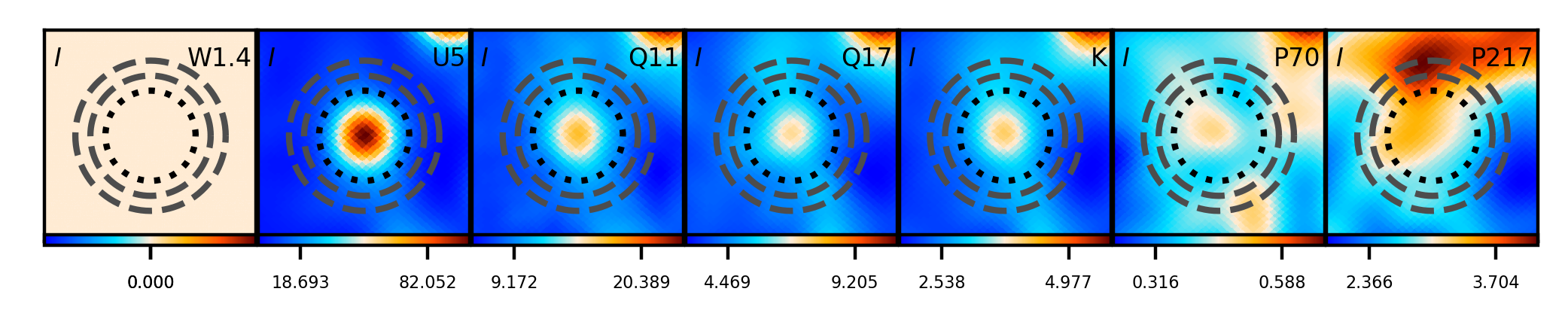}\\
\includegraphics[trim = 0.38cm 0.25cm 0.25cm 0.15cm,clip=true,height= 2.75 cm]{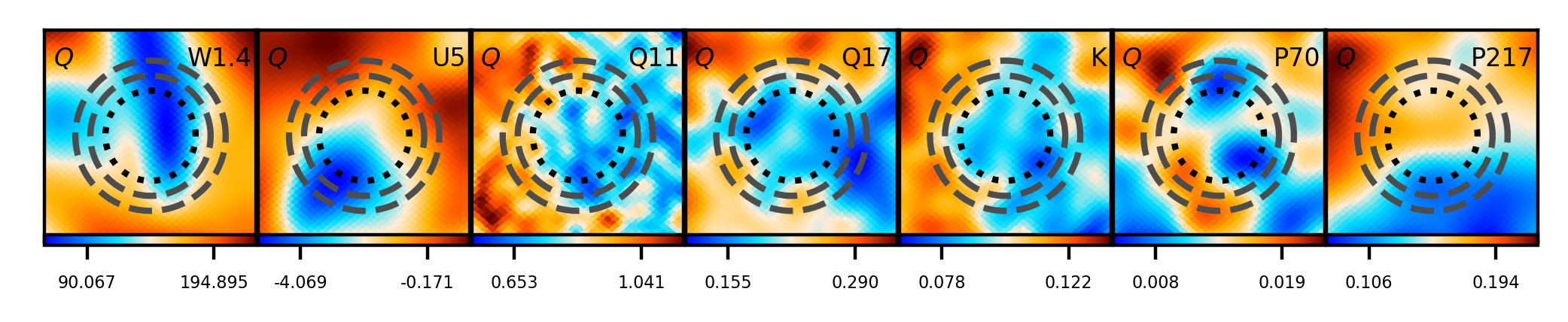}\\
\includegraphics[trim = 0.38cm 0.25cm 0.25cm 0.15cm,clip=true,height= 2.75 cm]{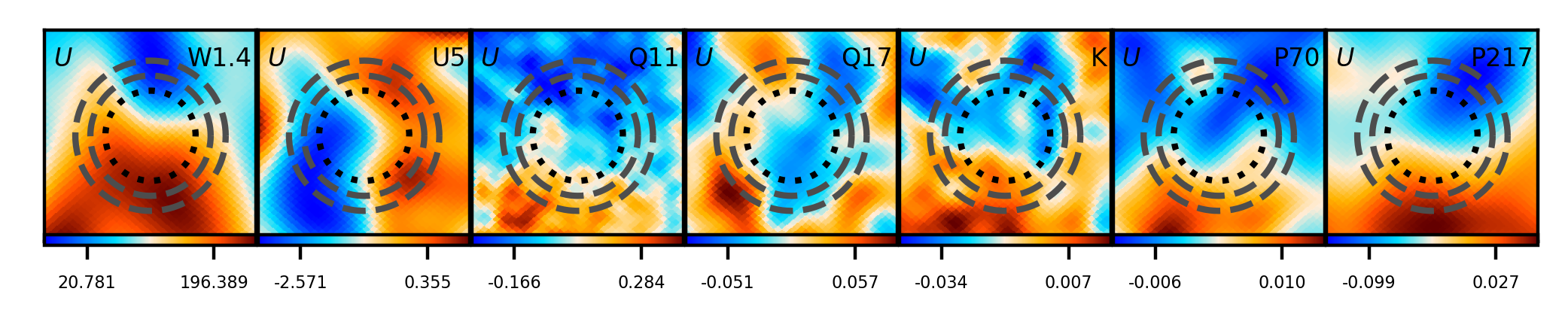}\\
\includegraphics[trim = 0.38cm 0.25cm 0.25cm 0.15cm,clip=true,height= 2.75 cm]{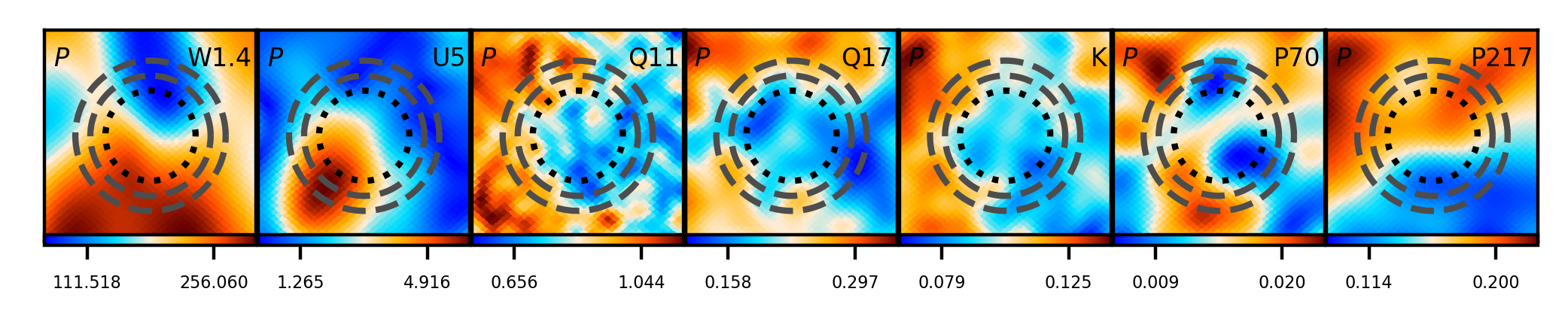}\\
\caption{Intensity and polarization maps of Tycho 
from the QUIJOTE, WMAP and \textit{Planck} data. 
Maps are smoothed to an angular resolution of 1\,deg, 
and centred on $l,b = 120.1^\circ,1.4^\circ$ with a 
field of view of $4.7^\circ \times 4.7^\circ$ (Galactic 
coordinate system).}
\label{ima:tycho}
\end{center}
\end{figure*}

\begin{figure*}
\begin{center}
\includegraphics[trim = 0.38cm 0.25cm 0.25cm 0.15cm,clip=true,height= 2.75 cm]{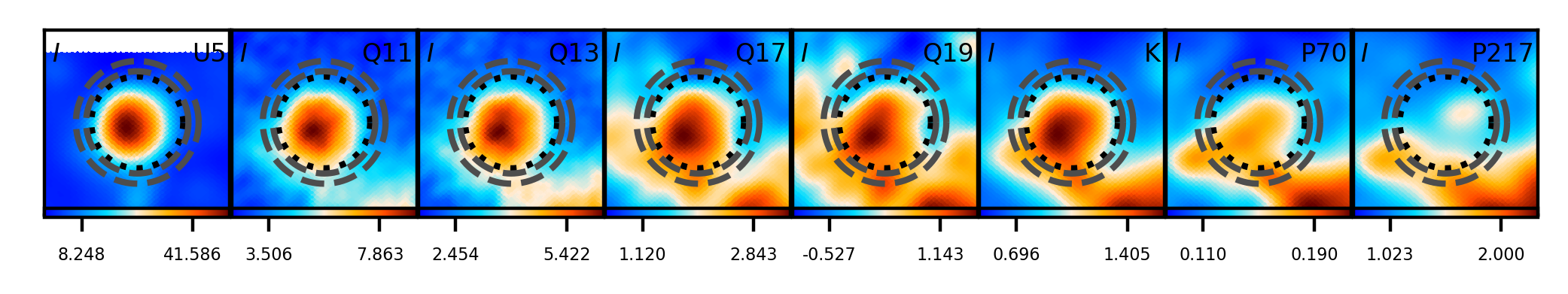}\\
\includegraphics[trim = 0.38cm 0.25cm 0.25cm 0.15cm,clip=true,height= 2.75 cm]{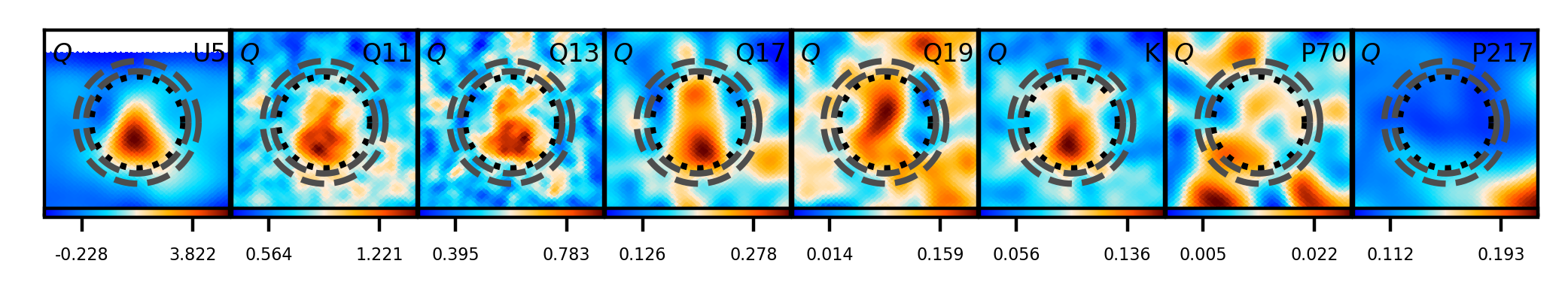}\\
\includegraphics[trim = 0.38cm 0.25cm 0.25cm 0.15cm,clip=true,height= 2.75 cm]{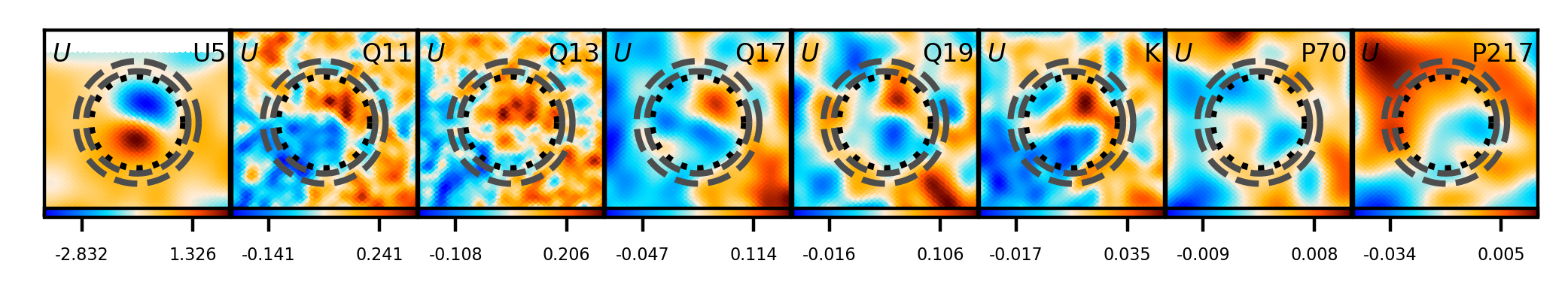}\\
\includegraphics[trim = 0.38cm 0.25cm 0.25cm 0.15cm,clip=true,height= 2.75 cm]{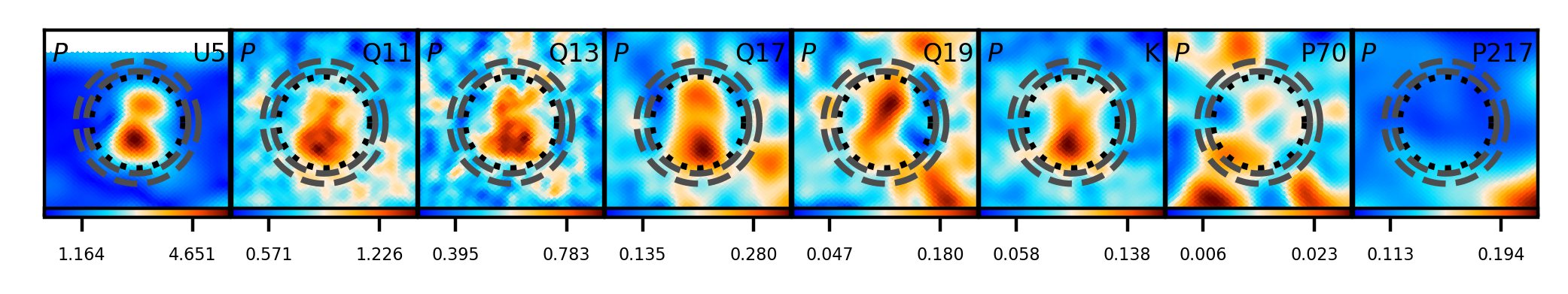}\\
\caption{Intensity and polarization maps of HB\,9 from 
the QUIJOTE, WMAP and \textit{Planck} data. Maps are 
smoothed to an angular resolution of 1\,deg, and centred 
on $l,b = 160.4^\circ,2.8^\circ$ with a field of view of 
$6.0^\circ \times 6.0^\circ$ (Galactic coordinate system).}
\label{ima:hb9}
\end{center}
\end{figure*}

This section provides the tables with the raw intensity 
and polarization flux densities (without colour correction) measured for our six SRNs, 
using the aperture photometry method described in 
section~\ref{sec:results_discussion}.
In addition, we also provide the intensity and polarization 
(Stokes $Q$, $U$ and $P$) maps of the six Galactic SNRs 
studied in this paper.
The selected maps range from the radio up  the far infrared in 
intensity and polarization, with the purpose of showing the 
morphology variations of SNRs across the whole frequency range.
In particular, we  consider maps from the QUIJOTE-MFI survey 
and other maps to be useful for the description of the SNRs (see 
details in Section~\ref{sec:results_discussion}) mainly from 
the WMAP and \textit{Planck}.
The apertures values (Table~\ref{tab:phot_params}) and masks 
used to recover the flux densities are drawn in each case.
Tables and maps are displayed for CTB\,80 (Table~\ref{tab:fluxes_ctb80} 
and Figure~\ref{ima:ctb80}), for Cygnus Loop 
(Table~\ref{tab:fluxes_cygnusloop} and Figure~\ref{ima:cygnusloop}), 
for HB\,21 (Table~\ref{tab:fluxes_hb21} and Figure~\ref{ima:hb21}), 
for CTA\,1 (Table~\ref{tab:fluxes_cta1} and Figure~\ref{ima:cta1}), 
for Tycho (Table~\ref{tab:fluxes_tycho} and Figure~\ref{ima:tycho}), 
and for HB\,9 (Table~\ref{tab:fluxes_hb9} and Figure~\ref{ima:hb9}).

%
%   Table and images for CTB80
%
\begin{table*}
	\centering
	\caption{Summary of recovered flux densities for CTB\,80. 
The $P$, $\Pi$ and $\gamma$ are obtained from the raw flux 
densities (second, third and fourth columns).% without colour correction.
}
	\begin{tabular}{l c c c c c c c} 
		\hline
		$\nu$ [GHz] & $I$ [Jy] & $Q$ [Jy] & $U$ [Jy]& $P$ [Jy] & $\gamma$ [deg]& $\Pi$ [\%]\\
		\hline
0.408 & 109$\pm$13 & - & - & - & - & -  \\ 
0.820 & 47.1$\pm$7.1 & - & - & - & - & -  \\ 
1.420 & 54.9$\pm$8.4 & - & - & - & - & -  \\ 
2.720 & 45.9$\pm$5.6 & - & - & - & - & -  \\ 
4.800 & 45.0$\pm$5.8 & -1.11$\pm$0.13 & 0.24$\pm$0.10 & 1.13$\pm$0.13 & -83.9$\pm$3 & 2.5$\pm$0.4 \\ 
11.1 & 21.4$\pm$3.4 & -0.16$\pm$0.15 & -0.78$\pm$0.17 & 0.78$\pm$0.17 & 50.8$\pm$6 & 3.6$\pm$1.0 \\ 
12.9 & 19.2$\pm$3.5 & 0.04$\pm$0.16 & -0.73$\pm$0.16 & 0.71$\pm$0.16 & 43.4$\pm$7 & 3.8$\pm$1.1 \\ 
16.8 & 15.5$\pm$3.5 & -0.33$\pm$0.23 & -0.72$\pm$0.20 & 0.76$\pm$0.21 & 57.3$\pm$8 & 5.1$^{1.8}_{1.9}$ \\ [0.5 ex]
18.8 & 11.7$\pm$3.5 & -0.38$\pm$0.22 & -0.80$\pm$0.37 & 0.85$\pm$0.36 & 57.7$\pm$12 & 7.7$^{3.4}_{3.6}$ \\ [0.5 ex]
22.8 & 13.9$\pm$3.6 & -0.02$\pm$0.06 & -0.67$\pm$0.13 & 0.67$\pm$0.13 & 45.9$\pm$6 & 5.1$^{1.3}_{1.4}$ \\ [0.5 ex]
28.4 & 11.7$\pm$3.4 & -0.07$\pm$0.07 & -0.57$\pm$0.09 & 0.57$\pm$0.09 & 48.5$\pm$5 & 5.2$^{1.6}_{1.4}$ \\ [0.5 ex]
32.9 & 11.8$\pm$3.3 & -0.09$\pm$0.13 & -0.53$\pm$0.11 & 0.52$\pm$0.11 & 49.8$\pm$6 & 4.8$\pm$1.6 \\ 
40.7 & 10.5$\pm$3.2 & -0.04$\pm$0.11 & -0.39$\pm$0.11 & 0.38$\pm$0.11 & 47.9$\pm$9 & 4.1$^{1.8}_{1.4}$ \\ [0.5 ex]
44.1 & 10.0$\pm$3.1 & 0.16$\pm$0.08 & -0.21$\pm$0.05 & 0.25$\pm$0.07 & 26.4$\pm$7 & 2.8$^{1.1}_{1.0}$ \\ [0.5 ex]
60.7 & 7.8$\pm$3.4 & 0.60$\pm$0.14 & -0.75$\pm$0.41 & 0.92$\pm$0.35 & 25.7$\pm$11 & 13.3$^{5.4}_{6.6}$ \\ [0.5 ex]
70.4 & 5.7$\pm$4.4 & 0.11$\pm$0.13 & -0.35$\pm$0.15 & 0.34$\pm$0.16 & 36.3$\pm$13 & $<$18.2 \\ 
93.5 & 4.3$\pm$8.3 & 0.51$\pm$0.69 & -0.65$\pm$0.78 & 0.59$\pm$1.04 & 25.9$\pm$50 & $<$115.9 \\ 
100 & 4$\pm$11 & 0.46$\pm$0.17 & -0.16$\pm$0.15 & 0.46$\pm$0.18 & 9.6$\pm$11 & $<$120.6 \\ 
143 & 4$\pm$23 & 2.18$\pm$0.56 & 0.15$\pm$0.36 & 2.16$\pm$0.57 & -2.0$\pm$8 & -  \\ 
217 & 0$\pm$75 & 7.76$\pm$2.30 & 0.87$\pm$1.38 & 7.68$\pm$2.33 & -3.2$\pm$9 & -  \\ 
353 & 8$\pm$280 & 35.04$\pm$11.13 & 2.60$\pm$6.44 & 34.54$\pm$11.30 & -2.1$\pm$9 & -  \\ 
545 & 36$\pm$829 & - & - & - & - & -  \\ 
857 & -298$\pm$2507 & - & - & - & - & -  \\ 
1249 & -1303$\pm$4998 & - & - & - & - & -  \\ 
2141 & 1938$\pm$7229 & - & - & - & - & -  \\ 
2997 & 2157$\pm$3779 & - & - & - & - & -  \\ 
\hline
	\end{tabular}
	\label{tab:fluxes_ctb80}
\end{table*}

%
%   Table and images for Cygnus
%
\begin{table*}
	\centering
	\caption{Summary of recovered flux densities for the Cygnus 
Loop. The $P$, $\Pi$ and $\gamma$ are obtained from the raw flux 
densities (second, third and fourth columns).% without colour correction.
}
	\begin{tabular}{l c c c c c c c} % four columns, alignment for each
		\hline
		$\nu$ [GHz] & $I$ [Jy] & $Q$ [Jy] & $U$ [Jy]& $P$ [Jy] & $\gamma$ [deg]& $\Pi$ [\%] \\
		\hline 
0.408 & 199$\pm$21 & - & - & - & - & -  \\ 
0.820 & 159$\pm$16 & - & - & - & - & -  \\ 
1.420 & 121$\pm$13 & - & - & - & - & -  \\ 
2.720 & 98$\pm$10 & - & - & - & - & -  \\ 
11.1 & 45.7$\pm$3.9 & -1.33$\pm$0.75 & 3.55$\pm$0.72 & 3.72$\pm$0.74 & 34.7$\pm$6 & 8.2$^{1.8}_{1.7}$ \\ [0.5 ex]
12.9 & 42.6$\pm$3.9 & -1.32$\pm$0.46 & 3.28$\pm$0.80 & 3.50$\pm$0.77 & 34.0$\pm$6 & 8.3$\pm$1.6 \\ 
16.8 & 30.1$\pm$3.8 & -0.31$\pm$0.58 & 2.37$\pm$0.62 & 2.32$\pm$0.64 & 41.3$\pm$8 & 7.8$^{2.4}_{2.2}$ \\ [0.5 ex]
18.8 & 21.3$\pm$8.4 & -0.12$\pm$0.65 & 2.11$\pm$0.44 & 2.01$\pm$0.46 & 43.4$\pm$7 & 10.5$^{4.3}_{4.8}$ \\ [0.5 ex]
22.8 & 34.0$\pm$2.2 & -0.44$\pm$0.20 & 1.76$\pm$0.18 & 1.80$\pm$0.18 & 38.0$\pm$3 & 5.3$\pm$0.7 \\ 
28.4 & 29.3$\pm$1.8 & -0.25$\pm$0.22 & 1.59$\pm$0.29 & 1.59$\pm$0.29 & 40.5$\pm$5 & 5.4$^{1.0}_{0.9}$ \\ [0.5 ex]
32.9 & 26.5$\pm$1.8 & -0.50$\pm$0.30 & 1.62$\pm$0.33 & 1.67$\pm$0.33 & 36.4$\pm$6 & 6.2$\pm$1.3 \\ 
40.7 & 24.6$\pm$1.6 & -0.73$\pm$0.34 & 1.71$\pm$0.28 & 1.83$\pm$0.29 & 33.4$\pm$5 & 7.5$^{1.4}_{1.3}$ \\ [0.5 ex]
44.1 & 23.9$\pm$1.6 & -0.77$\pm$0.42 & 0.64$\pm$0.34 & 0.93$\pm$0.42 & 19.9$\pm$13 & 3.9$^{1.8}_{1.6}$ \\ [0.5 ex]
60.7 & 21.0$\pm$1.8 & -0.79$\pm$0.65 & 1.56$\pm$0.59 & 1.63$\pm$0.65 & 31.6$\pm$11 & 7.7$^{3.4}_{3.1}$ \\ [0.5 ex]
70.4 & 20.2$\pm$2.1 & -1.14$\pm$0.75 & 1.22$\pm$0.65 & 1.52$\pm$0.77 & 23.5$\pm$14 & 7.3$^{4.1}_{3.5}$ \\ [0.5 ex]
93.5 & 17.6$\pm$4.2 & 4.20$\pm$2.29 & 2.95$\pm$2.84 & 4.45$\pm$2.86 & -17.5$\pm$18 & $<$54.8 \\ 
100 & 30.8$\pm$5.5 & -0.89$\pm$0.44 & 1.49$\pm$0.42 & 1.68$\pm$0.44 & 29.6$\pm$7 & 5.6$^{1.7}_{1.8}$ \\ [0.5 ex]
143 & 35$\pm$13 & 0.28$\pm$0.76 & 1.61$\pm$0.86 & 1.46$\pm$0.96 & -40.1$\pm$19 & 4.5$^{3.2}_{2.6}$ \\ [0.5 ex]
217 & 124$\pm$54 & 1.28$\pm$4.15 & 7.36$\pm$3.11 & 6.38$\pm$3.69 & -40.1$\pm$17 & 5.8$^{4.0}_{3.7}$ \\ [0.5 ex]
353 & 453$\pm$240 & 7.60$\pm$13.85 & 28.29$\pm$15.32 & 26.01$\pm$17.15 & -37.5$\pm$19 & $<$15.5 \\ 
545 & 1562$\pm$526 & - & - & - & - & -  \\ 
857 & 4616$\pm$1528 & - & - & - & - & -  \\ 
1249 & 9033$\pm$2978 & - & - & - & - & -  \\ 
2141 & 9989$\pm$3315 & - & - & - & - & -  \\ 
2997 & 2569$\pm$1271 & - & - & - & - & -  \\ 
		\hline
	\end{tabular}
	\label{tab:fluxes_cygnusloop}
\end{table*}

%
%   Table and images for HB21
%
\begin{table*}
	\centering
	\caption{Summary of recovered flux densities for HB\,21. 
The $P$, $\Pi$ and $\gamma$ are obtained from the raw flux densities 
(second, third and fourth columns).% without colour correction.
}
	\begin{tabular}{l c c c c c c c} % four columns, alignment for each
		\hline
		$\nu$ [GHz] & $I$ [Jy] & $Q$ [Jy] & $U$ [Jy]& $P$ [Jy] & $\gamma$ [deg]& $\Pi$ [\%] \\
		\hline 
0.408 & 255$\pm$26 & - & - & - & - & -  \\ 
0.820 & 191$\pm$20 & - & - & - & - & -  \\ 
1.420 & 189$\pm$20 & - & - & - & - & -  \\ 
11.1 & 55.6$\pm$3.5 & 3.33$\pm$0.22 & -0.08$\pm$0.22 & 3.32$\pm$0.22 & 0.7$\pm$2 & 5.9$\pm$0.6 \\ 
12.9 & 50.1$\pm$3.3 & 2.73$\pm$0.32 & -0.01$\pm$0.17 & 2.72$\pm$0.32 & 0.1$\pm$3 & 5.4$\pm$0.6 \\ 
16.8 & 42.1$\pm$3.1 & 2.05$\pm$0.23 & -0.03$\pm$0.26 & 2.03$\pm$0.23 & 0.4$\pm$3 & 4.8$\pm$0.7 \\ 
18.8 & 36.0$\pm$2.8 & 1.99$\pm$0.16 & -0.11$\pm$0.22 & 1.98$\pm$0.16 & 1.6$\pm$2 & 5.4$\pm$0.7 \\ 
22.8 & 32.6$\pm$2.4 & 1.53$\pm$0.09 & -0.45$\pm$0.11 & 1.59$\pm$0.09 & 8.2$\pm$2 & 4.9$\pm$0.5 \\ 
28.4 & 26.6$\pm$2.1 & 1.23$\pm$0.09 & -0.38$\pm$0.13 & 1.28$\pm$0.09 & 8.6$\pm$2 & 4.8$\pm$0.6 \\ 
32.9 & 24.2$\pm$2.0 & 1.00$\pm$0.13 & -0.13$\pm$0.12 & 1.00$\pm$0.13 & 3.7$\pm$4 & 4.1$\pm$0.6 \\ 
40.7 & 21.0$\pm$1.9 & 1.02$\pm$0.16 & -0.46$\pm$0.16 & 1.11$\pm$0.16 & 12.1$\pm$4 & 5.3$\pm$0.9 \\ 
44.1 & 19.6$\pm$2.2 & 0.67$\pm$0.13 & -0.23$\pm$0.11 & 0.70$\pm$0.13 & 9.5$\pm$5 & 3.6$^{0.8}_{0.7}$ \\ [0.5 ex]
60.7 & 14.8$\pm$4.2 & -0.03$\pm$0.56 & 0.08$\pm$0.40 & 0.04$\pm$0.84 & - & $<$6.7 \\ 
70.4 & 13.0$\pm$5.6 & 0.28$\pm$0.26 & -0.30$\pm$0.27 & 0.33$\pm$0.33 & 23.5$\pm$28 & $<$7.4 \\ 
93.5 & 13.7$\pm$8.8 & 3.30$\pm$1.17 & 0.12$\pm$1.78 & 2.84$\pm$1.36 & -1.0$\pm$14 & $<$62.8 \\ 
100 & 5$\pm$13 & 0.56$\pm$0.17 & 0.17$\pm$0.30 & 0.51$\pm$0.21 & -8.4$\pm$12 & $<$74.3 \\ 
143 & 31$\pm$18 & 0.83$\pm$0.78 & 0.70$\pm$0.96 & 0.80$\pm$1.16 & -20.1$\pm$41 & $<$9.7 \\ 
217 & 116$\pm$62 & 2.57$\pm$3.43 & 2.09$\pm$3.72 & 2.19$\pm$5.36 & -19.6$\pm$70 & $<$8.8 \\ 
353 & 640$\pm$271 & 10.63$\pm$16.21 & 5.12$\pm$17.04 & 7.13$\pm$27.08 & - & $<$6.4 \\ 
545 & 1896$\pm$892 & - & - & - & - & -  \\ 
857 & 4291$\pm$2485 & - & - & - & - & -  \\ 
1249 & 9658$\pm$4521 & - & - & - & - & -  \\ 
2141 & 8733$\pm$5144 & - & - & - & - & -  \\ 
2997 & -616$\pm$2056 & - & - & - & - & -  \\  
		\hline
	\end{tabular}
	\label{tab:fluxes_hb21}
\end{table*}

%
%   Table and images for CTA1
%
\begin{table*}
	\centering
	\caption{Summary of recovered flux densities for CTA\,1. 
The $P$, $\Pi$ and $\gamma$ are obtained from the raw flux densities 
(second, third and fourth columns).% without colour correction.
}
	\begin{tabular}{l c c c c c c c c} % four columns, alignment for each
		\hline
		$\nu$ [GHz] & $I$ [Jy] & $Q$ [Jy] & $U$ [Jy]& $P$ [Jy] & $\gamma$ [deg]& $\Pi$ [\%]\\
		\hline 
0.408 & 44.6$\pm$4.9 & - & - & - & - & -  \\ 
0.820 & 33.1$\pm$3.5 & - & - & - & - & -  \\ 
1.420 & 27.5$\pm$2.9 & - & - & - & - & -  \\ 
11.1 & 9.0$\pm$1.4 & -0.48$\pm$0.18 & -1.05$\pm$0.32 & 1.14$\pm$0.31 & -32.7$\pm$8 & 12.4$^{3.2}_{3.6}$ \\ [0.5 ex]
12.9 & 7.6$\pm$1.8 & -0.37$\pm$0.21 & -0.91$\pm$0.15 & 0.96$\pm$0.16 & -33.9$\pm$5 & 13.4$^{3.8}_{3.9}$ \\ [0.5 ex]
16.8 & 6.1$\pm$1.6 & 0.20$\pm$0.32 & -0.64$\pm$0.25 & 0.60$\pm$0.29 & 36.3$\pm$14 & 10.4$^{6.0}_{5.6}$ \\ [0.5 ex]
18.8 & 9.3$\pm$2.1 & -0.22$\pm$0.65 & 0.40$\pm$0.40 & 0.28$\pm$0.76 & 30.6$\pm$77 & $<$13.5 \\ 
22.8 & 6.5$\pm$1.4 & -0.16$\pm$0.07 & -0.47$\pm$0.09 & 0.49$\pm$0.09 & -35.6$\pm$5 & 7.7$^{1.9}_{2.1}$ \\ [0.5 ex]
28.4 & 5.8$\pm$1.2 & -0.34$\pm$0.05 & -0.44$\pm$0.08 & 0.55$\pm$0.07 & -26.1$\pm$4 & 9.8$^{2.2}_{2.3}$ \\ [0.5 ex]
32.9 & 5.7$\pm$1.4 & -0.27$\pm$0.12 & -0.36$\pm$0.15 & 0.43$\pm$0.15 & -26.6$\pm$10 & 7.9$\pm$3.2 \\ 
40.7 & 5.0$\pm$1.3 & 0.13$\pm$0.18 & 0.22$\pm$0.32 & 0.18$\pm$0.40 & -29.7$\pm$63 & $<$12.8 \\ 
44.1 & 5.1$\pm$1.6 & 0.35$\pm$0.13 & -0.15$\pm$0.18 & 0.34$\pm$0.15 & 11.6$\pm$13 & 7.3$^{4.1}_{3.8}$ \\ [0.5 ex]
60.7 & 5.4$\pm$3.1 & -0.36$\pm$0.42 & -0.98$\pm$0.37 & 0.96$\pm$0.41 & -34.9$\pm$12 & $<$46.4 \\ 
70.4 & 5.2$\pm$4.0 & -0.05$\pm$0.33 & 0.21$\pm$0.25 & 0.13$\pm$0.43 & - & $<$16.1 \\ 
93.5 & 6.0$\pm$6.1 & -0.91$\pm$1.29 & -1.17$\pm$1.58 & 1.03$\pm$2.12 & -26.1$\pm$59 & $<$96.3 \\ 
100 & 4$\pm$11 & -0.30$\pm$0.19 & -0.12$\pm$0.30 & 0.23$\pm$0.29 & -10.9$\pm$36 & $<$69.0 \\ 
143 & 5$\pm$16 & 0.08$\pm$0.86 & -0.03$\pm$0.89 & 0.04$\pm$1.72 & - & $<$39.7 \\ 
217 & 4$\pm$59 & -2.27$\pm$3.03 & 0.84$\pm$2.98 & 1.53$\pm$4.77 & 10.2$\pm$89 & -  \\ 
353 & -50$\pm$215 & -6.45$\pm$13.60 & 0.99$\pm$17.88 & 3.47$\pm$25.77 & - & -  \\ 
545 & -231$\pm$210 & - & - & - & - & -  \\ 
857 & -657$\pm$609 & - & - & - & - & -  \\ 
1249 & -585$\pm$1086 & - & - & - & - & -  \\ 
2141 & 281$\pm$1238 & - & - & - & - & -  \\ 
2997 & 250$\pm$545 & - & - & - & - & -  \\  
		\hline
	\end{tabular}
	\label{tab:fluxes_cta1}
\end{table*}

%
%   Table and images for Tycho
%
\begin{table*}
	\centering
	\caption{Summary of recovered flux densities for Tycho. 
The $P$, $\Pi$ and $\gamma$ are obtained from the raw flux 
densities (second, third and fourth columns).% without colour correction.
}
	\begin{tabular}{l c c c c c c c c} % four columns, alignment for each
		\hline
		$\nu$ [GHz] & $I$ [Jy] & $Q$ [Jy] & $U$ [Jy]& $P$ [Jy] & $\gamma$ [deg]& $\Pi$ [\%]\\
		\hline 
0.408 & 88.6$\pm$8.9 & - & - & - & - & -  \\ 
0.820 & 58.1$\pm$5.9 & - & - & - & - & -  \\ 
1.420 & 49.7$\pm$5.0 & - & - & - & - & -  \\ 
4.800 & 25.9$\pm$2.6 & -1.10$\pm$0.16 & -0.87$\pm$0.13 & 1.40$\pm$0.15 & -19.2$\pm$3 & 5.4$\pm$0.8 \\ 
11.1 & 12.7$\pm$1.3 & -0.25$\pm$0.13 & -0.04$\pm$0.13 & 0.22$\pm$0.15 & -4.5$\pm$19 & 1.6$^{1.2}_{0.9}$ \\ [0.5 ex]
12.9 & 11.9$\pm$1.4 & -0.23$\pm$0.17 & -0.26$\pm$0.15 & 0.31$\pm$0.18 & -24.2$\pm$16 & 2.6$^{1.6}_{1.4}$ \\ [0.5 ex]
16.8 & 8.8$\pm$1.5 & -0.35$\pm$0.15 & -0.18$\pm$0.19 & 0.35$\pm$0.18 & -13.6$\pm$15 & 4.1$\pm$2.2 \\ 
18.8 & 7.6$\pm$2.1 & -0.01$\pm$0.16 & 0.09$\pm$0.28 & 0.05$\pm$0.49 & - & -  \\ 
22.8 & 9.9$\pm$1.2 & -0.15$\pm$0.09 & -0.14$\pm$0.09 & 0.19$\pm$0.10 & -21.5$\pm$15 & 1.8$\pm$1.0 \\ 
28.4 & 7.7$\pm$1.1 & -0.11$\pm$0.06 & -0.04$\pm$0.06 & 0.10$\pm$0.07 & -10.0$\pm$19 & 1.2$^{0.9}_{0.7}$ \\ [0.5 ex]
32.9 & 6.7$\pm$1.1 & 0.03$\pm$0.04 & -0.01$\pm$0.10 & 0.02$\pm$0.09 & - & $<$2.0 \\ 
40.7 & 5.6$\pm$1.3 & 0.01$\pm$0.08 & 0.17$\pm$0.16 & 0.15$\pm$0.18 & -43.3$\pm$34 & $<$6.2 \\ 
44.1 & 5.0$\pm$1.5 & -0.08$\pm$0.19 & 0.17$\pm$0.11 & 0.13$\pm$0.18 & 32.4$\pm$40 & $<$8.6 \\ 
60.7 & 3.1$\pm$2.3 & -0.22$\pm$0.21 & -0.42$\pm$0.28 & 0.42$\pm$0.30 & -31.2$\pm$20 & $<$43.6 \\ 
70.4 & 1.4$\pm$2.9 & -0.29$\pm$0.31 & -0.28$\pm$0.18 & 0.33$\pm$0.31 & -22.0$\pm$27 & $<$176.0 \\ 
93.5 & -0.3$\pm$4.8 & -0.77$\pm$0.59 & -0.96$\pm$0.84 & 1.04$\pm$0.89 & -25.6$\pm$24 & -  \\ 
100 & 7$\pm$11 & 0.18$\pm$0.55 & -0.90$\pm$0.29 & 0.77$\pm$0.36 & 39.4$\pm$14 & $<$58.5 \\ 
143 & 16$\pm$15 & 0.54$\pm$1.84 & -2.06$\pm$1.03 & 1.56$\pm$1.50 & 37.7$\pm$28 & $<$43.4 \\ 
217 & 136$\pm$80 & 0.80$\pm$7.16 & -8.57$\pm$3.41 & 6.34$\pm$4.70 & 42.3$\pm$21 & $<$16.7 \\ 
353 & 582$\pm$284 & 0.14$\pm$33.84 & -37.44$\pm$14.84 & 26.64$\pm$20.85 & 44.9$\pm$22 & 6.3$^{5.6}_{3.7}$ \\ [0.5 ex]
545 & 1835$\pm$374 & - & - & - & - & -  \\ 
857 & 4696$\pm$1013 & - & - & - & - & -  \\ 
1249 & 7732$\pm$1765 & - & - & - & - & -  \\ 
2141 & 8326$\pm$1797 & - & - & - & - & -  \\ 
2997 & 2779$\pm$688 & - & - & - & - & -  \\ 
		\hline
	\end{tabular}
	\label{tab:fluxes_tycho}
\end{table*}

%
%   Table and images for HB9
%
\begin{table*}
	\centering
	\caption{Summary of recovered flux densities for HB\,9. 
The $P$, $\Pi$ and $\gamma$ are obtained from the raw flux densities 
(second, third and fourth columns).% without colour correction.
}
	\begin{tabular}{l c c c c c c} % four columns, alignment for each
		\hline
		$\nu$ [GHz] & $I$ [Jy] & $Q$ [Jy] & $U$ [Jy]& $P$ [Jy] & $\gamma$ [deg]& $\Pi$ [\%]\\
		\hline 
0.408 & 102$\pm$10 & - & - & - & - & -  \\ 
0.820 & 79.3$\pm$8.0 & - & - & - & - & -  \\ 
1.420 & 58.9$\pm$5.9 & - & - & - & - & -  \\ 
4.800 & 35.0$\pm$3.5 & 2.12$\pm$0.23 & -0.82$\pm$0.10 & 2.27$\pm$0.22 & 10.6$\pm$3 & 6.6$\pm$0.8 \\ 
11.1 & 21.2$\pm$1.4 & 1.84$\pm$0.24 & 0.05$\pm$0.18 & 1.83$\pm$0.24 & -0.8$\pm$4 & 8.7$^{1.1}_{1.2}$ \\ [0.5 ex]
12.9 & 19.5$\pm$1.5 & 1.43$\pm$0.22 & 0.39$\pm$0.24 & 1.46$\pm$0.22 & -7.6$\pm$4 & 7.4$^{1.3}_{1.4}$ \\ [0.5 ex]
16.8 & 16.4$\pm$1.9 & 0.69$\pm$0.31 & -0.08$\pm$0.21 & 0.66$\pm$0.32 & 3.3$\pm$14 & 4.0$^{1.8}_{1.7}$ \\ [0.5 ex]
18.8 & 16.2$\pm$3.6 & 0.63$\pm$0.48 & -0.33$\pm$0.18 & 0.66$\pm$0.47 & 13.8$\pm$20 & $<$7.8 \\ 
22.8 & 14.6$\pm$1.4 & 0.79$\pm$0.15 & -0.01$\pm$0.08 & 0.79$\pm$0.15 & 0.4$\pm$5 & 5.3$\pm$0.9 \\ 
28.4 & 12.7$\pm$1.4 & 0.88$\pm$0.12 & 0.02$\pm$0.08 & 0.88$\pm$0.12 & -0.7$\pm$4 & 6.9$\pm$1.1 \\ 
32.9 & 12.5$\pm$1.8 & 0.42$\pm$0.18 & -0.23$\pm$0.19 & 0.44$\pm$0.20 & 14.3$\pm$13 & 3.5$^{1.7}_{1.6}$ \\ [0.5 ex]
40.7 & 12.3$\pm$2.5 & 0.21$\pm$0.25 & -0.10$\pm$0.42 & 0.13$\pm$0.50 & - & $<$6.1 \\ 
44.1 & 12.2$\pm$2.9 & 0.08$\pm$0.29 & 0.50$\pm$0.23 & 0.43$\pm$0.27 & -40.5$\pm$18 & 3.2$^{2.5}_{2.1}$ \\ [0.5 ex]
60.7 & 12.6$\pm$5.7 & -0.66$\pm$0.59 & -0.24$\pm$0.53 & 0.53$\pm$0.77 & 80.0$\pm$41 & $<$14.3 \\ 
70.4 & 14.9$\pm$7.3 & 0.35$\pm$0.49 & -0.64$\pm$0.38 & 0.59$\pm$0.50 & 30.7$\pm$24 & $<$11.8 \\ 
93.5 & 15$\pm$12 & 0.32$\pm$1.60 & -0.27$\pm$1.79 & 0.22$\pm$3.27 & - & $<$24.3 \\ 
100 & 30$\pm$12 & -0.50$\pm$0.71 & -0.42$\pm$0.39 & 0.48$\pm$0.82 & 70.0$\pm$49 & $<$5.4 \\ 
143 & 36$\pm$21 & -1.18$\pm$1.83 & -2.10$\pm$1.30 & 1.88$\pm$1.85 & 59.7$\pm$28 & $<$17.3 \\ 
217 & 147$\pm$49 & -5.04$\pm$5.08 & -11.23$\pm$4.85 & 11.28$\pm$5.34 & 57.1$\pm$14 & 9.1$^{5.4}_{3.8}$ \\ [0.5 ex]
353 & 580$\pm$230 & -20.30$\pm$19.15 & -45.34$\pm$23.20 & 45.71$\pm$24.54 & 57.1$\pm$15 & 9.0$^{5.4}_{4.8}$ \\ [0.5 ex]
545 & 1726$\pm$380 & - & - & - & - & -  \\ 
857 & 4388$\pm$1000 & - & - & - & - & -  \\ 
1249 & 5424$\pm$1685 & - & - & - & - & -  \\ 
2141 & 5269$\pm$1625 & - & - & - & - & -  \\ 
2997 & 892$\pm$515 & - & - & - & - & -  \\ 
		\hline
	\end{tabular}
	\label{tab:fluxes_hb9}
\end{table*}

%-----------------------
%%---------- Sub-section
\section{Synchotron models with curved spectra}
\label{appendix:curved_models}

As mentioned in section~\ref{sec:metodology_sed_modelling}, 
we selected a smooth broken power-law function (SBPL, 
equation~\ref{eq:model_brokenpowerlaw_smooth}) to characterize the 
curved spectra observed for CTB\,80 and HB\,21.
Here, we present the parametric representation of other 
functions to describe the curved spectrum observed in SNRs 
based on a power-law function (with spectral index $\alpha$, 
equation~\ref{eq:model_powerlaw}) that dominates at lower frequencies, 
and then decreases further at high frequencies.

The power law with an exponential cut-off function has been considered in 
previous SED analyses in the microwave range 
\citep[e.g.][]{2013ApJ...779..179P,2021MNRAS.500.5177L,mfi_widesurvey_w51}. 
This behaviour is expected when the energy losses exceed the rate 
of energy gain from shock acceleration \citep{1984A&A...137..185W}, 
although the break frequency is commonly at high energies.
This function can be written as:
\begin{equation}
S_{\rm PLcutoff}(\nu;A_{\rm syn},\alpha,\nu_{\rm c}) = A_{\rm syn} \; 
\left( \dfrac{\nu}{\nu_{\rm ref}} \right)^{\alpha} \; \mathrm{e}^{-\dfrac{\nu - \nu_{\rm ref}}{\nu_c}},  
\label{eq:model_cutoff}
\end{equation}
where $\nu_c$ is the cut-off frequency and $A_{\rm syn}$ is the 
amplitude normalised to the reference frequency $\nu_{\rm ref}$. 
Alternatively, the Rolloff function also describes a smoothed
 decay from low to high frequencies. \cite{2011ApJS..192...19W} used 
this function to model the SED of 3C\,58, which is parametrized as 
follows:
\begin{equation}
S_{\rm Rf}(\nu;A_{\rm syn},b,c,d) = A_{\rm syn} 
\left( \dfrac{\nu}{\nu_{\rm ref}} \right)^b  \: \dfrac{1 +c}{ 1+ c  
 \left( \nu / \nu_{\rm ref} \right)^{d}} , 
\label{eq:model_rolloff}  
\end{equation}
where $b$, $c$ and $d$ define the shape at high and low frequencies 
(it is also normalised to $\nu_{\rm ref}$).

\begin{figure}
\begin{center}
\includegraphics[trim = 0cm 0cm 0cm 0.cm,clip=true,width= 8.5 cm]{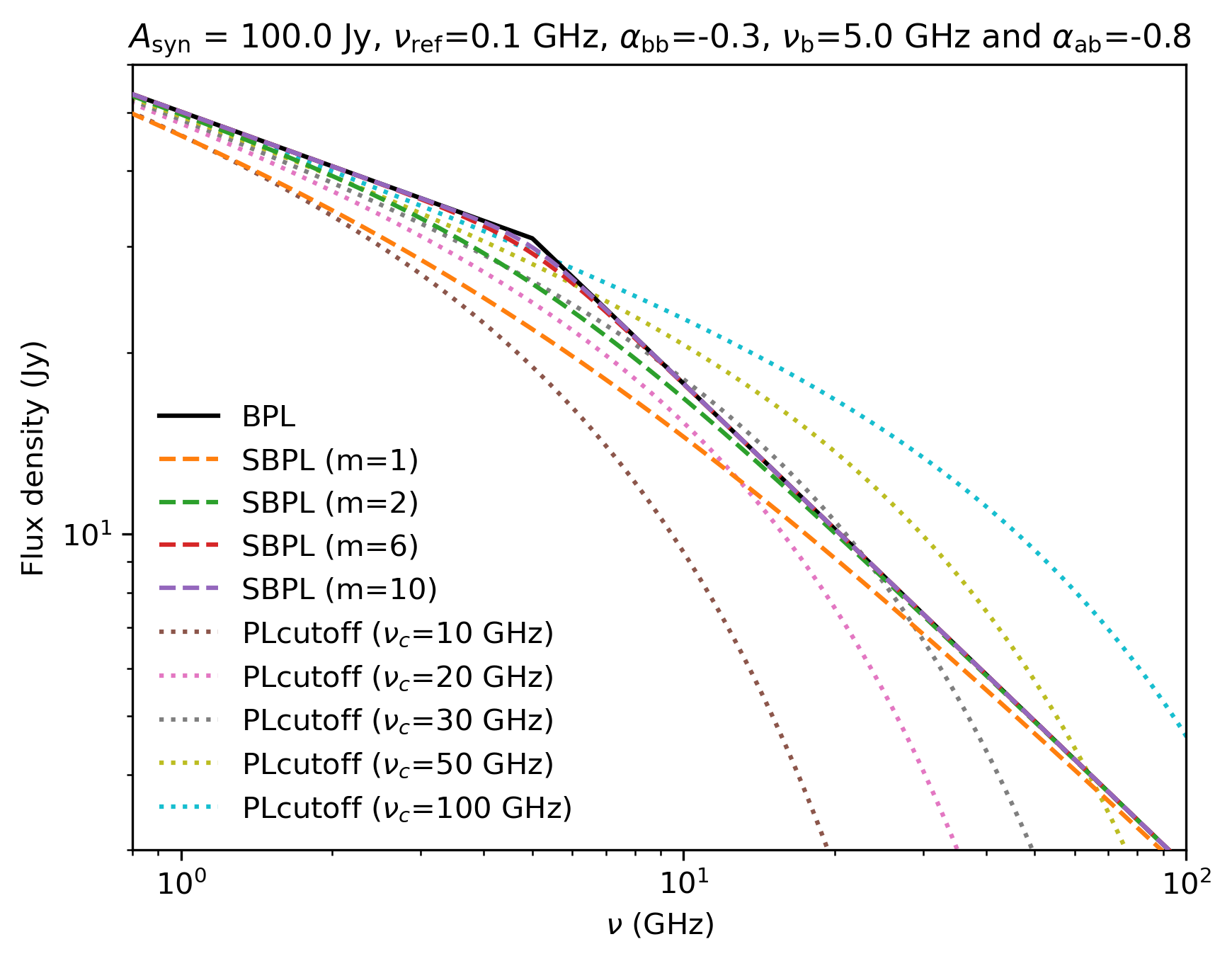}
\caption{Parametric models that were explored to characterize 
the curved spectra of CTB\,80 and HB\,21. These are displayed 
according to the broken power law (solid line, 
equation~\ref{eq:model_brokenpowerlaw_sharp}), to the smooth broken 
power law (dashed lines, equation~\ref{eq:model_brokenpowerlaw_smooth}), 
and to the power law with exponential cut-off (dotted lines, 
equation~\ref{eq:model_cutoff}). 
Overall, the amplitude is fixed to $A_{\rm syn} = 100$\,Jy at 
$\nu_{\rm ref}=0.1$\,GHz, with a radio spectral index of 
$\alpha = \alpha_{\rm bb} = -0.3$, a break frequency 
$\nu_{\rm b}=5$\,GHz and a spectral index above the break 
frequency of $\alpha_{\rm ab}=-0.8$. For the $S_{\rm PLcutoff}$
 model, the coloured dotted lines identify the cut-off frequency 
$\nu_{\rm c}$. Similarly, the coloured dashed lines represent the 
$m$ value considered for the $S_{\rm SBPL}$ model.}
\label{fig:sed_powerlaw_examples}
\end{center}
\end{figure}

Since the observed spectra of CTB\,80 and HB\,21 follow a clear
 power-law behaviour in the QUIJOTE-MFI frequency range, we also 
explored the broken power-law function with a sudden spectral index 
variation at $\nu_{\rm b}$: 
\begin{equation}
S_{\rm BPL}(\nu;A_{\rm syn},\alpha_{\rm bb},\alpha_{\rm ab},\nu_{\rm b}) = A_{\rm syn}
  \begin{cases}
    \left( \dfrac{\nu}{\nu_{\rm ref}} \right)^{\alpha_{\rm bb}} &, \quad \nu \leq \nu_b\\ \\
   \left( \dfrac{\nu_{\rm b}}{\nu_{\rm ref}} \right)^{\alpha_{\rm bb}}  \left( \dfrac{\nu}{\nu_{\rm b}} \right)^{\alpha_{\rm ab}}  &, \quad \nu > \nu_b,
  \end{cases}
\label{eq:model_brokenpowerlaw_sharp}
\end{equation}
where $\alpha_{\rm bb}$ and $\alpha_{\rm ab}$ are respectively 
the spectral indices below and above of the break frequency.
We choose this function as the final parametric shape 
to model our curved spectra, although with a smoothed spectral 
break (SBPL, equation~\ref{eq:model_brokenpowerlaw_smooth}).
For the SBPL spectral model, the parameter $m$ drives the level of 
smoothing around the break frequency (see 
Figure~\ref{fig:sed_powerlaw_examples}). In this figure, we observe 
that the SBPL model converges to the BPL model as $m$ increases. 
For instance, $m=10$ provides a close concordance between the SBPL 
and BPL spectral models.

%-----------------------
%%---------- Sub-section
\section{Measurements from the literature: CTB\,80 and HB\,21}
\label{appendix:ctb80_hb21}
For HB\,21 and CTB\,80, we use literature measurements of 
the intensity flux densities for frequencies below the Q11 
band, which allow us to improve the fit of their broken spectrum.
For CTB\,80, we take into account flux densities between 0.01 
and 11\,GHz compiled by \cite{Gao2011A&A...529A.159}, with 
frequencies up to 10.2\,GHz \citep{1983PASJ...35..437S}.
We excluded measurements at frequencies lower than 200\,MHz 
to avoid modelling the low-frequency turnover due to the 
free--free thermal absorption \citep[proposed by][]{2005A&A...440..171C}.
For HB\,21, we consider the flux densities used by 
\cite{2006A&A...457.1081K} and \cite{2013ApJ...779..179P} covering 
frequencies below 5\,GHz  \citep{Gao2011A&A...529A.159}.
In general, the literature information must be treated carefully in 
the SED fitting, mainly, owing to the wide range of systematic effects 
affecting each measurement, which also encompass the differences of the beam 
resolution of maps and methodologies for recover the flux densities.
Therefore, we consider conservative uncertainties for those flux 
densities taken from the literature, which consists in increasing 
the uncertainty up to 10\% of the intensity signal when the literature 
error bars are less than this threshold of 10\%.
Table~\ref{tab:literature_ctb80_hb21} shows the flux densities from 
the literature used in the SED fits of CTB\,80 (section~\ref{sec:snr_ctb80}) 
and HB\,21 (section~\ref{sec:snr_hb21}).
\begin{table}
	\centering
	\caption{Integrated radio flux densities of CTB\,80 and 
HB\,21 taken from the literature.}
	\begin{tabular}{c c | c c} % four columns, alignment for each
		\hline
		$\nu$ [MHz] & $I$ [Jy] & $\nu$ [MHz] & $I$ [Jy]\\
		\hline 
\multicolumn{4}{c}{\textbf{CTB\,80}}\\ 
240 & 106.0 $\pm$ 16.0 &1400 & 75.0 $\pm$ 8.0\\ 
324 & 91.0 $\pm$ 10.0 &1410 & 62.0 $\pm$ 9.0\\ 
408 & 67.0 $\pm$ 13.0 &1420 & 80.0 $\pm$ 10.0\\ 
408 & 67.6 $\pm$ 10.5 &1720 & 66.0 $\pm$ 6.6\\ 
408 & 83.0 $\pm$ 15.0 &2695 & 51.0 $\pm$ 5.1\\ 
750 & 127.0 $\pm$ 15.0 &2700 & 42.2 $\pm$ 4.2\\ 
750 & 75.0 $\pm$ 8.0 &4750 & 44.0 $\pm$ 4.4\\ 
1000 & 121.0 $\pm$ 15.0 &10200 & 18.8 $\pm$ 1.9\\ 
1380 & 56.0 $\pm$ 6.0 & - & -\\
\hline
\multicolumn{4}{c}{\textbf{HB\,21}}\\ 
38 & 790.0 $\pm$ 200.0 &960 & 170.0 $\pm$ 60.0\\ 
81 & 530.0 $\pm$ 120.0 &1420 & 183.0 $\pm$ 18.3\\ 
158 & 280.0 $\pm$ 80.0 &1430 & 190.0 $\pm$ 30.0\\ 
178 & 375.0 $\pm$ 40.0 &2700 & 148.0 $\pm$ 16.0\\ 
178 & 440.0 $\pm$ 100.0 &2965 & 160.0 $\pm$ 30.0\\ 
400 & 190.0 $\pm$ 60.0 &4170 & 160.0 $\pm$ 40.0\\ 
408 & 259.0 $\pm$ 25.9 &4800 & 107.3 $\pm$ 10.7\\ 
611 & 308.0 $\pm$ 60.0 &5000 & 103.0 $\pm$ 18.0\\ 
863 & 228.0 $\pm$ 22.8 & - & -\\ 
		\hline
	\end{tabular}
	\label{tab:literature_ctb80_hb21}
\end{table}



%%%%%%%%%%%%%%%%%%%%%%%%%%%%%%%%%%%%%%%%%%%%%%%%%%

% Don't change these lines
\bsp	% typesetting comment
\label{lastpage}
\end{document}